# Electron Trapping by Polar Molecules in Alkane Liquids: Cluster Chemistry in Dilute Solution. [1]

Ilya A. Shkrob [a)] and Myran C. Sauer, Jr.

*Chemistry Division , Argonne National Laboratory, 9700 S. Cass Ave, Argonne, IL 60439*

**Abstract**

Experimental observations are presented on condensed-phase analogs of gas phase dipole-bound anions and negatively charged clusters of polar molecules. Both monomers and small clusters of such molecules can reversibly trap conduction band electrons in dilute alkane solutions. The dynamics and energetics of this trapping have been studied using pulse radiolysis - transient absorption spectroscopy and time-resolved photoconductivity. Binding energies, thermal detrapping rates, and absorption spectra of excess electrons attached to monomer and multimer solute traps are obtained and possible structures for these species are discussed. "Dipole coagulation" (stepwise growth of the solute cluster around the cavity electron) predicted by Mozumder in 1972 is observed. Acetonitrile monomer is shown to solvate the electron by its methyl group, just like the alkane solvent does. The electron is dipole-bound to the CN group; the latter points away from the cavity. The resulting negatively charged species has a binding energy of 0.4 eV and absorbs in the infrared. Molecules of straight-chain aliphatic alcohols solvate the excess electron by their OH groups; at equilibrium, the predominant electron trap is a trimer or a tetramer; the binding energy of this solute trap is ca. 0.8 eV. Trapping by smaller clusters is opposed by the entropy which drives the equilibrium towards the electron in a *solvent* trap. For alcohol monomers, the trapping does not occur; a slow proton transfer reaction occurs instead. For acetonitrile monomer, the trapping is favored energetically but the thermal detachment is rapid (ca. 1 ns). Our study suggests that a composite cluster anion consisting of a few polar molecules imbedded in an alkane "matrix" might be the closest gas phase analog to the core of solvated electron in a *neat* polar liquid.

[a] Author to whom correspondence should be addressed; electronic mail: shkrob@anl.gov.

# 1. Introduction.

The way in which the excess electron localizes in a dielectric fluid strongly depends on the nature of the fluid. [1] Henceforward, only liquids constituted of polyatomic molecules that have no electron affinity are considered. In many such liquids, the excess electron occupies a void (the solvation cavity) lined by polar (or polarizable) groups of the solvent molecules; the spreading of the electron density onto the solvent is minor. In water and aliphatic alcohols, the ground state (*s*-) electron is confined in a small, nearly spherical cavity lined by the solvent hydroxyl groups. [2-4] This structure ($e^-_{solv}$) is stabilized by Coulomb attraction of the electron to positive charges on hydroxyl protons. [4] The binding energy of the solvated electron in such liquids is 1-2 eV [2,4] and thermal re-excitation of this electron to the conduction band (CB) does not occur (though such a process may occur for excited-state *p*-electrons). [5] Naturally, nonpolar liquids localize the excess electron differently, as permanent dipoles in polar groups are absent. In saturated hydrocarbons, the electrons are trapped in large cavities of ca. 7 Å in diameter (the so-called "electron bubbles"). [6,7] The electron is stabilized by interaction with polarized C-H and C-C bonds in 6-8 methyl groups that form the solvation cavity; [8] additional stabilization is provided by electron exchange [9] and/or sharing of the electron density with the solvent molecules. [10] A similar arrangement exists for the solvated electron in the polar liquid acetonitrile (*MeCN*). [10-13] Although the dipole moment of acetonitrile is large (3.9-4.1 D vs. 1.6-1.9 D for alcohols), [12] the positive charge resides on the CN carbon which is not accessible to $e^-_{solv}$. The CN groups point *away* from the cavity; the latter is lined by methyl groups, just like in alkane liquids. [10,13] Since the $(MeCN)_2^-$ anion in neat acetonitrile is more stable by 450 meV than $e^-_{solv}$, there is a rapid equilibrium between these two electron states. [10,11,14] Similar equilibria exist for other liquids where more than one type of electron is present at any time. In a typical alkane, the binding energy $E_t$ of the electron is only 180-200 meV and thermal excitation to the CB readily occurs at room temperature.[6,7] Thus, the electron spends some time in a quasifree state at the bottom of the CB (for which the drift mobility is as high as 10-100 cm$^2$/Vs),[6,7,15,16] whereas most of the time it dwells in a trapped state (for which the mobility $\mu_f$ is only 10$^{-3}$-10$^{-2}$ cm$^2$/Vs).[17] For room-temperature *n*-hexane, the probability of finding the electron in a quasifree state is low, ca. 3x10$^{-3}$.[6] For hydrocarbons composed of spherical molecules, e.g., 2,2,4-trimethylpentane (*iso*-octane), the binding energy is 50-60 meV (vs. the thermal energy of 25 meV),[7] and this probability is two orders of the magnitude greater.[7,15,17] In the two-state model for electron conduction in nonpolar liquids,[6,7,16,17] the apparent drift mobility $\langle \mu \rangle \approx \mu_f \tau_f \langle \tau_t^{-1} \rangle$ of the electron depends on the equilibrium fraction of quasifree electrons ($e_{qf}^-$) that is reached in reversible reaction (1)

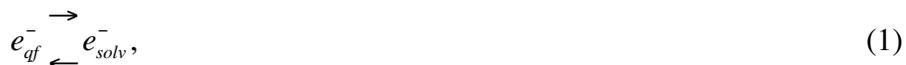
$$e_{qf}^- \underset{\leftarrow}{\overset{\rightarrow}{\phantom{xx}}} e_{solv}^-, \tag{1}$$

where $\tau_f$ is the localization time for $e_{qf}^-$ (ca. 20-30 fs for *n*-hexane)[6] and $\langle \tau_t^{-1} \rangle$ is the mean rate of thermal emission from traps ($\langle \tau_t \rangle \approx$ 8-9 ps for *n*-hexane).[6] The contribution from trapped/solvated electrons ($e_{solv}^-$) to the average mobility is negligible.[6,17] It is usually assumed that the product $\mu_f \tau_f$ exhibits weak temperature dependence,[6,7,15-17] i.e. the activation energy for $\langle \mu \rangle$ is close to the binding energy $E_t$ of electron traps.

What happens to the excess electron in a dilute solution of polar molecules (e.g., alcohol molecules) in a typical nonpolar solvent (e.g., *n*-hexane)? This question, originally posed by Mozumder,[18] still lacks complete resolution. Since the excess electron is strongly attracted to permanent dipoles in the solute molecules, replacement of nonpolar *solvent* molecules by these polar molecules in the solvation shell of the cavity electron ("dipole coagulation") is energetically favorable. This trend is countered by entropy preventing the substitution. Over time, an equilibrium is reached, and a new type of (solute-)trapped electron emerges. Hereafter, electron "trapping" or "attachment" refers to the formation of a $\{e^- : S_n\}_{solv}$ species, in which $n \geq 1$ *solute* molecules ($S$) are included in the first solvation shell of the cavity electron; the *solvent* molecules are still included in the cavity. No $S_n^-$ anions in which the electron occupies a molecular orbital of the *solute* molecule are involved.

Mozumder's paper [18] outlining this scenario stimulated a brief flurry of experimental activity. [19-27] It was expected that small electron clusters, as opposed to solvated electrons in neat liquids, would be simple to study and to model. (Similar expectations were later nurtured for gas phase cluster anions). Electron localization in dilute solutions of water and alcohols in liquid [19-26] and vitreous [e.g., 27] alkanes was studied using pulse radiolysis - transient absorbance (TA) spectroscopy, [19,20,23,24,27] time-resolved conductivity, [21-25] and optically detected magnetic resonance (ODMR). [26] The results obtained in these studies hinted at a complex picture of electron dynamics and the interest in mixed solvents quickly waned. Nevertheless, a consensus has been reached as to the mechanism for electron localization in such systems (section 2.1). As shown in the present study, this consensual picture requires revision and clarification. In retrospect, alcohols were an inopportune choice for the initial studies due to their tendency to form strongly bound multimers. Our results and analyses suggest that the interest in mixed solvents was fully justified: such systems do provide a new vista on electron solvation in molecular liquids.

To reduce the length of the paper, some figures and the Appendix are placed in the Supplement. The figures with the designator "S" (e.g., Figure 1S) are placed therein.

**2. Background.**

5.

## 2.1. Polar solute traps in alkane solvents.

The studies in the 1970s and 1980s [19-26] showed that in dilute solutions of hydroxylic molecules (such as alcohols and water) in normal and branched alkanes, the electrons attach to *pre-existing* clusters $S_n$ (multimers) of these solute ($S$) molecules.

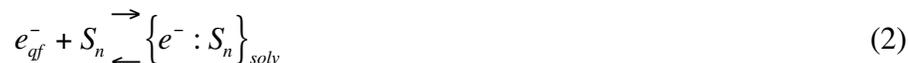

$$e^-_{qf} + S_n \rightleftarrows \{e^- : S_n\}_{solv} \qquad (2)$$

Note that due to the occurrence of reaction (1), this electron trapping reaction can also be represented by reaction (3)

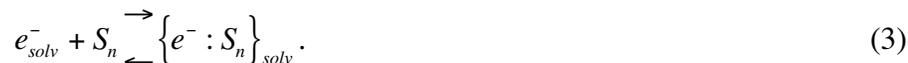

$$e^-_{solv} + S_n \rightleftarrows \{e^- : S_n\}_{solv}. \qquad (3)$$

Individual alcohol and water molecules (the "monomers", $n = 1$) do not trap the excess electrons, and in very dilute (< 1-5 mM) alkane solutions, localized electrons still reside in solvent traps. By contrast, alcohol clusters present in more concentrated solutions [19-26] bind the electrons quite strongly. H-bonded dimers and higher multimers of hydroxylic molecules form spontaneously in alkanes by reactions (4)

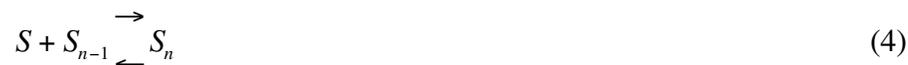

$$S + S_{n-1} \rightleftarrows S_n \qquad (4)$$

when the mole fraction $\chi$ of the solute exceeds $10^{-3}$. [19,21,25,28] Let $K_n$ be the equilibrium constant of reaction (4) for solute concentrations given in mole fractions. For open chain $n$-mers of normal alcohols, Stokes [28] obtained $K_2=11$ (with a standard enthalpy of -21.2 kJ/mol) and $K_3=122.7$ and $K_{n>3}=76$ (with a standard heat of -23.5 kJ/mol). Using these equilibrium constants and enthalpies, it is easy to obtain the concentrations of various multimers in solution (typical speciation plots are given in Fig. 1S in the Supplement).

6.

Less known is that *acetonitrile* molecules also form clusters in nonpolar solvents, albeit less efficiently than the hydroxylic molecules,[29,30] as the heat of dimerization of *MeCN* in $CCl_4$ is only -6 kJ/mol.[29] In CCl4 mixtures with $\chi$ <0.05, there is an equilibrium between the antiparallel $(MeCN)_2$ dimer and the monomer.[30b] For $\chi$ >0.2, multimers in which several *MeCN* molecules couple in the antiparallel fashion to a central one are observed.[29,30e] These are mainly pentamers with a typical size of 11 Å.[30e] For $\chi \cong 0.01$-0.1, both trimers (ca. 9 Å)[30e] and pentamers were observed by NMR.[30d]

For small alcohol multimers, electron trapping in the $\{e^- : S_n\}_{solv}$ species is reversible (as thermal emission of electrons into the CB is still possible); for larger clusters, the traps are too deep (> 1 eV) to observe this emission within the lifetime of the electron species.[20,21,25] The minimum size *n* of the $S_n$ cluster capable of reversible electron trapping is uncertain. Some authors reached the conclusion that alcohol dimers can trap the electrons.[20,21,26] Others concluded that only tetramers or higher multimers can trap these electrons (reversibly and irreversibly, respectively).[19,25] For $\chi$ <5x10$^{-2}$, this reversible trapping decreases the apparent electron mobility $\langle \mu \rangle$ (by reducing the equilibrium fraction of $e^-_{qf}$ via reaction (2)); however, this trapping has almost no effect on the absorption spectrum of the excess electron.[19,20,23] At higher concentration, the absorption peak of the solvated electron shifts towards the blue and the TA signal increases several fold.[19,20,23] At the onset of this spectral transformation ($\chi \approx 5x10^{-2}$)[19,23] almost no quasifree electrons are left in the solution.[20,21] For $\chi$ >0.2-0.3, the TA spectrum resembles that for the electron in neat alcohol:[19,20,23] the substitution of solvent molecules by the solute in the first solvation shell of the cavity electron is complete. The lifetime of this electron species (which is a few microseconds) appears to be limited by proton transfer from the alcohol molecule in the solvation shell, as is also the case in neat

7.

alcohols. [23] The resulting $\{e^-:S_n\}_{solv}$ species should be at least a tetramer. [19] The same applies to solvated electron observed in water-saturated alkanes, [20,23] though the corresponding spectrum more closely resembles that of the excess electron in dense water vapor [31] and supercritical water [32] than hydrated electron in liquid water. [2] For $\chi > 0.1$, there is no evolution of the TA spectra after the 30 ps electron pulse. [19] At lower alcohol concentrations (for $\chi$ between 0.05 to 0.1) the main effect of the alcohol addition is a decrease in the decay rate of the TA signal on the sub-nanosecond time scale, which is the expected result of lowering the electron mobility. It appears that electrons are scavenged by large alcohol clusters very rapidly (< 30 ps), [19] the rate constant of electron attachment may be > $10^{12}$ M$^{-1}$ s$^{-1}$. [25] No further growth of the $\{e^-:S_n\}_{solv}$ species by "dipole coagulation" reaction (5)

$$\{e^-:S_n\}_{solv} + S_m \rightleftarrows \{e^-:S_{n+m}\}_{solv} \qquad (5)$$

was observed within the first 500 ps after the formation. [19] In fact, this reaction has not been observed even on a longer time scale. [20-25] The only observation of a cluster-growth reaction similar to reaction (5) is by Ahmad et al. [22] who observed, albeit indirectly, slow (ca. 7x10$^9$ M$^{-1}$ s$^{-1}$) complexation of methanol with the tentative water tetramer in *iso*-octane.

Recently, it has been demonstrated [33] that the electron in supercritical CO$_2$ (in which the excess negative charge is trapped as C$_2$O$_4^-$ anion) forms dipole-bound complexes with the dimers and *monomers* of the alcohols and acetonitrile. In another publication (see sections 4.1 and 1S in ref. 10) we observed that the electron in *n*-hexane can be trapped by a *single* acetonitrile molecule. The electron binding energy of the resulting $\{e^-:MeCN\}_{solv}$ species was estimated to be only 200 meV lower [10] than that of

8.

the intrinsic solvent trap. [6,7] Ahmad et al. [22] reported reversible trapping of the electron in *iso*-octane by monomers of two other nonhydroxylic polar solutes, trimethylamine and diethyl ether, though the equilibrium constants for electron attachment were very low (ca. 150 and 3.5 $M^{-1}$, respectively, vs. > 400 $M^{-1}$ for acetonitrile). [10] Thus, the electron can be trapped by polar *monomers* in solvents other than alkanes. *Acetonitrile* monomers (as well as some other polar molecules missing OH groups) can trap the electron in the alkanes, but *alcohol* molecules have to form clusters. In the present paper, we further explore these patterns.

## 2.2. *Pulse radiolysis of alkanes: some basics*.

One of the techniques used to characterize $\{e^- : S_n\}_{solv}$ species in this study is nanosecond pulse radiolysis - TA spectroscopy. It is appropriate to make some general remarks concerning the radiolysis of neat alkanes and alkane solutions, as such an insight is needed to interpret the results given in section 4.1 correctly.

In neat alkanes, such as *n*-hexane, two radiolytic products are observed several nanoseconds after the 20 MeV electron pulse: trapped electrons, which absorb in the near- and mid- infrared (IR), and olefin cations, which absorb mainly in the blue and ultraviolet (UV) (the $\pi\pi^*$ band) but also have a spectral extension to the visible (the $\sigma\pi^*$ band). [34-36] This TA signal can be observed most distinctively in $CO_2$-saturated solution (see Fig. 2S(a), trace (i) in section 4.1) since the TA signal from trapped electrons is removed rapidly by $e^-_{qf}$ scavenging (see below); the cation signal is enhanced (Fig. 2S(b)) due to the slowing of geminate recombination (the mobility of $CO_2^- << \langle \mu \rangle$). The mechanism for formation of these "satellite cations" (observed as early as 30 ps after the electron pulse) [36] is still uncertain. [34] Fragmentation of vibrationally excited solvent holes (that is, alkane radical cations) and intraspur reactions of these holes and olefins

9.

generated via fragmentation of excited solvent molecules are the two most likely routes. [34,37] For cycloalkanes (whose holes are very mobile, due to rapid resonant charge transfer), [34,38] dimer olefin cation is also observed in the red, [35,38] but this species is not formed in *n*-alkanes. [35] The excited state of *n*-hexane (with a yield of 1.6 molecules per 100 eV absorbed energy) [34,39] is short-lived (ca. 300 ps) [39] and its radical cation rapidly deprotonates (in ca. 2 ns). [40] For *iso*-octane, the excited state and the hole fragment in < 40 ps, [39,41] which is much shorter than the duration of the 4 ns fwhm electron pulse used in this study; trapped-electron absorbance in the infrared is also missing as the binding energy is only 50-60 meV. [7] Recent results from Barbara's group [42] suggest that the TA signal from *photoionized iso*-octane observed on the femtosecond time scale [44] is likely to originate from an *excitonic* species rather than a trapped electron.

For neat *n*-hexane, the Onsager radius $r_c$ (at which the Coulomb interaction between the geminate partners equals the thermal energy, $k_B T \approx 25$ meV) is ca. 300 Å, whereas the distribution width for $e^-$ thermalization path is ca. 70 Å. [44] Thus, < 3% of the electrons escape their geminate partner's Coulomb field (become "free") in isolated pairs (ca. 4 pairs per 100 eV). [44] Consequently, the yield of free solvated electrons is low (ca. 0.13 per 100 eV), and since their molar absorptivity ($\varepsilon_{1000} \approx 8{,}300$ M$^{-1}$ cm$^{-1}$) [24,45] is also low, relatively large dose is needed to observe $e^-_{solv}$ on the nanosecond time scale (to obtain an absorbance > 10$^{-3}$). The time scale of geminate recombination is given by the Onsager time $t_c = r_c^2/D_e$, where $D_e = (k_B T/e)\langle\mu\rangle$ is the mean diffusion coefficient of the electron (which is much greater than that of the cation). For *n*-hexane at 23 °C, $\langle\mu\rangle \approx 0.092$ cm$^2$/s and $t_c \approx 3.8$ ns. Under the same conditions, the Debye constant $k_D = 4\pi D r_c$ for bimolecular charge neutralization in the bulk is ca. 5.3x10$^{13}$ M$^{-1}$ s$^{-1}$ and the critical concentration $C_{cr} = (4\pi r_c^3)^{-1}$ [46] of electrons at which $k_D C_{cr} t_c \approx 1$ (i.e., cross and geminate recombination occur on the same time scale) is ca. 5 μM. The observed

10.

concentration of (*free*!) electrons at the end of the electron pulse was ca. 0.6 μM (section 4.1). Since the electron concentration for $t < t_c$ is at least an order of the magnitude higher than that of the free electrons, the loss of $e^-_{solv}$ to bulk and intraspur cross recombination during the geminate stage is substantial. This is not the case in the photoconductivity experiments discussed in section 4.2, as the yield of free electrons in these experiments is very low (< 10 nM). For *isolated* electron-hole pairs, the yield of free electrons does not depend on electron mobility; this yield is entirely determined by the initial electron distribution around the parent hole. [44] Our conductivity measurements suggest that for dilute solutions of acetonitrile or alcohol in alkanes the free electron yield indeed does not change due to the occurrence of reactions (2) and (3) (see sections 4.2.1 and 4.2.2, respectively). However, in the dose regime typical of pulse radiolysis - TA studies [19-24] (as opposed to pulse radiolysis - d.c. conductivity studies), [21-23,25] this is not the case. Since for alcohols, electron attachment reaction (2) occurs with a rate constant > $10^{12}$ $M^{-1}$ $s^{-1}$, [25] electron trapping in 0.1 M ethanol occurs well within the geminate stage. As the mobility of the resulting $\{e^- : S_n\}_{solv}$ species (< $10^{-2}$ $cm^2/Vs$) is much lower than $\langle \mu \rangle$, both cross and geminate recombination are arrested, and the end-of-the-pulse electron yield increases (contrary to the claims made at the end of ref. 20). This effect introduces ambiguity in the estimates for molar absorptivity for the $\{e^- : S_n\}_{solv}$ species. For in-depth discussion of this effect (in a different system) see ref. 46.

## 3. Experimental.

*Materials:* *n*-Hexane (99+%, Aldrich) and *iso*-octane (99+%, Baker) were passed through activated silica gel to remove olefin impurity. Biotech grade acetonitrile (99.93+%) stored under $N_2$ and highest grade alcohols (99.9+%) and their deuterated analogs (>98+ atom % D) were obtained from Aldrich and were used without purification, but without exposure to air. The alcohol and acetonitrile solutions were deoxygenated by purging with dry nitrogen or argon. All measurements of the electron mobility and TA were carried out in these $N_2$- or Ar- saturated solutions. Purging these solution or even moving the liquid between the containers causes substantial loss of the polar solute to the head space. Gas chromatography was used (samples were taken from the exit of the cell, with no exposure of the sample to a "head space") to determine solute concentrations. This monitoring was absolutely necessary: due to the extreme volatility of polar molecules in dilute hydrocarbon solutions, reproducible results cannot be obtained otherwise. Previous researches [25] reported similar problems; from our experience, the irreproducibility is always traceable to the solute loss.

*Electron mobility.* The conductivity setup was the same as described in our previous publication. [6] Fifteen ns fwhm pulses of 248 nm photons from a KrF excimer laser were used to ionize the solutions via biphotonic excitation. Room temperature solutions were photolyzed in a cell with a 4 cm optical path. To obtain the temperature dependence, 5 µM anthracene solutions were photolyzed in a 2 cm path cell (photoionization of the anthracene gives higher electron yield compensating for a shorter path). Both cells have two planar *Pt* electrodes spaced by 6.5 mm operated at 4 kV. For time-of-flight conductivity experiments, the electrode spacing was reduced to 800 µm and a 100 µm slit was used to generate the charges near the electrodes. A 1064 nm beam from a Nd:YAG laser passed thru the cell in the opposite direction to the 248 nm beam, and completely enveloped the 248 nm beam inside the cell. The maximum fluence *J* of 1064 nm photons through the cell was ca. 1.5 J/cm$^2$ (9x10$^{18}$ photons/cm$^2$); the fluence of 248 nm light was < 0.1 J/cm$^2$. The typical (free) electron concentration in our conductivity experiments was 5-10 nM. The lifetime of the electron (< 1 µs) is controlled by an electron-scavenging impurity. Under our excitation conditions, cross recombination of charges in the bulk and their movement towards the electrodes were negligible for $t < 1$ μs. The transient photocurrent signal was amplified and recorded with a time resolution < 2 ns. The delay time $t_L$ of the 1064 nm pulse relative to the 248 nm pulse was 25-800 ns; the time jitter between these two pulses was < 3 ns. To determine the conductivity signal $\Delta\kappa(t)$ induced by the 1064 nm laser pulse, this laser was pulsed on and off while the 248 nm laser was pulsed for every shot, and the corresponding signals $\kappa_{on}(t)$ and $\kappa(t)$ were subtracted. If not specified otherwise, the measurements were carried out at 23 °C. The conductivity is given in units of nS/cm (= $10^{-7}$ $\Omega^{-1}$ $m^{-1}$).

*Pulse Radiolysis - Transient Absorbance (TA)*. Room temperature solutions were radiolyzed using electron pulses from the Argonne LINAC (20 MeV, 4 ns fwhm, 21.5 nC per pulse). The solutions were placed in a 2 cm optical path cell with Suprasil windows. The analyzing light from a pulsed Xe arc lamp was coaxial with the electron beam and traveled in the opposite direction. A set of 10 nm fwhm band pass interference filters (50 nm interval) was used for wavelength ($\lambda$) selection between 0.5 and 1.6 μm. A fast Ge detector with flat spectral response was used to detect the TA signal on the nanosecond time scale. Cerenkov light and radiation-induced TA signal from the cell windows (< $10^{-3}$) were subtracted from the kinetic traces giving the TA signal $\Delta OD_\lambda(t)$ from the irradiated sample vs. the delay time $t$.

**4. Results.**

*4.1. Pulse Radiolysis.*

A typical TA spectrum observed at the end of 4 ns fwhm electron pulse in neat *n*-hexane is shown in Fig. 1, trace (i). In the first 200 ns, the electron rapidly decays via a scavenging reaction with impurity and by homogeneous recombination; only the TA signal from the olefin cation (trace (ii) in Fig. 1; see section 2.2) persists at longer delay times (Fig. 2S(a)). Addition of 4-50 mM of *MeCN* increases the TA signal in the near infrared ($\lambda \approx 0.8$-$2$ μm), whereas the relative fraction of the olefin cation at $t \cong 50$-$100$ ns decreases (Figs. 3S(a) and 4S). The end-of-pulse spectra obtained for 10-50 mM solutions are very similar (Fig. 4S(a)). For $t > 100$ ns, some spectral evolution is observed in the visible (Fig. 4S(b)) where the olefin cation absorbs. (The dimer anion of acetonitrile may also contribute to this TA signal; see section 4.3). As trapped electron decays, the relative contribution from the olefin cation increases. The plot of the end-of-pulse TA signal at 1 and 1.55 μm vs. [*MeCN*] is given in Fig. 1(b); the slight discrepancy between these two plots is due to the interference from the olefin cation that absorbs at 1 μm (Fig. 1(a)). The TA signal first increases linearly with the increasing [*MeCN*], then "saturates". For *iso*-octane solution, the plot of the 1.55 μm absorbance is linear with [*MeCN*] to 0.18 M (Fig. 5S(a)). Otherwise, the spectral evolution is similar to that in acetonitrile/*n*-hexane solutions (compare Figs. 4S and 6S). In both alkane liquids, as the concentration increases, the decay of the TA signal in the infrared becomes slower. (Fig. 3S(a) and 5S(b)).

Fig. 1(a) shows a comparison between the electron spectra in neat *n*-hexane, trace (i) and 50 mM and 75 mM acetonitrile solutions in *n*-hexane and *iso*-octane, respectively (traces (iv) and (v), respectively; the spectral profile does not change further at higher concentrations). To facilitate the comparison, all of these spectra are normalized at 1.55 μm. For neat *n*-hexane, the olefin cation signal interferes with the electron signal (see Fig. 2, trace (ii)), so the direct comparison is difficult. Some compensation can be made by subtracting the spectrum of olefin cation, trace (ii), from the composite spectrum, trace (i), assuming that 500 nm absorbance is only from the cation (the difference trace (iii)). Still it is clear from Fig. 2 that the TA spectrum in acetonitrile solution is less broad than that in neat *n*-hexane. The increase in the TA signal from the electron with increasing

14.

$[MeCN]$ can be accounted for by the formation of $\left(e^-:MeCN\right)_{solv}$ species. As shown in section 4.2.1, this trapping becomes nearly irreversible on the observation time scale for $[MeCN] > 10$ mM. As it greatly decreases the electron mobility, the efficiency of cross (and, to a lesser degree, geminate) recombination is reduced, which results in a greater electron yield at the end of the 4 ns electron pulse (section 2.2). The magnitude of this effect in our dose regime (ca. 2 times) may be estimated by comparing the yield of the cation absorbance at 500 nm in Ar- and $CO_2$- saturated *n*-hexane solutions (Fig. 2S(b)). $CO_2$ rapidly (< 200 ps) scavenges the electron yielding a slowly migrating $CO_2^-$ anion. The TA signal from the electron at $\lambda = 1$ μm increases ca. 5 times in 50 mM acetonitrile solution. Since there is always a parity between the yields of charges of different sign, it is possible to crudely estimate that the molar absorptivity of the electron at 1 μm is 2-2.5 times greater than that in neat *n*-hexane.

For alcohols ($ROH$), the spectral evolution is more complex. In Fig. 2(a), end-of-pulse TA spectra for 40-260 mM ethanol in *n*-hexane are shown. (This system has previously been studied, [20] although the solute concentrations were not reported). No spectral evolution was observed for $t < 1$ μs, for all ethanol concentrations (e.g., Figs. 7S(a) and 7S(b)). Alcohol monomers and multimers are rapidly protonated by the solvent holes and the formation of olefin cations is thereby suppressed. Consequently, there is no interference from these cations in the TA spectra. On the other hand, there is also no spectroscopic evidence for gradual "dipole coagulation" (reaction (5)) on the nanosecond time scale, i.e., all electron equilibria settle in < 10 ns (at 23 ºC). The increase of the TA signal at $\lambda = 1$ μm vs. $[ROH]$ (Fig. 2(b)) is a sigmoid curve similar to that for acetonitrile except for low alcohol concentrations. This difference is due to the fact that only *multimers* can trap the electron in alcohol solutions (sections 2.1 and 4.2.2), whereas even *single* acetonitrile molecules can trap these electrons (section 4.2.1). A comparison with the TA data in ref. 23 (filled triangles in Fig. 2(b)) suggests that the absorbance vs. concentration plots for ethanol and 1-propanol are identical when $[\text{Pr}OH]$ is scaled down by a factor of two. As shown in section 4.2.2, the plots for $\langle \mu \rangle$ vs. $[ROH]$ are also similar for all alcohols studied once their concentrations are appropriately scaled. As $[ROH]$ increases, the decay kinetics (like those for acetonitrile, Fig. 3S(a)) observed on the sub-microsecond time scale slow down (Fig. 3S(b)). For both solutes, this decrease is significant in dilute solutions (< 20-50 mM); at higher concentrations (when the trapping becomes irreversible, see section 4.2,1 and 4.2.2) further slowing down of these kinetics does not occur. The magnitude of the increase in $\Delta OD_\lambda$ for $\lambda = 1$ μm is ca. 9 times for $[EtOH] \approx 0.26$ M (Fig. 2(b)). Since, as noted above, part of this increase (ca. 2 times) is due to the suppression of recombination, the rough estimate for the increase in the $\lambda = 1$ μm absorptivity of the trapped electron is 4-5 times.

The crucial difference between the acetonitrile and ethanol solutions is that for the latter solute not only the amplitude of the TA signal increases as $[S]$ increases but also the spectral profile evolves continuously (Fig. 2(a)). The band maximum shifts from 1.5 μm at 40 mM to 1.1 μm at 73 mM to 0.95-1 μm at 120 mM to 0.8 μm at 262 mM. One can inquire whether the TA spectra observed at the intermediate ethanol concentrations can be obtained by addition of weighted spectra observed at the highest and the lowest ethanol concentration (dotted lines in Fig. 2(a)). Such would be the case if only one kind of the $(e^- : S_n)_{solv}$ species was present in the reaction mixture. Although these weighted sums can be made close to the spectra observed, this does not seem to be the case (Fig. 2(a)). This argues that only a few (perhaps, 2-3) types of trapped electron species are present in the solution at equilibrium, as previously suggested by Gangwer et al. [25] and others. [20-24] We will return to these observations in section 5.2.

*4.2. D.C. photoconductivity studies.*

*4.2.1. Acetonitrile.*

Figs. 3(a) and 3(b) show the typical photoconductivity signals $\kappa(t)$ from the electrons generated by 248 nm photon ionization of dilute acetonitrile solutions in Ar-saturated, room-temperature *n*-hexane. Qualitatively, very similar kinetics were observed using 5 μm anthracene (added to increase the photoionization yield). The conductivity signal ($t < 2$ μs) decays exponentially to a plateau as $\kappa(t) = \kappa_0 \exp(-kt) + \kappa_i$ (this constant offset is subtracted from traces shown in Fig. 3) The exponential decay is due to the reaction of the electron with impurity or the polar solute (see below); the plateau conductivity $\kappa_i$ is from ions that decay slowly (on the millisecond time scale) by recombination in the bulk and neutralization at the electrodes. This residual ion signal does not depend on the acetonitrile concentration (< 60 mM), suggesting that *addition of acetonitrile has no effect on the electron yield*. The conductivity signal $\kappa_0$ from (free) electrons can be obtained by exponential extrapolation to the zero time; this quantity is plotted vs. $[MeCN]$ in Fig. 4(a) (open circles, to the left). Note that for the electron concentrations generated in our conductivity experiments (a few nM) second-order recombination on the nano- and micro- second time scales is very minor.

Addition of acetonitrile creates a new kind of electron trap in a dynamic equilibrium with $e_{qf}^-$. Following the approach of ref. 10, and further developed in the Appendix, we introduce the equilibrium constant $K_{eq}$ of reaction (3) between the electrons in the solvent trap, $e_{solv}^-$, and in this monomer solute trap, $\{e^- : MeCN\}_{solv}$. As the *net* molar concentration $c$ of the solute increases, the equilibrium fraction of $e_{solv}^-$ and the apparent mobility $\langle \mu \rangle$ of the electron decrease as

17.

$$\langle\mu\rangle/\langle\mu_n\rangle = \left(1 + K_{eq}c\right)^{-1} \tag{6}$$

where $\langle\mu_n\rangle$ is the apparent electron mobility in *neat n*-hexane. Provided that the electron yield does not change with the solute concentration (as is the case in acetonitrile solutions, see above), the ratio of the corresponding conductivity signals $\kappa_0$ is given by eq. (6). The decay rate of the electron decreases in proportion to the mobility $\langle\mu\rangle$ for $c <$ 0.03 M, as can be seen from the correlation plot given by Fig. 4S(b) in ref. 10. At higher concentrations, the rate constant $k$ gradually begins to increase, possibly due to dimer formation (section 4.3). Formula (6) may be conveniently expressed as $K = K_{eq}c$, where $K = \langle\mu_n\rangle/\langle\mu\rangle - 1$ (the concentration plot for this quantity is given in Fig. 4(a), solid circles to the right). In either form, eq. (6) can be used to fit the plot of $\kappa_0 \propto \langle\mu\rangle$ vs. [*MeCN*], both for *n*-hexane and *iso*-octane solutions. The slope of the van't Hoff plot for the resulting constants $K_{eq}$ obtained at different temperatures yields the standard heat $-\Delta H_{eq}^0$ of reaction (3) for *MeCN* monomer. In ref. 10 (see section 1S in the Supplement and Fig. 7(b) in section 4.2.2 of that paper) we have carried out just such an analysis and obtained $K_{eq} \approx 440\pm20$ M$^{-1}$ (at 25ºC) and $-\Delta H_{eq}^0 \approx 19.6\pm0.9$ kJ/mol (ca. 200 meV). [10] For dilute *iso*-octane solutions ([*MeCN*] < 30 mM), eq. (1) also holds (Fig. 8S(a)) and $K_{eq} \approx 950\pm50$ M$^{-1}$ at 25 ºC.

Since for *iso*-octane $\langle\mu\rangle$ is 6,700 times greater than the combined anion and cation mobility $\mu_i$ (ca. $10^{-2}$ cm$^2$/Vs) [7b,c] whereas for *n*-hexane $\langle\mu\rangle$ is only ca. 56 times greater than $\mu_i$ (ca. 1.5x$10^{-2}$ cm$^2$/Vs), [7c] the former solvent gives a better opportunity to follow the decrease in the electron mobility for $\chi > 5\times10^{-3}$ since even at this high solute concentration $\langle\mu\rangle > \mu_i$ (the typical kinetics are shown in Fig. 9S) As shown in Fig. 8S(b), where $\kappa_0$ is plotted vs. [*MeCN*] on the logarithmic scale, $\langle\mu\rangle$ decreases by a factor of

18.

300 between 32 and 175 mM, in the concentration range where eq. (6) is no longer applicable. It is precisely this concentration range that was explored in section 4.1. Thus, the equilibrium was completely shifted towards the acetonitrile traps and the spectra shown in Fig. 1 are from such traps (the same pertains to *n*-hexane solutions). Given the constancy of the spectrum as a function of $[MeCN]$, it may be expected that $\{e^- : MeCN\}_{solv}$ is the predominant species in both solvents. Thus, it is not presently clear what causes the decrease of $\langle \mu \rangle$ in concentrated acetonitrile/*iso*-octane solutions. The entire plot in Fig. 8S(b) can be fit using an empirical formula $\langle \mu_n \rangle / \langle \mu \rangle = 1 + K_{eq} c + K_n c^n$, with $n \approx 3.6$. This suggests that a multimer species in equilibrium with $\{e^- : MeCN\}_{solv}$ might be involved at high *MeCN* concentrations (see section 4.3). Vis-absorbing, covalently bound $(MeCN)_2^-$ anion occurs in neat acetonitrile [10,11,14] and this species may also occur in the *iso*-octane solutions. Another possible rationale is that at high concentration, solute molecules scatter quasifree electrons, thereby changing $\langle \mu \rangle$. As our interest is mainly in the cavity electrons, hereafter we focus on *dilute* solutions (< 30 mM) for which the multimer anion formation and/or electron scattering may be safely neglected.

For sufficiently concentrated acetonitrile solutions (yet still within the range of applicability of eq. (6)), $K \gg 1$ and $\log\langle\mu\rangle - \log\langle\mu_n\rangle \propto -\Delta H_{eq}^0$. Thus, the enthalpy $-\Delta H_{eq}^0$ can be estimated as the difference of activation energies $E_\mu^*$ and $E_{\mu n}^*$ of the electron mobility (and, therefore, $\kappa_0$) in the acetonitrile solution and neat solvent, respectively. Fig. 5 demonstrates the Arrhenius plots for $\kappa_0$ in 17.3 mM MeCN in *n*-hexane (the corresponding kinetics are shown in Fig. 10S) and 22 mM MeCN in *iso*-octane. [48] At these concentrations, $\langle\mu\rangle$ decreases > 10 times vs. neat solvents. The activation energies for *n*-hexane and *iso*-octane are 4.9±0.7 and 32.6±0.4 kJ/mol (for neat

19.

solvents) and 52.7±2.2 and 45.1±1.3 kJ/mol (for MeCN solutions), respectively, from which $-\Delta H_{eq}^0$ is estimated as 20.1±3 and 48±3 kJ/mol, respectively. Note that in neat *n*-hexane, the binding energy for the electron trap is ca. 200 meV, i.e. the binding energy of acetonitrile trap vs. the CB edge is ca. 400 meV. A better, more reliable estimate for the same parameter (that does not require making an assumption that the product $\mu_f \tau_f$ is temperature-independent) [6,7] can be obtained using 1064 nm (1.17 eV) photon induced electron detachment, [6] as described below.

Fig. 3 shows the effect of 1064 nm photoexcitation on the conductivity signal from the electron (see also Figs. 9S(a) and 10S(b)). The temporal profile of the difference signal $\Delta \kappa$ (section 3) follows the Gaussian profile of the 1064 nm pulse. At all solute concentrations and temperatures, the amplitude of the $\Delta \kappa$ signal decreases with the delay time $t_L$ of the 1064 nm laser pulse in the same way as $\kappa_e(t_L) = \kappa(t_L) - \kappa_i$, the conductivity signal from the electron induced by 248 nm light decays. When the electrons are scavenged (as $\kappa(t)$ decays to a plateau within 1-5 μs), 1064 nm light does not produce any increase in the conductivity. [49] This behavior suggests that the $\Delta \kappa$ signal originates from 1064 nm photons detaching the electron from the traps and promoting these electrons into the CB of the solvent. The equilibrium is rapidly reestablished, but the conductivity increases significantly during the 1064 nm laser pulse. Observe that all $\Delta \kappa(t)$ kinetics eventually approach zero, i.e., there is no inhibition of geminate recombination due to the photodetrapping. As shown by Lukin et al, [50] such a process is significant only for $t_L < t_c/10$, i.e., on the time scale that is much shorter than the duration of the 248 nm pulse.

It has been demonstrated [6] that the ratio $r$ of the area $\Delta A(t_L) = \int dt\, \Delta\kappa(t)$ under the $\Delta\kappa(t)$ kinetics to the electron conductivity $\kappa_e(t_L)$ prior to the photoexcitation is given by

$$r = \Delta A(t_L)/\kappa_e(t_L) \approx \langle \sigma_t \tau_t \rangle J \tag{7}$$

where $\sigma_t$ is the cross section for electron photodetachment from a given trap, $\tau_t$ is the life time of the electron in this trap, and $\langle ... \rangle$ stands for the average over all such traps weighted by their equilibrium fractions. As seen from eq. (7), this ratio does not depend on the delay time of the 1064 nm pulse (in accordance with Fig. 3), and is independent of electron mobilities and yields as well. Eq. (7) is valid only for low fluence $J$ of the 1064 nm light, i. e., when the deviation from the equilibrium is relatively small. At high fluence, the equilibrium is shifted during the pulse, and a phenomenon akin to saturation sets in (see the Appendix for more discussion). The typical plots of $r$ vs. $J$ are shown in Fig. 11S(a). In Fig. 11S(b), the ratio $r$ is normalized by its value $r_{max}$ attained at the maximum fluence $J_{max}$ of 1064 nm photons (ca. $8.5 \times 10^{18}$ photons/cm²). As seen from the latter plot, the ratio $r$ first increases linearly with increasing fluence (for $J < 10^{18}$ photons/cm²) with a slope that weakly depends on $[MeCN]$ and then "saturates" (Fig. 11S). In Fig 4(b), the ratios $r$ are plotted for $J \approx 5.4 \times 10^{17}$ photon/cm² *(squares)* and $8.1 \times 10^{18}$ photon/cm² *(circles)* as functions of $K$ ($\approx K_{eq}c$). The higher fluence corresponds to the "saturation" regime whereas the lower fluence corresponds to the linear regime, eq. (7). The two curves exhibit the same initial slope but diverge at higher concentrations, as the equilibrium shifts towards $\{e^- : MeCN\}_{solv}$. Very similar behavior is obtained using the two-trap model in the Appendix. The plateau value ($r \approx 30$ ns) at the lower fluence is attained when reaction (3) is completely shifted towards the right side. Using eq. (7), $\sigma_t \tau_t \approx 5.6 \times 10^{-26}$ cm² s is obtained for this trap. For electrons in neat *n*- hexane, $\sigma_t \tau_t \approx 2.5 \times 10^{-28}$ cm$^2$ s, [24] i.e., for electrons in the acetonitrile traps, the product $\sigma_t \tau_t$ is ca. 225 times greater. In section 4.1, we estimated that the absorption cross section at $\lambda = 1$ μm for an electron captured by the acetonitrile trap is ca. 2 times greater than that of $e^-_{solv}$. Using the estimate of 8.3 ps for $\tau_t$ in $n$-hexane [6] and assuming unity quantum yield for electron photodetachment (justified in section 5.2), we estimate that the residence time $\tau_i$ for $\{e^- : MeCN\}_{solv}$ is ca. 1 ns. In Fig. 5, Arrhenius plots of the ratio $r$ (see Figs. 9S(b) and 10S(b) for kinetic traces) are shown for the same two concentrations used to estimate $-\Delta H_{eq}$. At these concentrations, the equilibrium is shifted towards $\{e^- : MeCN\}_{solv}$ so that the activation energy for this ratio approximately equals that of the detrapping rate. The latter activation energy is commonly identified with the binding energy $E_t$ of the trapped electron with respect to the CB edge. [6,7,16] For $n$-hexane and $iso$-octane, estimates of 39±2 and 39±1 kJ/mol were obtained, respectively. These two estimates agree perfectly with the estimate of $E_t \approx 400$ meV given above. The closeness of binding energies for the two alkanes qualitatively accounts for the similarity of the TA spectra shown in Fig. 2.

*4.2.2. Alcohols in n-hexane.*

Fig. 12S shows a family of $\kappa(t)$ kinetics for 248 nm photoexcitation of 12 mM ethanol in $n$-hexane solutions for several temperatures between 2 and 42 °C. As is observed for acetonitrile, these kinetics decay exponentially (except for the lowest temperatures; see below), otherwise the two systems show quite different behavior.

*First,* in alcohol solutions the conductivity signal $\kappa_i$ from the ions depends on $[EtOH]$ (Fig. 13S). The concentration dependence of $\kappa_i$ is the same in solutions with and without anthracene. Other photosensitizers, such as benzene and triethylamine, exhibit

22.

the same $\kappa_i$ dependence. Moreover, the same $\kappa_i$ vs. $[EtOH]$ plots are observed in $SF_6$- and $CO_2$-saturated solutions, where the electrons are promptly converted to anions. Almost the same dependencies are observed at lower temperatures (Fig. 13S). Time of flight experiments (Fig. 14S(a) shows a few typical traces) indicate that addition of ethanol *decreases the mobility of fluoride anions* in $SF_6$-saturated hexane solution (with 0.65 mM triethylamine added as a photosensitizer). For the cation, no such decrease was observed. The decrease in the anion mobility (ca. 2 times for 0.12 M ethanol, Fig. 14S(b)) is very substantial, and it readily accounts for the observed decrease in $\kappa_i$ (Fig. 13S), suggesting that the ionization yield *does not* depend on the alcohol concentration, as is also the case for acetonitrile. [51]

*Second*, for alcohols the electron conductivity $\kappa_0$ decreases with increasing $[S]$ in a qualitatively different way than for acetonitrile, as it does not follow eq. (6). This dependence does not change upon the deuteration: it is exactly the same for $CH_3OH$ and $CD_3OD$ and $C_2H_5OH$ and $C_2D_5OD$ (Fig. 6(a)). By appropriately scaling $[ROH]$, the same $\kappa_0$ vs. $[ROH]$ dependence was observed for methanol, ethanol, and 1-propanol (Fig. 6(a)). Since all these dependencies are the same, in the following only ethanol solutions are considered; this solute is representative of other *normal* alcohols. Baxendale and Sharpe [21] reported that qualitatively different behaviors were observed for different *n*-alcohols. Our results suggest otherwise. The likely problem with the previous measurement was inadequate control of alcohol concentration (section 3).

For ethanol solution, $\kappa_0$ does not decrease until $\chi > 0.01$; at higher concentration, $\langle \mu \rangle$ decreases rapidly, and for $\chi \approx 0.1$ almost no $e_{qf}^-$ remain in the solution due to the occurrence of reaction (2). The decay rate of $\kappa(t)$ decreases as $\langle \mu \rangle$ decreases, however, unlike in acetonitrile solutions, addition of a 1-5 mM of alcohol actually *increases* the

23.

rate constant $k = k_n + \Delta k$ of the exponential decay ($k_n$ is the constant observed in neat *n*-hexane) with little change in the apparent electron mobility (Fig. 15S). This behavior (for 1-propanol) is in agreement with the pulse radiolysis - TA study by Baxendale and Rasburn.[23] Gangwer et al.[25] reported that the bimolecular rate constant $k_{eff} \approx \Delta k/c$ for electron scavenging by methanol in *iso*-octane decreases from $4 \times 10^8$ M$^{-1}$ s$^{-1}$ (the same limiting rate constant was obtained by Baxendale and Rasburn)[23] at 0.1 mM to $10^8$ M$^{-1}$ s$^{-1}$ at 5 mM. Their time-of-flight measurement suggested (in agreement with our results) that $\langle \mu \rangle$ did not change significantly at these low concentrations.[52] As shown in Figs. 16S(a) and 16S(b), in the course of the scavenging reaction occurring in this low-concentration regime (at 2 °C and 21.5 °C, respectively), the ratio *r does not* change with the delay time $t_L$. This argues against the occurrence of slow reaction (3) in these very dilute solutions (the trapping would increase the retention time of the electron in the trap, thereby increasing ratio *r* at longer delay times $t_L$). In principle, the initial increase in the decay rate *k* may be explained by a scavenging reaction involving electron-attaching impurity present in the *solute*, such as an aldehyde.[25] Our control experiments as well as gas chromatography suggest that the aldehyde concentration is too low to account for the effect observed. A crucial observation is that the initial bimolecular rate $k_{eff} \approx \Delta k/c$ decreases by ca. 20% when deuterated alcohols are used instead of protiated ones, both for ethanol and methanol (Figs. 15S(a) and 15(b), respectively). Only a proton transfer would exhibit such a considerable isotope effect (observe that no isotope effect was observed for electron mobility, Fig. 6(a)). All these observations point to a slow ($< 5 \times 10^9$ M$^{-1}$ s$^{-1}$ vs. a typical electron attachment constant of $(1-3) \times 10^{12}$ M$^{-1}$ s$^{-1}$),[19,25] inefficient reaction of trapped electron with ethanol monomer via a proton transfer reaction

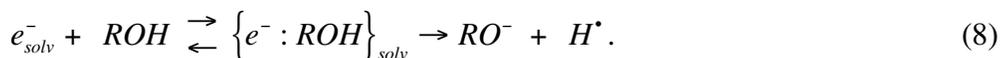

$$e^-_{solv} + ROH \rightleftarrows \{e^- : ROH\}_{solv} \rightarrow RO^- + H^\bullet. \qquad (8)$$

24.

(The hydrogen atom subsequently reacts with the solvent). Proton transfer reaction on the microsecond time scale analogous to eq. (8) also occurs [23,53] for solvated electrons in *neat* alcohols, where the cavity electron is stabilized against the proton transfer by strong electrostatic interactions with several OH groups. [e.g., 54] Such a stabilization mechanism is lacking for $\{e^- : ROH\}_{solv}$ and after the formation this species either promptly dissociates or undergoes proton transfer. Direct, prompt deprotonation of alcohol monomers in their encounter complex with the solvated electrons was previously postulated by Baxendale and Rasburn; [23] our results further support their suggestion.

At higher alcohol concentrations ($\chi$ >0.01), the electron mobility rapidly decreases with the net solute concentration $c$ of the alcohol (Figs. 6(a) and 6(b)). Following Baxendale and Sharpe, [21] the decrease in $\langle \mu \rangle$ can be described by an empirical formula

$$\langle \mu_n \rangle / \langle \mu \rangle = 1 + (c/c_0)^m \tag{9}$$

which generalizes eq. (6), where $m$ is the *mean* number of solute molecules per $\{e^- : S_m\}_{solv}$ cluster and $c_0$ is a (temperature-dependent) characteristic concentration. The data in Fig. 6(b) can be fit using this equation assuming temperature-independent $m \approx 3.5$ ($c_0$ is ca. 18±1 mM at 23 °C). Thus, in agreement with Gangwer et al. [25] we conclude, contrary to Baxendale and co-workers, [20-24] that *only higher multimers* trap the electron in alcohol solutions at equilibrium. To make more quantitative estimates, the following model was used: [25] The alcohol molecules were assumed to cluster according to reactions (4) with the equilibrium constants $K_n$ given in section 2.2, and solvated electrons $e^-_{solv}$ were assumed to attach to these multimers via reaction (3) with the equilibrium constants $K_{eq}^{(n)}$. The tacit assumption of this scheme is that reactions (3) are much faster than

25.

reactions (4). From eq. (4), we have $[S_n] = K'_n[S]^n$, where $K'_n = K_1 K_2 ... K_n$. The mass balance is given by

$$c = [S] + \sum_{n>1} n K'_n [S]^n \qquad (10)$$

By solving eq. (10) numerically, the concentration of free solute molecules and the $n$-mers can be found. Since the apparent electron mobility $\langle \mu \rangle$ is proportional to the equilibrium fraction of $e^-_{solv}$, we obtain

$$\langle \mu_n \rangle / \langle \mu \rangle = 1 + \sum_{n>1} K^{(n)}_{eq} [S_n] \qquad (11)$$

Two models were examined in which the electron is attached (exclusively) to (i) a trimer or (ii) a tetramer. As seen from Fig. 7(a) both of these models fit the data well, except for the room-temperature data, which are better accounted for by the tetramer attachment (this figure also shows an equilibrium fraction $f_4 = 4[S_4]/c$ of the tetramers). For methanol at 23 °C, the tetramer also provides the best fit to the data (not shown). The latter result is in agreement with Gangwer et al. [25] who suggested tetramer or pentamer as the predominant electron trapping $(ROH)_n$ cluster for solutions of methanol in *iso*-octane and tetramethylsilane. Van't Hoff plots for the equilibrium constants of reaction (3) obtained using these two models are shown in Fig. 7(b). From these plots the enthalpy of reaction (3) for $n = 3$ and 4 would be -58±2 and -64.5±2.5 kJ/mol, respectively. Thus, the $(ROH)_n$ trap binds the electron ca. 0.6 eV deeper than the intrinsic solvent trap, and the binding energy $E_t$ is ca. 800 meV. Similar estimates were obtained using the approach described in section 4.2.1, by comparison of the activation energies for $\kappa_0$ at different ethanol concentrations (Fig. 8(a)). As [EtOH] increases from 0 to 6 to 12 to 25 to 45 mM, these activation energies increase from 32.5±0.5 to 36±0.5 to 45±1.4 to 76±4 to 90±2 kJ/mol. The difference between the first and the last of these energies is ca. -57±3 kJ/mol which is in a reasonable agreement with the heat of reaction (3) obtained using van't Hoff analysis. For methanol tetramer in *iso*-octane, Gangwer et al. estimated this heat as -63±14 kJ/mol. [25] At 23 °C, $K_{eq}$ for ethanol tetramer is ca. 65 times greater than for acetonitrile monomer (Fig. 7(b)).

We turn to electron detachment experiments in which a 1064 nm laser pulse was used to promote the electron from a $\{e^-:(ROH)_n\}_{solv}$ cluster to the CB and observe the subsequent relaxation of the conductivity signal $\Delta\kappa$. The time profile of the $\Delta\kappa$ kinetics does not depend on the delay time $t_L$ of the 1064 nm pulse and the maximum amplitude of the $\Delta\kappa$ signal follows $\kappa_e(t_L)$, as is the case for acetonitrile (two examples are given in Fig. 16S). In other respects, the $\Delta\kappa$ kinetics are very different from those observed in acetonitrile/*n*-hexane solutions. The most remarkable feature is that the $\Delta\kappa(t)$ kinetics *do not* follow the time profile of the 1064 nm excitation pulse (Figs. 9, 10, 17S, and 18S). Even at room temperature (e.g., Fig. 17S), there is a "slow" component with a time constant $\tau_s$ of a few nanoseconds. To analyze these data, the $\Delta\kappa$ kinetics were fit by a weighted sum of (i) a Gaussian with the same $J(t) = (J/\tau_p\sqrt{\pi})\exp(-[(t-\tau_L)/\tau_p]^2)$ profile as that of the 1064 nm excitation pulse (the "spike") and (ii) the same Gaussian convoluted with $\exp(-t/\tau_s)$. Several examples of such fits are given in Figs. 9 and 10. Such an approach can be justified for $J \to 0$ using the two-trap model (see the Appendix). At high EtOH concentration, the "slow" component decays too fast and/or has too low weight to be analyzed in this fashion. At lower temperature (Figs. 9 and 10), the "slow" component (with a relative weight approaching 10-50% of the total signal) entirely separates from the initial "spike". For $[EtOH] < 10$ mM, $\tau_s$ increases over 20 ns (e.g., Fig. 9). With increase in $[EtOH]$ (Figs. 9 and 18S) or in the temperature (Figs. 10

27.

and 17S(b)), $\tau_s$ decreases until the "slow" component is no longer discernible. The observation of the "slow" component indicates that the equilibration of the electron between different traps occurs on a time scale that is longer than the duration of the 1064 nm pulse ($\tau_p \approx 4$ ns). At 1.7 °C, this slow equilibration can already be seen in the $\kappa(t)$ kinetics (Figs. 12S and 17S(a)) which become biexponential (the $\kappa_0$ values given in Fig. 6 were extrapolated to $t \to 0$ using the *slower* component, as shown in Fig. 12S).

There are two general ways to interpret these observations. First, it can be assumed that the slow component corresponds to settling of equilibrium reaction (3) that involves a (unique) electron-trapping cluster $S_n$ (e.g., the trimer). As shown in the Appendix, in such a case the time constant $\tau_s \approx (\tau_2 + K\tau_1)/(1+K)$, where $\tau_1$ and $\tau_2$ are the time constants for thermal emission of the electron from $e^-_{solv}$ and $\{e^-:S_n\}_{solv}$ to the CB, respectively, and $K = \langle \mu_n \rangle / \langle \mu \rangle - 1 \approx K^{(n)}[S_n]$. For $K \ll 1$, $\tau_s \to \tau_2$; for $K \gg \tau_2/\tau_1 \gg 1$, $\tau_s \to \tau_1$; in the intermediate regime, $\tau_s \approx \tau_2/K$. Thus, if the duration $\tau_p$ of the 1064 nm pulse is such that $\tau_1 \ll \tau_p < \tau_2$, for sufficiently low net concentration $c$ of the solute ($K < 10$) the time constant $\tau_s$ would behave much as is observed experimentally, provided that $\tau_2$ is 5-30 ns ($\tau_2$ becomes shorter with the increasing temperature). However, further analysis leads to contradiction. While there is always a "slow" component for $K \ll 1$, its relative weight is quite small unless $\sigma_2 \ll \sigma_1$ (see the Appendix), where $\sigma_{1,2}$ are the cross sections for photodetachment. If that were the case, the power dependence of ratio $r$ vs. $J$ would be almost linear, whereas experimentally it "saturates" around $J \approx 10^{18}$ photon/cm$^2$ (Fig. 19S). Furthermore, as shown in section 4.1, the *absorptivity* of $\{e^-:S_n\}_{solv}$ is certainly *greater* than that of $e^-_{solv}$ and it may be expected that the detachment cross section would also be larger (unless there is a side photoreaction, such as proton transfer). Lastly, it seems counterintuitive that $\tau_2$ increases by no more than an order of the magnitude as the binding energy increases from ca. 400 meV (for acetonitrile) to ca. 800 meV (for alcohol multimer). As pointed out in section 4.1, the simple two-trap model cannot account completely for the observed spectral transformations for electrons in alcohol solutions as a function of solute concentration. The analysis of concentration plots for $\langle \mu \rangle$ given above also suggests that more than one kind of electron-attaching solute trap exists in the solution. The "slow" component is due to the equilibration between these *solute* traps. Such an equilibration may occur via coupled reactions (3) even if the conversion between $\{e^- : S_n\}_{solv}$ species via reaction (5) does not occur, however, as shown below, the experimental observations can be readily rationalized assuming that this "dipole coagulation" reaction does occur and $\tau_s$ is the measure of the corresponding reaction rate. We stress that the treatment is qualitative by necessity since all pertinent parameters for the several interrelated equilibria involved (reactions (3), (4), and (5)) cannot be extracted from the data unambiguously.

Using our estimates for $\tau_s$, the effective bimolecular constant $k_{eff} = (\tau_s c)^{-1}$ can be calculated (Figs. 11(a) and 11(b)). Due to the solute speciation via reaction (4), such a "constant" does not relate to any particular reaction (5). Still, for all temperatures between 2 and 23 °C the rate constants $k_{eff}$ for $c > 30$ mM converge to $(3-5) \times 10^9$ M$^{-1}$ s$^{-1}$ (Fig. 11(a)). The activation energy for the solvent viscosity is low (ca. 6.3±0.4 kJ/mol) so that diffusion controlled reactions in *n*-hexane are weakly activated. The typical rate of such reactions is ca. $(1-5) \times 10^{10}$ M$^{-1}$ s$^{-1}$, i.e., the apparent rate of reaction that may be responsible for the slow component is a fraction of what one would expect for reaction (5) involving a (normally diffusing) $\{e^- : (ROH)_n\}_{solv}$ cluster and a free *ROH* molecule or a small $(ROH)_m$ cluster. Given that in this concentration range, the fraction of the monomers is only 20-50% of the nominal concentration $c$ (Fig. 1S), the observed equilibration would be consistent with the occurrence of reaction (5) in the solution. As

29.

[EtOH] decreases below 20 mM, $k_{eff}$ rapidly increases, reaching $4 \times 10^{10}$ M$^{-1}$ s$^{-1}$ (for 5 mM ethanol at 15 °C). This increase becomes larger with increasing temperature (Fig. 11(a)). Note that reaction (5) can occur at a higher rate if the $\{e^- : S_n\}_{solv}$ species on the left side exhibits high (apparent) mobility.

The Arrhenius plots for $k_{eff}$ at fixed $c$ exhibit a large decrease of the activation energy with increasing $c$ (Fig. 11(b)). As [EtOH] increases from 6 to 12 to 25 to 45 mM, the activation energy decreases from 50±10 to 42±3 to 34±4 to 15±1 kJ/mol. The low activation energy obtained at the higher end of this concentration range can be explained by low mobility of the partners in reaction (5): as the reaction rate becomes controlled by (normal) diffusion, the activation energy approaches the activation energy for solvent viscosity. The dramatic increase in the reaction rate in very dilute solutions suggests that a mobile $\{e^- : (ROH)_n\}_{solv}$ species is involved. Thus, a large reaction barrier exists for reaction (5) involving the smallest $\{e^- : (ROH)_n\}_{solv}$ species (apparently, monomers or dimers) that has large apparent mobility. While the evidence for occurrence of direct reaction (5) is not clear-cut, our results clearly suggest that in very dilute solutions the electron equilibria are multistage and take considerable time to settle at the low temperature. This relatively slow settling is chiefly responsible for anomalies in the reaction rates first noticed by Gangwer et al. [25]

In Appendix A of ref. 6 we showed that eq. (7) would hold regardless of whether reactions (5) occur, provided that the residence time $\tau_t$ in a given trap is appropriately defined. On the strength of this result, one may inquire how this parameter (and the ratio $r$) change as a function of temperature and concentration. As shown in Fig. 6(b), for $\chi > 0.01$, $r \propto c^m$ with $m \approx 2.8$-2.9, at all temperatures. Note that as a function of $K$ (plotted in Fig. 20S(b)), the ratio $r$ in ethanol solutions behaves in a very similar fashion as this ratio in acetonitrile solutions (Fig. 4(b)). For $\chi \approx 0.14$ ethanol solution, $r$ increases by three orders of the magnitude relative to neat *n*-hexane. This huge increase swamps the (relatively small) effect of the change in the absorption (photodetachment) cross section with the ethanol concentration (section 4.1), i.e., most of the increase is due to the increase in the residence time $\tau_t$. As the exponential parameter $m$ is close to 3, the trimer is likely to be the prevalent $\{e^- : (ROH)_n\}_{solv}$ species in the solution that is photoexcited by 1064 nm light. The activation energy for $r$ does not depend on the ethanol concentration for $c > 20$ mM (Fig. 7(b)). From the Arrhenius plots, the activation energy is ca. 74±5 kJ/mol (770±50 meV). This energy is reasonably close to the binding energy $E_t$ of ca. 800 meV for the trimer (or, possibly, the tetramer) estimated from the van't Hoff plot in Fig. 8(b).

### *4.3. Irreversible trapping.*

The conductivity measurements presented in section 4.2 are not sensitive to electrons that are *irreversibly* trapped by higher multimers (for which the thermal detrapping time is longer than electron lifetime). Still, the electron could be photodetached even from such a species provided that the latter absorbs laser light.[6] An additional $\Delta \kappa$ signal with the amplitude increasing with the delay time $t_L$ [6] would be expected from such a photoexcitation. The time profile of the resulting $\Delta \kappa$ signal would be identical with that of $\kappa(t)$, save for the delay time of the 1064 nm pulse.[6,10] Surprisingly, no such additional signal was observed in either photosystem. Since these stable $\{e^- : S_n\}_{solv}$ species should absorb in the near infrared (as suggested by TA spectra in section 4.1) we are forced to conclude that the electron cannot be detached from large ethanol clusters by 1064 nm photons. As the TA signals observed in pulse radiolysis experiments decay in 1 μs, it is possible that electrons in such clusters are unstable,

31.

decaying by proton transfer to $RO^-$ anions, as happens for $e^-_{solv}$ in neat alcohols. [53,55] This slow reaction, however does not explain the lack of $\Delta\kappa$ signal from such a stable $\{e^- : S_n\}_{solv}$ species at shorter delay times (50-200 ns). Thus, we are left with the conclusion that the quantum yield for photodetachment from such clusters must be very low, either due to the predominance of bound-to-bound ($s \to p$) transitions and/or to rapid proton transfer in the excited state (as occurs in neat alcohols). [55]

For acetonitrile in *n*-hexane, a similar problem exists since no conclusive evidence for IR-absorbing $\{e^- : [MeCN]_n\}_{solv}$ multimer species was found in our photodetachment experiments, though there is some evidence for such a species (or a molecular anion) in concentrated solutions in *iso*-octane (section 4.2.1). The possible explanation for inefficient "dipole coagulation" is that the reaction of $\{e^- : MeCN\}_{solv}$ with acetonitrile yields the dimer anion, $(MeCN)_2^-$. [10] This covalently-bound anion absorbs poorly in the infrared, as its absorption band is centered in the visible. [10,11,14] Upon excitation in the vis and infrared, it photodissociates to $CH_3 + CN^- + MeCN$, although for 1064 nm photoexcitation the quantum yield is very low (<0.01). [10] The increase in the TA signal in the visible at longer delay time observed in pulse radiolysis experiments (Figs. 4S and 6S) can be accounted for by the slow generation of dimer anion via reaction (5), although a persistent cation (sections 2.2 and 4.1) can also explain these observations. The photodetachment experiments using 532 nm light [10] indicate that shorter-wave photoexcitation does induce the expected long-lived $\Delta\kappa$ signal with a pattern typical of a molecular anion or a stable $\{e^- : [MeCN]_n\}_{solv}$ species (see Figs. 7S and 8S in the Supplement of ref. 10). The results for *iso*-octane suggest that electron trapping by such a species may in fact be reversible; however, reaction (5) is strongly shifted towards the anion in the concentration range where eq. (6) does not hold.

## 5. Discussion.

### 5.1. Synopsis.

The following picture emerges from our results. In dilute solutions of acetonitrile in *n*-hexane and *iso*-octane ($\chi < 0.01$), a new electron species, $\{e^- : MeCN\}_{solv}$, is formed. The binding energy of this species is ca. 0.4 eV (relative to the mobility edge of the CB) which is ca. 0.2 eV greater than the binding energy for the intrinsic electron trap in neat *n*-hexane. The trapping reduces the rate of thermally-activated emission to the CB by ca. 200 times. In the specified concentration range, the solute trap involves a *single* acetonitrile molecule. For *iso*-octane, there is some evidence for nearly-irreversible electron attachment to larger solute clusters and/or delayed formation of molecular anions at higher solute concentration. For $\lambda < 1.6$ µm, the absorption spectra of trapped electrons in acetonitrile solutions are qualitatively similar to those in neat *n*-hexane, save for less prominent extension towards the visible. As explained in ref. 10, the formation of a molecular anion, $MeCN^-$, with a bent C-C-N chain is unlikely, given the low entropy of trapping and unfavorable energetics; furthermore, both theoretical calculations [10,11] and experimental observations [56] indicate that such an anion would not exhibit electron transitions in the infrared. Thus, the electron in the "acetonitrile trap" still resides in the interstitial cavity. The electron is dipole-bound to the CN group of acetonitrile molecule in the first solvation shell; [10,13] this interaction is mainly electrostatic.

It is commonly assumed [6,7,16] that the rate of thermal emission from electron traps is given by

$$\tau_t^{-1} \approx \nu_t \exp(-E_t/k_B T), \qquad (12)$$

where $v_t$ is the attempt-to-escape frequency ($10^{12}$-$10^{15}$ Hz or $\approx E_t/h$),[16] $E_t$ is the binding energy, and $k_B T$ is the thermal energy (ca. 25.4 meV at 23 °C). Simpleminded use of this formula indicates that the thermal emission from $\{e^- : MeCN\}_{solv}$ should be 3,000 times slower than that from $e^-_{solv}$, whereas experimentally it is only 200 times slower. Furthermore, for both of these electron traps, the frequency $v_t$ ($7 \times 10^{15}$ and $3 \times 10^{14}$ Hz, respectively) is unrealistically large. We conclude that eq. (12) overestimates the stability of electron traps in mixed solvents: something other than energetics controls this stability. The most likely cause is the loss of a solute molecule via backward reaction (5): the entropy factor prevents substitution in the solvation shell of the cavity electron. As shown by the previous workers [20-26] and confirmed in this study (section 4.2.2), the $\{e^- : ROH\}_{solv}$ species, for which the interaction with the electron is relatively weak, is thermodynamically unstable, while the binding energy for this species can only be higher than that for $e^-_{solv}$. Apparently, the entropy term prevails for this species: it either loses the $ROH$ molecule or undergoes proton transfer.

For larger alcohol clusters, the situation is different because the electron can interact with several OH groups and the binding energies are large. The same behavior was observed for methanol, ethanol, and 1-propanol. Our data point to alcohol trimer or tetramer as the predominant form of the cluster that reversibly traps electron via reactions (2,3) for $\chi < 0.015$, with a binding energy of 770 or 800 meV (section 4.2.2). Despite this large binding energy, the driving force of reaction (3) for these multimers is only 0.1 eV more negative than that for acetonitrile monomer with $E_t \approx 400$ meV. Surprisingly, photodetachment experiments using infrared light give no evidence for a larger $\{e^- : (ROH)_n\}_{solv}$ species postulated to account for *irreversible* electron scavenging. [20-25] Instability of the ground and/or excited state of these species towards proton transfer is a possible explanation for this unexpected observation. The size of the dominant electron trapping solute cluster appears to be tightly constrained, both from below and above, over a wide concentration range. Previous studies seem to support this conclusion. E.g., Smirnov et al. [26] studied ODMR spectra from dilute ethanol/squalane solutions at 23 °C and observed no change in the shape of the resonance line of solvated electrons as [EtOH] increased from 10 to 100 mM. As seen from Fig. 2(a) in section 4.1 the TA spectra of electrons in ethanol/*n*-hexane solutions can be understood, to a first approximation, in terms of just two species contributing to the spectrum. All of these observations point to a relative uniformity of trapped-electron species, at thermodynamic equilibrium in dilute alcohol solutions. The "dipole coagulation" and electron attachment do not yield metastable $\{e^- : (ROH)_n\}_{solv}$ species beyond a certain size.

Another intriguing observation is the possible occurrence of reaction (5) which could account for the slow settling of electron equilibria on the nanosecond time scale in very dilute solutions ($\chi < 5\times 10^{-3}$) at low temperature. Previous researchers believed that electron equilibria are settled very rapidly, and this is indeed the case for concentrated, room temperature alcohol solutions. E.g., Kenney-Wallace and Jonah [19] concluded that for $\chi > 0.03$-$0.05$ (which is well above the concentration range where "dipole coagulation" on the nanosecond time scale occurs), all equilibria (3) are settled within 30 ps. Gangwer et al. [25] estimated that methanol multimers ($n \geq 4$) in *iso*-octane attach electron with a rate constant $> 10^{12}$ M$^{-1}$ s$^{-1}$. However, neither group considered the possibility that the trap, once filled, can incorporate more solvent molecules or exchange the electron with a larger cluster. Our results suggest that such reactions ("dipole coagulation") [18] do occur in very dilute solutions. Apparently, reaction (5) involves a monomer or a dimer. The reaction is slow and thermally activated, suggesting a

35.

substantial barrier towards the inclusion of the alcohol molecule in the solvation shell of the $\{e^- : (ROH)_n\}_{solv}$ species.

## 5.2. The structure of the $\{e^- : S_n\}_{solv}$ species.

Since the interaction of the electron and polar molecules in the solvation shell of $\{e^- : S_n\}_{solv}$ is mainly electrostatic it would be natural to use the so-called "dielectric continuum" models for electron solvation [2c,57] to model its structure. In this class of models a few fixed dipoles are treated explicitly in the interaction Hamiltonian; the rest of the solvent, beyond some cutoff radius, is treated as a continuum with bulk dielectric properties. Gangwer et al. [25] used such an approach to estimate the energetics of electron trapping by methanol multimers in *iso*-octane. Unfortunately, this model makes no provision for the involvement of *solvent* molecules, whereas such an involvement is certainly important.

Since a self-consistent theory of solvated electron in alkanes is presently lacking, [6] below we use the simplest (Wigner-Seiz cell) model of such an electron species: [6,58,59] the *s* electron wave function $\Psi_s(r)$ occupying a spherical potential well with a hard core radius $a$ and depth $U$ (measured relative to the CB edge at $V_0$). The binding energy $E_t$ of the electron is a function of the depth $U$; the plot of this function for $a = 3.5$ Å is shown in Fig. 12(a). The latter estimate for the radius is supported by simulations of optical spectra in liquid [6,45] and vitreous [58] alkanes, magnetic resonance spectroscopy, [8b] and d.c. conductivity measurements at high pressure; [7a] this estimate is also compatible with the current microscopic theories of electron trapping in dense simple liquids [60] and amorphous polyethylene. [61] The critical well depth $U_c = \pi^2 \hbar^2 / 8 m_e a^2$ (where $m_e$ is the electron mass and $\hbar$ is the Planck constant) for $a = 3.5$ Å is ca. 770 meV; for $U < 4 U_c$, only one bound state exists; for $U < U_c$, no bound state exists. [59] In neat *n*-alkane liquids, $E_t/U_c \approx 0.2\text{-}0.3$ and only *bound-to-continuum* transitions are possible. The well depth $U \approx V_0 - E_{pol}$, where $E_{pol}$ is the polarization energy and $V_0$ is the energy of $e_{qf}^-$ vs. vacuum. [6,7] The polarization energy can be crudely estimated using the Born equation, $E_{pol} \approx -(1-\varepsilon^{-1})e^2/2a$, where $e$ is the elementary charge and $\varepsilon$ is the bulk dielectric constant. [6] Apparently, this formula gives too low an estimate since the polarizability of $C-C$ bonds in the groups lining the solvation cavity appears to be several times greater than that in the bulk liquid. [8a] For $r > a$, the radial wave function $\Psi_s(r)$ of the ground state decreases exponentially towards the bulk as $r\Psi(r) \propto \exp(-r/r_{loc})$, where $r_{loc} \approx 2a\pi^{-1}[U_c/E_t]^{1/2}$ is the localization radius. [58a,59] In neat alkanes, this radius is 4-5 Å, i.e., the extension of the electron density onto the solvent is very significant. In this respect, $e_{solv}^-$ in alkanes is different from the electron in polar solvents, where the binding energy is large, the electron wavefunction is confined inside the cavity, and bound-to-bound $s \to p$ transitions dominate in the visible and in the infrared. [2,4,54] In polar solvents, the cavity radius $a$ rapidly increases with increasing temperature. [31,32] For alkanes, this radius is nearly constant; [6,7a,58] $E_{pol}$ also depends weakly on the temperature. However, $U$ (and, consequently, $E_t$) decreases rapidly with increasing temperature since the energy $V_0$ of $e_{qf}^-$ increases greatly with solvent density. [6,7a] The latter changes substantially as a function of temperature for alkane solvents. In ref. 6 the absorption spectra of $e_{solv}^-$ in liquid *n*-hexane were fit using the spherical well model and binding energies $E_t$ were estimated at several temperatures; these estimates are in good agreement with the activation energies of thermal emission obtained from d.c. conductivity experiments. For neat *n*-hexane at 23 ºC, $E_t \approx 200$ meV, $U \approx 1.39$ eV and the absorption maximum for the bound-to-continuum transition is at $\lambda \approx 2.95$ μm (Fig. 12(b), trace (i)). [6,45]

The same spherical well model can be used to estimate the binding energies for $\{e^-:S_n\}_{solv}$ species. To this end, we assumed that the "mean" well depth $U$ increases stepwise relative to the same quantity $U_n$ for the intrinsic solvent trap by the energy $V_{dip}$ of interaction with a polar molecule. The latter energy is estimated using the point dipole approximation. All solute molecules are equivalent, so that $U = U_n + \langle n \rangle V_{dip}$. We further assumed that the cavity size does not change upon the inclusion of these solute molecules.

Let us consider first the acetonitrile monomer. As suggested in section 5.1, the cavity electron is solvated by the methyl group with the CN group pointing away from the cavity center. Assuming that the methyl protons are at the hard core radius $r \approx a$, the center of the CN bond is ca. 6 Å away from the cavity center. The interaction energy $V_{dip}$ for the 3.92 D dipole in the cyano group is ca. 330 meV, which gives $U \approx 1.72$ eV and $E_t \approx 390$ meV. This binding energy compares favorably with the 400 meV obtained experimentally (section 4.2.1). Fig. 12(b), trace (v) shows the simulation of the absorption spectrum for the 390 meV trap. The absorption maximum is at $\lambda \approx 1.68$ μm which is just beyond the observation range of our pulse radiolysis - TA setup ($\lambda < 1.6$ μm, section 4.1). The absorptivity at $\lambda = 1$ μm is ca. 2 times that for the $E_t \approx 200$ meV trap, in reasonable agreement with experiment. For $\lambda < 1.6$ mm, the spectral profiles for 200 and 400 nm traps are similar, with the shallower trap exhibiting less sloping towards the blue. This is also in agreement with experiment (Fig. 1(a)).

For alcohol clusters, the same method can be used to solve the inverse problem: estimating the mean distances to OH dipoles from the energetics. The binding energy $E_t$ for the prevalent ethanol multimer is ca. 770 meV (from the activation energy of electron photodetachment in Fig. 8(b)) or 800 meV (from van't Hoff plots in Fig. 7). Using the plot in Fig. 12(a) we find that $U$ of 2.27 or 2.31 eV would correspond to these binding energies, so that $V_{dip} \approx 880/\langle n \rangle$ meV, where $\langle n \rangle$ is the mean number of solute molecules in the solvation shell of the cavity electron. Placing the center of a radially aligned OH dipole (ca. 1.7 D) at a distance $r_{OH}$ from the cavity center, we find $r_{OH} \approx 4.2$ Å for $\langle n \rangle = 3$ and $r_{OH} \approx 4.8$ Å for $\langle n \rangle = 4$. Since the O-H bond length is ca. 1 Å, the trimer gives a better match, with the OH dipoles at an angle to the radial direction. Such an arrangement is in accord with quantum mechanical - molecular dynamics models of the solvated electron in neat water and alcohols. [4,54] The monomer and the dimer are predicted to have binding energies of 365 and 555 meV, respectively, with their absorption bands centered at 1.78 and 1.39 μm, respectively (Fig. 12(b), traces (ii) to (iv)). The trimer ($E_t \approx 770$ meV) is calculated to have maximum absorbance at $\lambda \approx 1.18$ μm; the molar absorptivity at $\lambda = 1$ μm is ca. 6.2 times greater than that for $e^-_{solv}$ ($E_t \approx 200$ meV; see Fig. 12(c)). This estimate is in agreement with the factor of ca. 5 times obtained experimentally (section 4.1). At the higher end of the concentration range explored (ca. 120 mM), at which the conductivity signal is dominated by a single reversibly trapped species, the TA spectrum peaks at 0.95-1 μm (Fig. 2(a)). This position is in a reasonable agreement with the estimate given above. While our approach is obviously crude, it yields reasonable estimates for the energetics observed. Improving this model is hindered by the lack of microscopic insight in the nature of *solvent* traps in liquid alkanes.

## 6. Conclusion.

Electron localization in dilute ($\chi < 0.015$) solutions of polar molecules in nonpolar liquids has been studied using TA spectroscopy and conductivity. In the conductivity experiments, 1.17 eV photon excitation was used to detach the electron from a $\{e^- : S_n\}_{solv}$ species and observe relaxation dynamics on the nanosecond time scale. In acetonitrile solutions, the electron can attach to a single solute molecule forming the $\{e^- : MeCN\}_{solv}$ species. The dynamics of this attachment can be understood using a simple two-trap model. The binding energy for this $\{e^- : MeCN\}_{solv}$ species is ca. 400 meV and its lifetime (limited by thermal emission to the conduction band) is ca. 1 ns at 23 °C. The properties of this electron species can be rationalized assuming that *MeCN* substitutes for the solvent molecules in the first solvation shell of $e^-_{solv}$. The methyl group of *MeCN* is at the cavity wall, while the C-N group is ca. 6 Å away from the center of the solvation cavity and points outwards. The resulting structure is midway between the solvated electron in neat alkanes [6,7,16,17,58] and in liquid acetonitrile. [10-14] Further inclusion of acetonitrile molecules does not occur, most likely due to the formation of a covalently bound dimer anion with lower energy. [10,11,14,56] Interestingly, thermal emission from the acetonitrile monomer trap (as well as from the related ethanol monomer trap) appears to be much faster than expected from the energetics alone (section 5.1). Apparently, entropy plays as much a role as the binding energy in determining the stability of these solute traps. E.g., while the binding energy for the $\{e^- : ROH\}_n$ species is ca. 165 meV greater than this energy for $e^-_{solv}$ (section 5.2), the electron equilibrium is completely shifted towards the shallower trap. The driving force of electron attachment to ethanol *tetramer* is only 0.1 eV more negative than that for acetonitrile *monomer*.

Electron trapping in dilute alcohol solutions is more involved, as "dipole coagulation" (reaction (5)) occurs concurrently with several electron and H-bonding equilibria (reactions (3) and (4), respectively). The resulting dynamics are rather complex and we were unable to disentangle all reactions involved. Still, several conclusions can be reached: The electron does not attach to the alcohol monomer. This is due both to unfavorable thermodynamics (as the resulting species is unstable towards the reverse reaction (3)) and to the occurrence of proton transfer reaction (8). The latter reaction probably occurs for other $\{e^-:(ROH)_n\}_{solv}$ species. Its rapid occurrence (either in the ground or in the excited states) for higher $\{e^-:(ROH)_n\}_{solv}$ multimers which attach the electron *irreversibly* might account for the lack of IR-light induced electron detachment from these species. The lower multimers ($n<5$) attach the electron *reversibly*. Following the initial attachment reaction (3), "dipole coagulation" reaction (5) is observed on the nanosecond time scale. When equilibrium is reached, the prevalent $\{e^-:(ROH)_n\}_{solv}$ species ($n \approx 3\text{-}4$) binds the electron by ca. 800 meV. These energetics and the TA spectra observed are consistent with the OH groups of solute molecules lining the solvation cavity.

This study was conceived as a search for condensed-matter analogs of dipole-bound [62] and cluster [2b,12,13,63,64] anions occurring in the gas phase. In recent years, cluster anions of polar molecules, such as $(H_2O)_n^-$, have been extensively studied (see refs 2b, 63, and 64 and references therein). Such clusters are interesting in their own right, but also as model systems for electron solvation in the bulk liquid. Since surface trapping prevails in small and even medium-size clusters ($n<20$), [2b,64] the direct comparison between these cluster anions and $e_{solv}^-$ in liquids is difficult, although possible. [64] On the other hand, electron trapping in large clusters ($n>25\text{-}50$) where internal localization prevails is as difficult to model as that in neat liquids. The $\{e^-:S_n\}_{solv}$ species occurring in alkanes provide what these small gas-phase $S_n^-$ clusters do not: a model system for $e_{solv}^-$ in a neat polar liquid -- with few polar molecules directly involved. This, in turn, suggests that a species whose anion core closely resembles the first solvation shell of $e_{solv}^-$ in a polar liquid may be achieved in the gas phase by making a composite cluster anion in which several polar molecules are embedded in a large number of nonpolar molecules. Perhaps, "solvents" for which $V_0$ (the energy of quasifree electron relative to the vacuum)

41.

is small and electron trapping is facile (e.g., *n*-alkanes other than methane) would be most suitable. The resulting $\{e^- : S_n\}_{solv}$ species has the same structure and energetics regardless of the alkane "solvent" (section 4).

Our study also hints at the possibility of a fixed-geometry molecular cage "solvating" the electron in an alkane solution. Such a cage (in analogy with alcohol multimers in *n*-hexane) would include several radial groups assembled around the central cavity. Crown ethers, cryptands, and cyclosiloxanes provide a structural motif conducive to electron trapping in this fashion. There are precedents for fixed-geometry $e^-_{solv}$ species in low-temperature crystalline solids: single crystals of sugars [65] and hexagonal ice [66] are known to trap electrons due to the fortuitous orientation of OH groups at certain interstitial sites. Another example is electrides (e.g., $Cs^+[\text{18-crown-6}] \bullet e^-$) [67] that (presumably) trap electrons in cavities and channels. Arguably, electrons "solvated" by well-defined cages would constitute an ideal condensed-phase model for solvated electrons in polar liquids since their fixed geometry would make *ab initio* modeling much easier and their properties would be less dependent on solvent fluctuations.

## 7. Acknowledgement.

IAS thanks C. D. Jonah, R. A. Holroyd and A. Mozumder for many useful discussions. This work was supported by the Office of Science, Division of Chemical Sciences, US-DOE under contract number W-31-109-ENG-38.

*Supporting Information Available:* A single PDF file containing (1.) Appendix: Two-trap model. (2.) Figs. 1S to 22S with captions. This material is available free of charge via the Internet at http://pubs.acs.org.

**Figure captions.**

**Fig. 1.**

(a) Normalized end-of-pulse TA spectra observed in pulse radiolysis of Ar-saturated (i) *n*-hexane (open triangles), (iv) 47 mM MeCN in *n*-hexane (filled squares) and (v) 166 mM MeCN in *iso*-octane (filled circles). A 4 ns fwhm, 21.5 nC, 20 MeV electron pulse was used to obtain all of these traces. To facilitate the comparison, the spectra were normalized at 1.55 µm, where only trapped electron absorbs. Trace (ii) (open diamonds) shows the spectrum of solvent olefin cation observed in $CO_2$-saturated *n*-hexane (see also Fig. 2S); this spectrum is normalized at 0.5 µm, where most of the absorbance is from this cation. Trace (iii) is the difference trace (i.e., electron absorbance). Solid lines are guides for the eye. The dashed line (vi) is scaled down trace (iv) drawn to illustrate that TA spectra for $(e^- : MeCN)_{solv}$ in the visible are similar for both solvents. (b) End-of-the pulse TA signal at $\lambda = 1$ µm (open squares, to the right) and 1.55 µm (filled circles, to the left) vs. $[MeCN]$ in Ar-saturated *n*-hexane. The kinetics are shown in Fig. 3S).

**Fig. 2.**

(a) End-of-pulse TA spectra observed in pulse radiolysis of Ar-saturated *n*-hexane containing 0 mM (open circles), (i) 39 mM (filled diamonds), (ii) 73 mM (filled triangles), (iii) 121 mM (filled squares) and (iv) 262 mM EtOH (filled circles). The solid lines are Lorentzian-Gaussian plots. Dashed lines (vi) and (v) are weighted sums of traces (iv) and (i). (b) End-of-pulse 1 µm absorbance vs. alcohol concentration for the same system (molar concentration is given at the bottom, mole fraction $\chi$ at the top). Open circles are for ethanol; filled triangles are for 1-propanol (plotted from Fig. 4 in ref. 23); for the latter solute the concentrations were scaled down by a factor of two.

**Fig. 3.**

Decay kinetics of d.c. photoconductivity signal $\kappa$ from Ar-saturated solutions of n-hexane containing (a) 21.3 mM and (b) 54.8 mM MeCN (to the left). The signal from ions ($\kappa_i$) is subtracted; the residual signal is from the electron. The solution was photoionized using a 15 fwhm pulse of 248 nm light; trapped electrons were subsequently photoexcited using a 6 ns fwhm, $9 \times 10^{18}$ photons/cm$^2$ pulse of 1064 nm light. The 1064 nm photon induced signal ($\Delta\kappa$) plotted to the right has the same temporal profile as the excitation pulse; the decrease in the amplitude as a function of the delay time $t_L$ of the 1064 nm pulse faithfully follows the $\kappa - \kappa_i$ kinetics. See also Figs. 9S(a) and 16S.

**Fig. 4.**

(a) Concentration dependence of the extrapolated electron conductivity $\kappa_0$ (open circles) for 248 nm photoexcitation of MeCN in Ar-saturated room-temperature *n*-hexane (to the left). The concentration dependence of parameter $K = \langle\mu_n\rangle/\langle\mu\rangle - 1$ (filled circles; to the right) is shown for the same system. The solid line is calculated using eq. (6), and the dashed line is a linear fit; the slope of this line gives the equilibrium constant of reaction (3). (b) Ratio $r$ (for $t_L \approx 50$ ns) for the same system determined using $8.1 \times 10^{18}$ photon/cm$^2$ (circles, to the left) and $5.4 \times 10^{17}$ photon/cm$^2$ (squares, to the right) pulse of 1064 nm light. See Fig. 11S for power dependencies at different MeCN concentrations. Open symbols indicate numerical integration of $\Delta\kappa$ signals; filled symbols indicate the integrals of least squares optimized Gaussian fits. Compare with $r$ vs. $K$ plot for ethanol/*n*-hexane solutions given in Fig. 20S(b).

**Fig. 5.**

Arrhenius plots for extrapolated conductivity signals $\kappa_0$ ($t \rightarrow 0$) from electrons (filled symbols and solid lines, to the left) and ratio $r$ for $t_L =50$ ns and $J =9 \times 10^{18}$ photon/cm$^2$ (open symbols and dashed lines, to the right) in 17.3 mM solution of MeCN in *n*-hexane (squares) and 22 mM solution of MeCN in *iso*-octane (circles). The corresponding activation energies for $\kappa_0$ and $r$ are 52.7±2.2 and 39±1.8 kJ/mol (*n*-hexane) and 45.1±1.3 and 39±0.9 kJ/mol (*iso*-octane).

**Fig. 6.**

(a) The plot of normalized electron mobility (i.e., normalized $\kappa_0$ signal) vs. $const \times [ROH]$ for solutions of methanol (filled circles), methanol-d$_4$ (open circles), ethanol (filed squares), ethanol-d$_1$ (open squares), and 1-propanol (filled diamonds) in Ar-saturated *n*-hexane at 23 °C. The molar concentration is given at the bottom; mole fraction $\chi$ is given at the top. The scaling constants are 1, 0.54, and 0.44 for methanol, ethanol, and 1-propanol, respectively. The solid line is a fit to eq. (9) with $m \approx 3.5$ and $c_0 \approx 18$ mM. Observe the universality of the behavior for all alcohols and their isotopes.

(2) *To the left:* Normalized electron mobility (filled symbols) vs. net [EtOH] for 1.7 °C (diamonds), 8.1 °C (triangles), 14.9 °C (squares) and 23 °C (circles) plotted on a logarithmic concentration scale. The solid lines are optimum fits obtained using eq. (9*). *To the right:* concentration plots for ratio $r$ (open symbols, for the same excitation conditions as in Fig. 5); note the logarithmic scale. The same symbol shapes are used as in (a). The straight lines drawn through the symbols correspond to the exponential power of ca. 2.8.

**Fig. 7.**

(a) *To the left:* Normalized electron mobility vs. nominal $c = [EtOH]$ (the same data and symbol shapes as in Fig. 6(b)). The solid and dashed lines drawn through the symbols are fits obtained using eq. (11), for a tetramer and a trimer, respectively. To the left:

estimated mole fraction $f_4 = 4[S_4]/c$ of tetramer at equilibrium vs. $c$ for the four temperatures given in Fig. 6(b); the higher fraction corresponds to the lower temperature. (b) van't Hoff plots for the equilibrium constant of reaction (3) for the tentative trimer (circles) and tetramer (squares) of EtOH and for MeCN monomer (diamonds). Vertical bars indicate 95% confidence limits of the least squares fit to eqs. (11) and (6) for ethanol and acetonitrile, respectively.

**Fig. 8.**

Arrhenius plots for (a) electron conductivity and (b) ratio $r$ (the same excitation conditions as in Fig. 5) for ethanol solutions in Ar-saturated $n$-hexane containing 5 μm anthracene as a photosensitizer. The same symbols are used for the same concentrations in each plot: $[EtOH]$ is 0 mM (upturned triangle), 6 mM (diamonds), 12 mM (triangles), 25 mM (circles) and 45 mM (squares). Filled and open symbols in (b) indicate different integration procedures (the same convention used in Fig. 5). In (a), the activation energy increases with the given concentrations as 32.6±0.4, 36±0.5, 44.8±1.4, 76.1±3.6 and 89.7±2 kJ/mol, respectively. The activation energies for the four traces shown in (b) are 20.9±0.4, 25.9±1.7, 73.4±2.4 and 74±5 kJ/mol, respectively (in ascending order with increasing [EtOH]).

**Fig. 9.**

Decay kinetics $\Delta\kappa$ of the conductivity signal induced by 1064 nm photoexcitation of trapped electrons in Ar-saturated $n$-hexane solution containing 6.1 mM (solid diamonds), 11 mM (open triangles), 15.6 mM (filled squares), 24.9 mM (open squares) and 35 mM (filled circles) ethanol at 1.7 C, using the same excitation conditions as in Fig. 5. The kinetics are spaced vertically to facilitate the comparison. The solid lines are least squares fits to the weighted sum of a Gaussian and the Gaussian convoluted with an exponential.

**Fig. 10.**

Same as Fig. 9, for normalized $\Delta\kappa$ kinetics in 12 mM ethanol in $n$-hexane at three solution temperatures: 1.7 °C (circles), 8.1 °C (squares) and 14.9 °C (triangles).

**Fig. 11.**

(a) Concentration plots (for fixed temperature) and (b) Arrhenius plots (for fixed concentration) of the effective reaction constant $k_{eff} = (\tau_s c)^{-1}$ corresponding to the slow components in the $\Delta\kappa$ kinetics (see Figs. 9 and 10). The concentrations and temperatures are indicated in the plots. Filled and open symbols in (a) correspond to different least-squares fitting protocols which differ in the statistical weight given to the slow component.

**Fig. 12.**

Theoretical simulations using a spherical well model with a hard core well radius $a = 3.5$ Å. (a) The dependence of the binding energy $E_t$ of the trap on the interaction potential $U$ inside the well. (b) Simulated absorption spectra for (i) $e^-_{solv}$ in neat $n$-hexane, (ii-iv) ethanol clusters in $n$-hexane ($(e^- : S_n)_{solv}$ with $n = 1$-$3$, respectively) and (v) $(e^- : MeCN)_{solv}$. Simulation parameters are given in the text. (c) The plot of molar extinction coefficient for trapped electron at 1 μm vs. the number of "attached" EtOH (filled circles) and MeCN (open squares) molecules in the first solvation shell (the same calculation as in (b)).

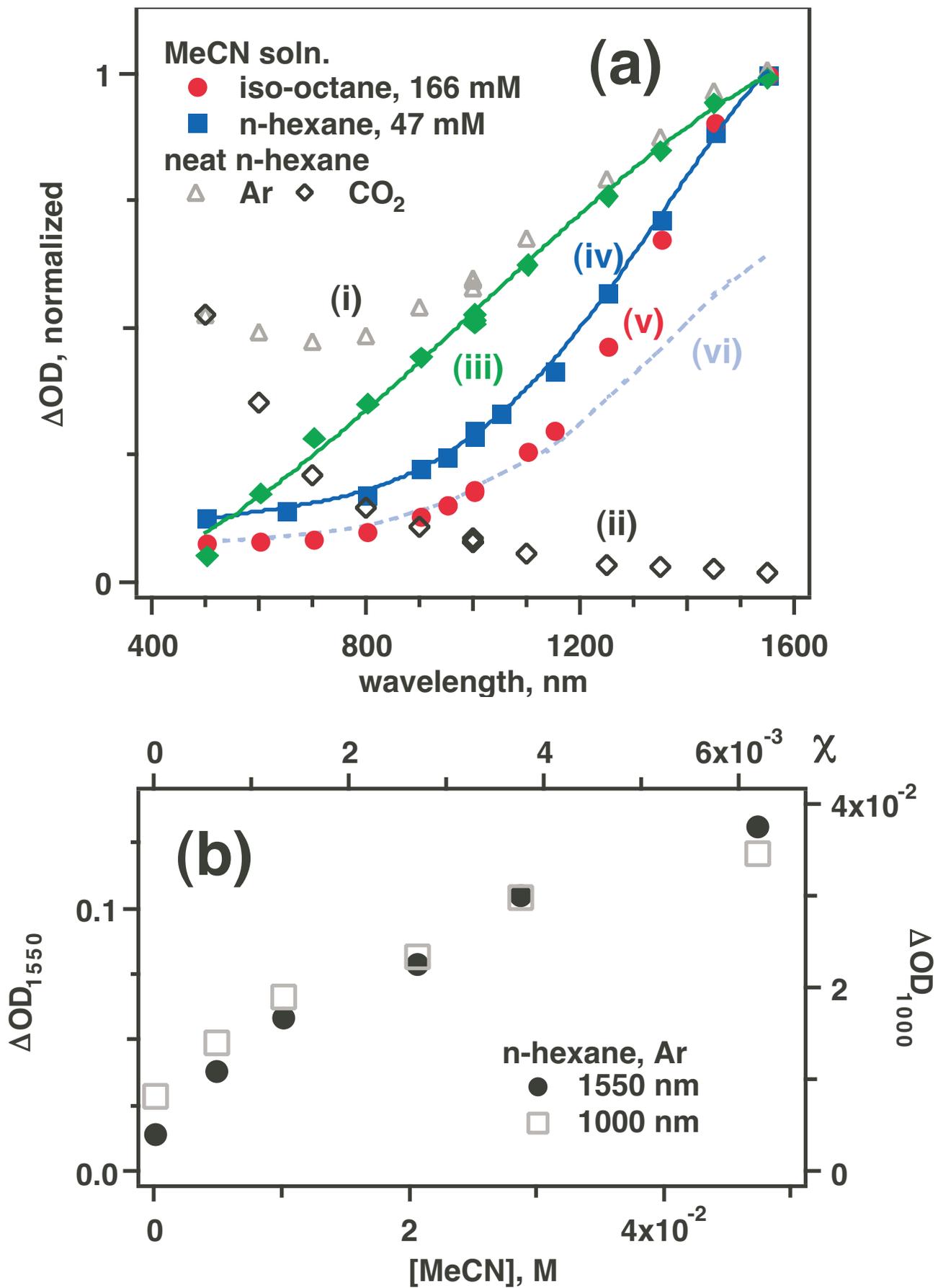

Figure 1; Shkrob & Sauer

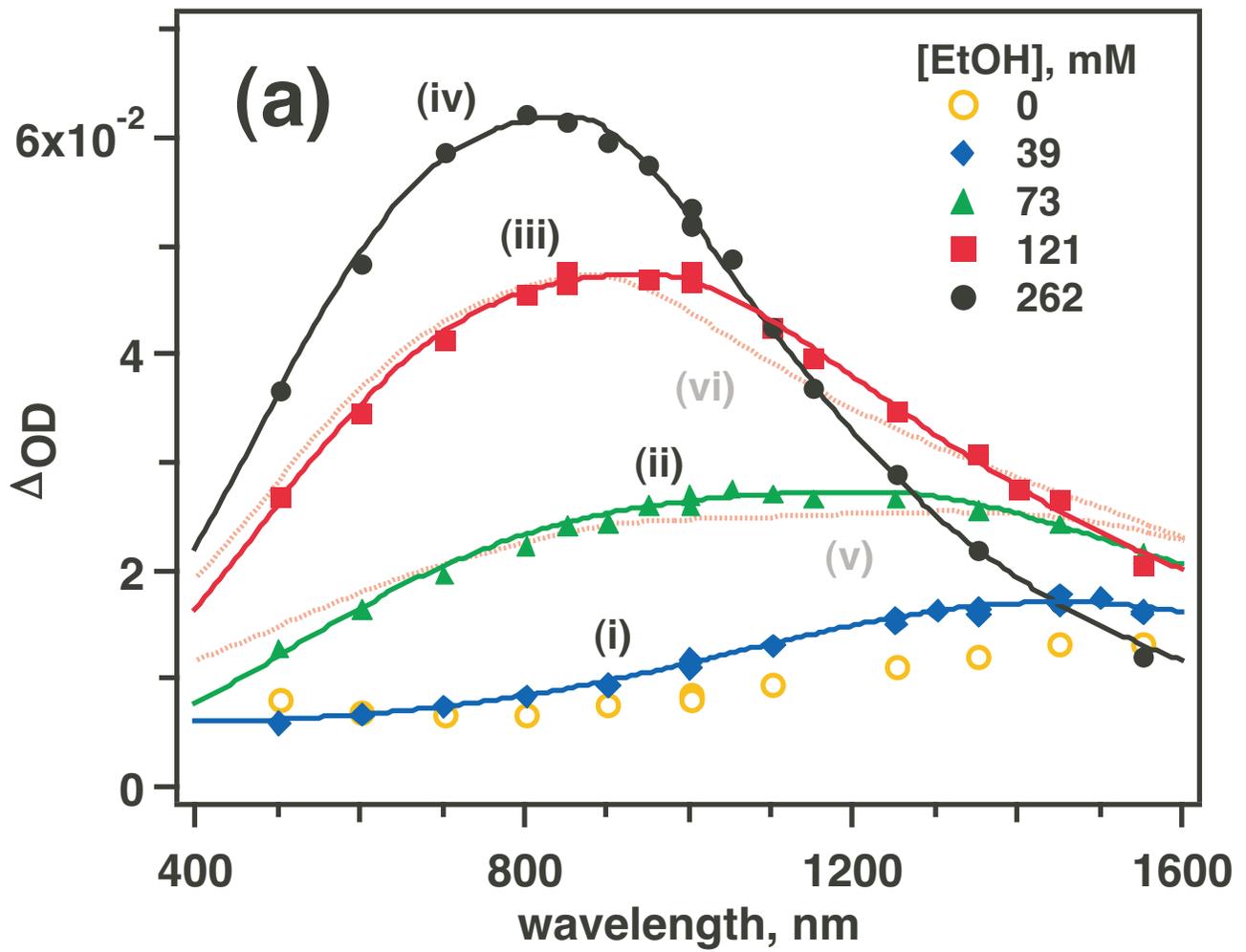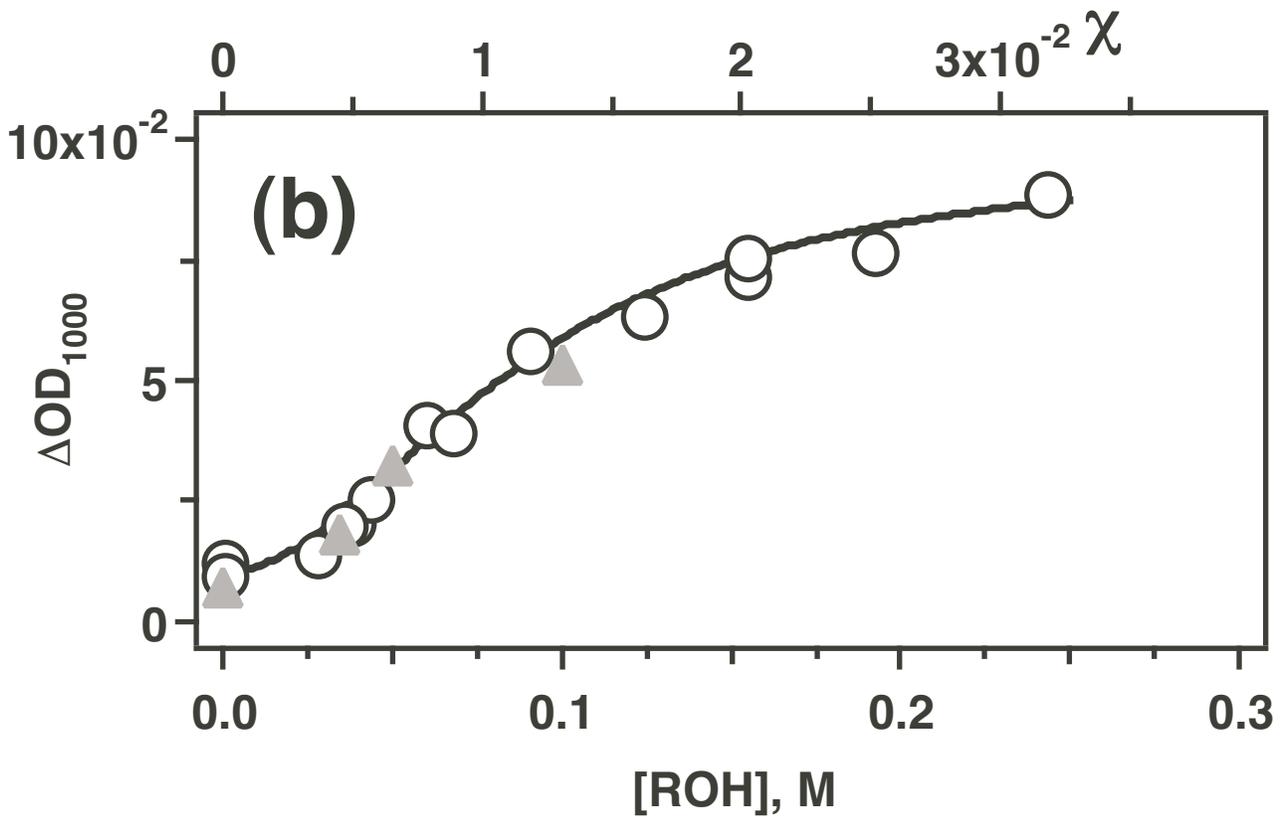

Figure 2; Shkrob & Sauer

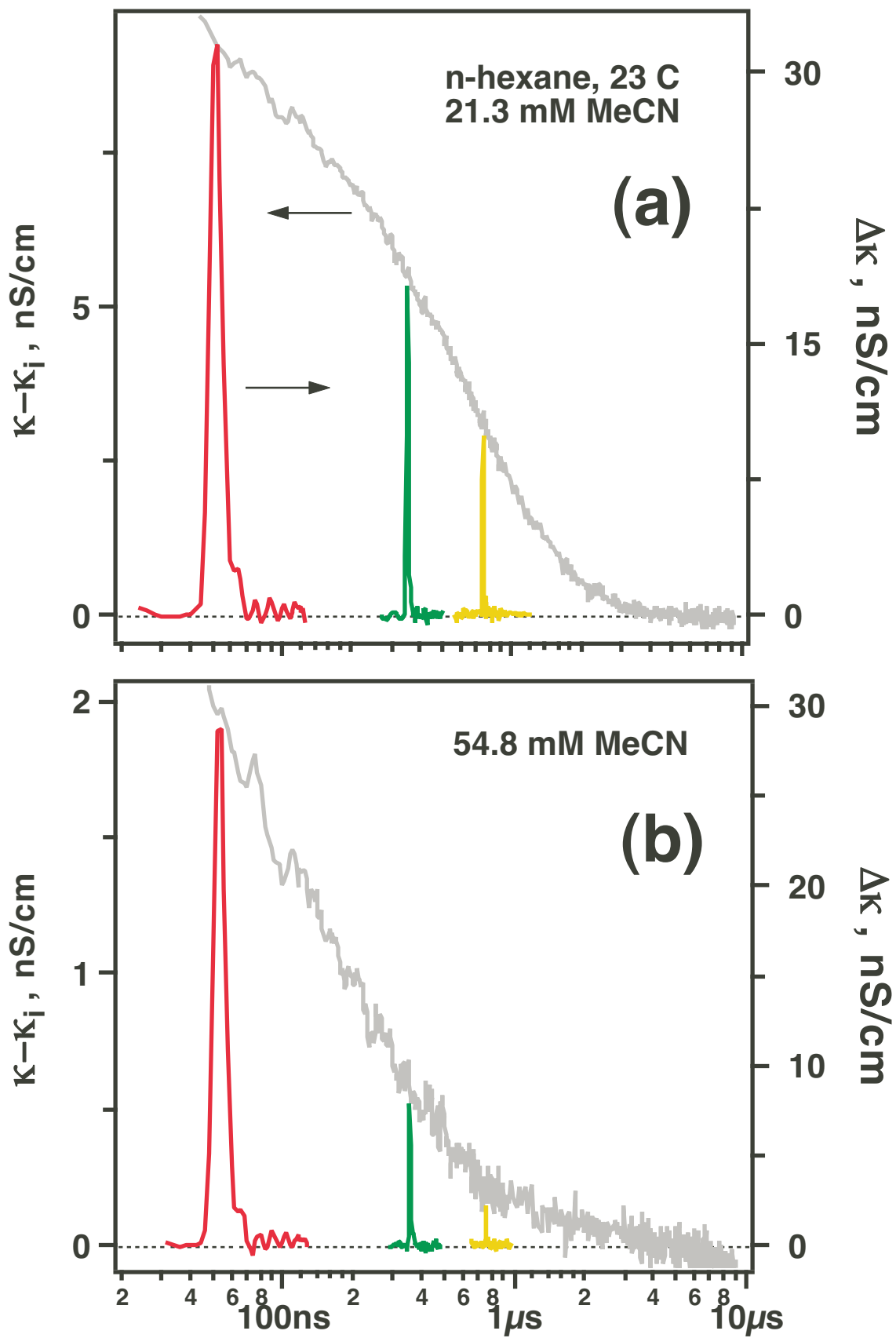

Figure 3; Shkrob & Sauer

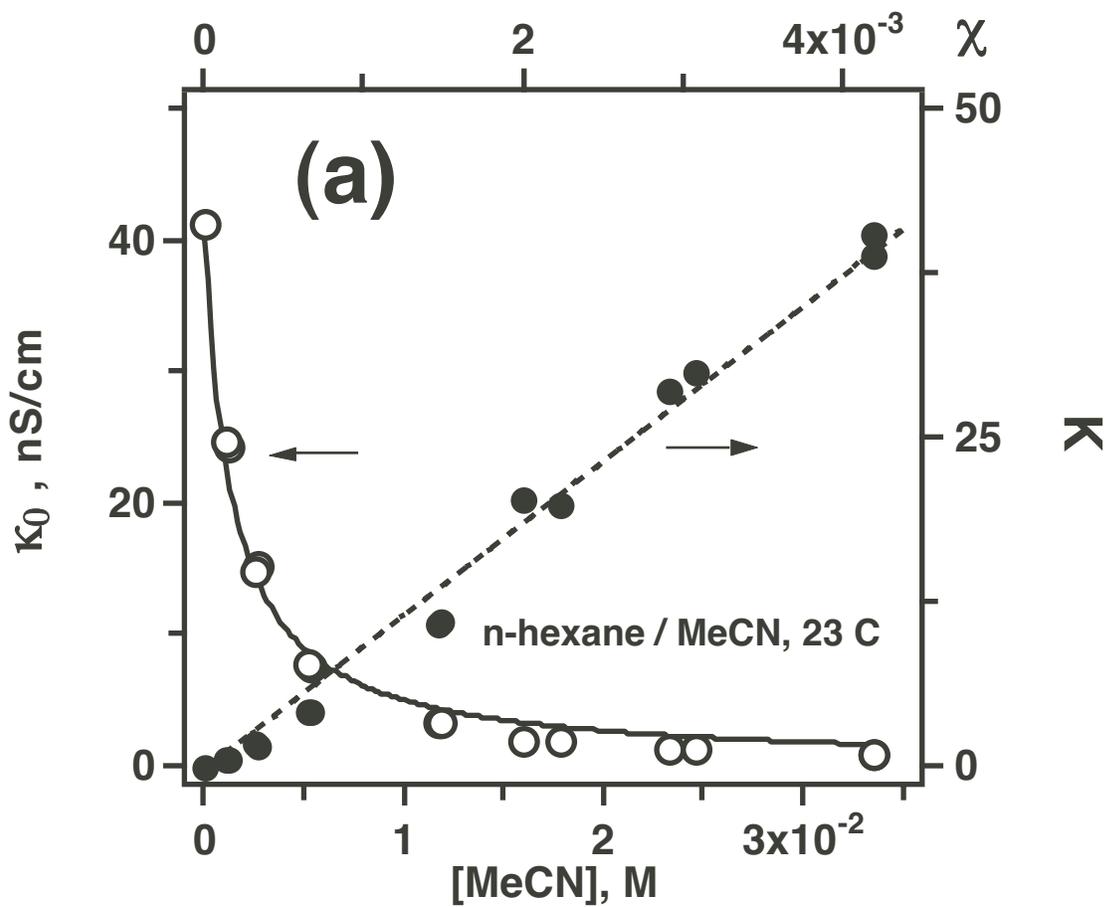
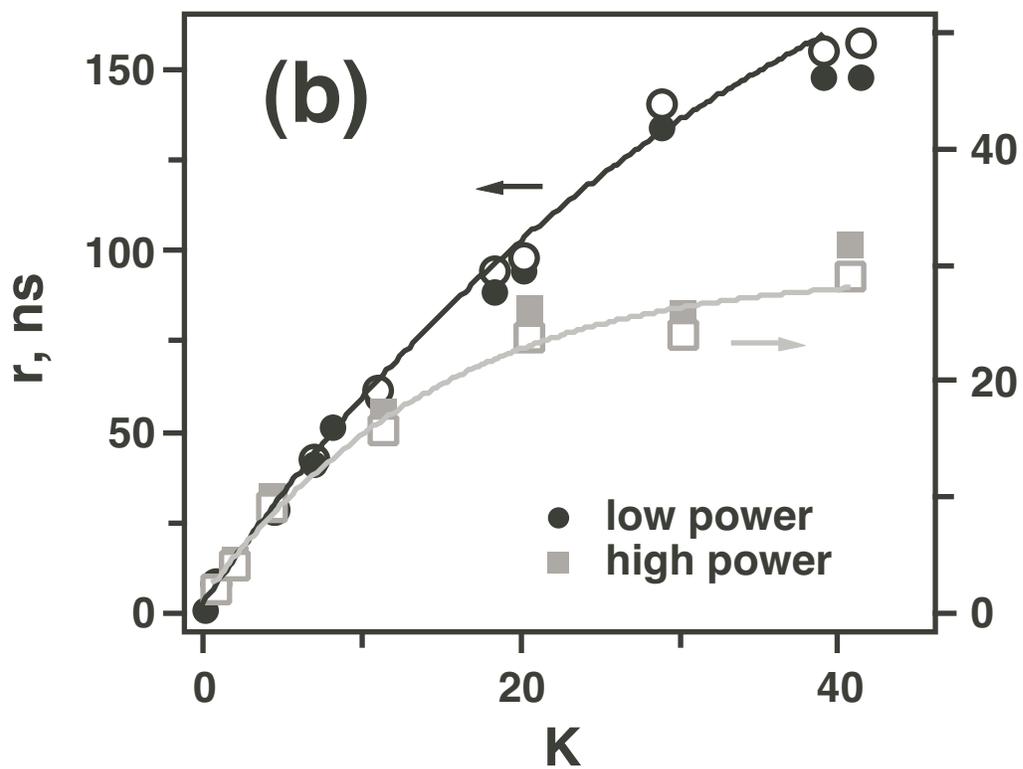

**Figure 4; Shkrob & Sauer**

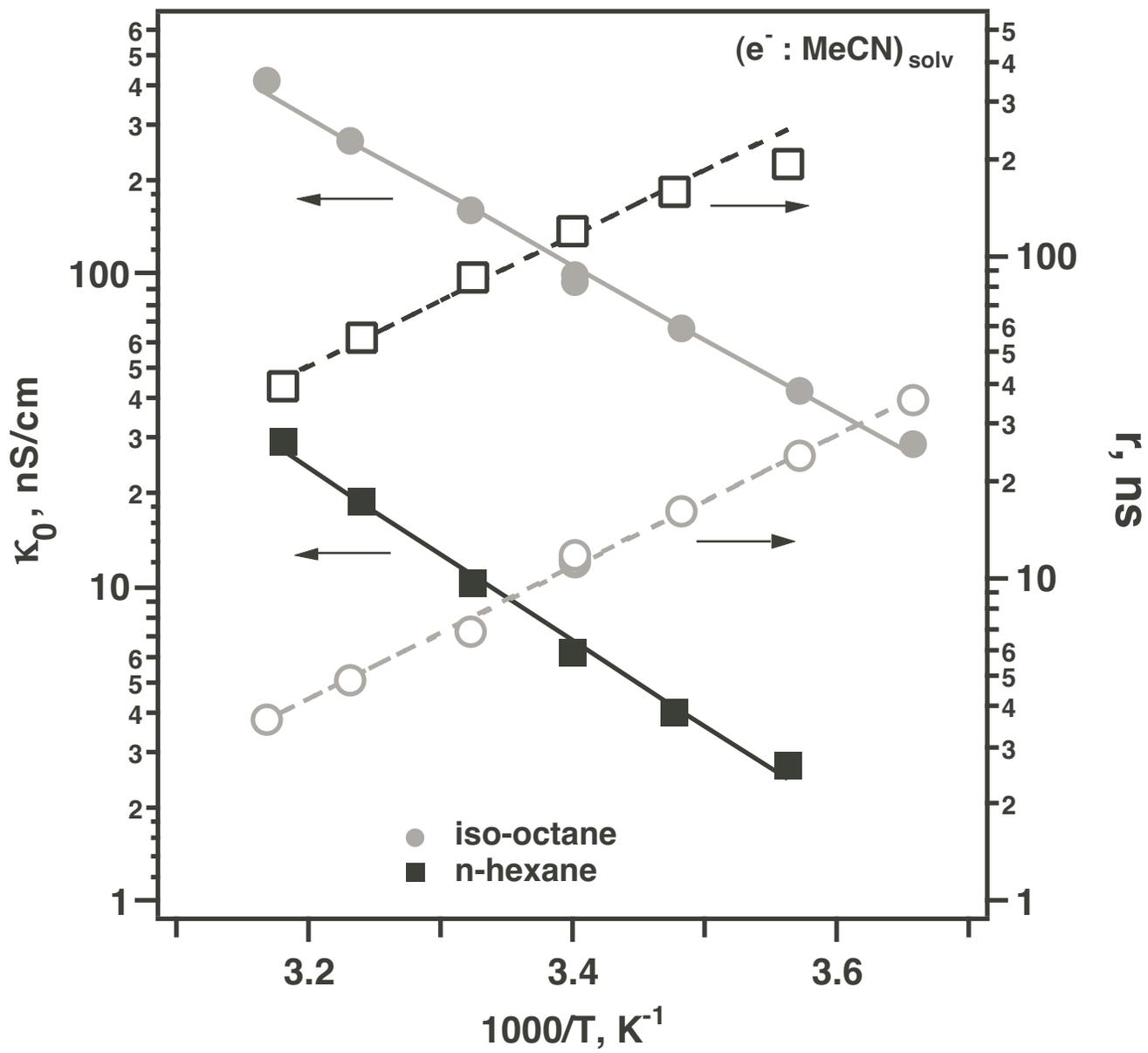

**Figure 5; Shkrob & Sauer**

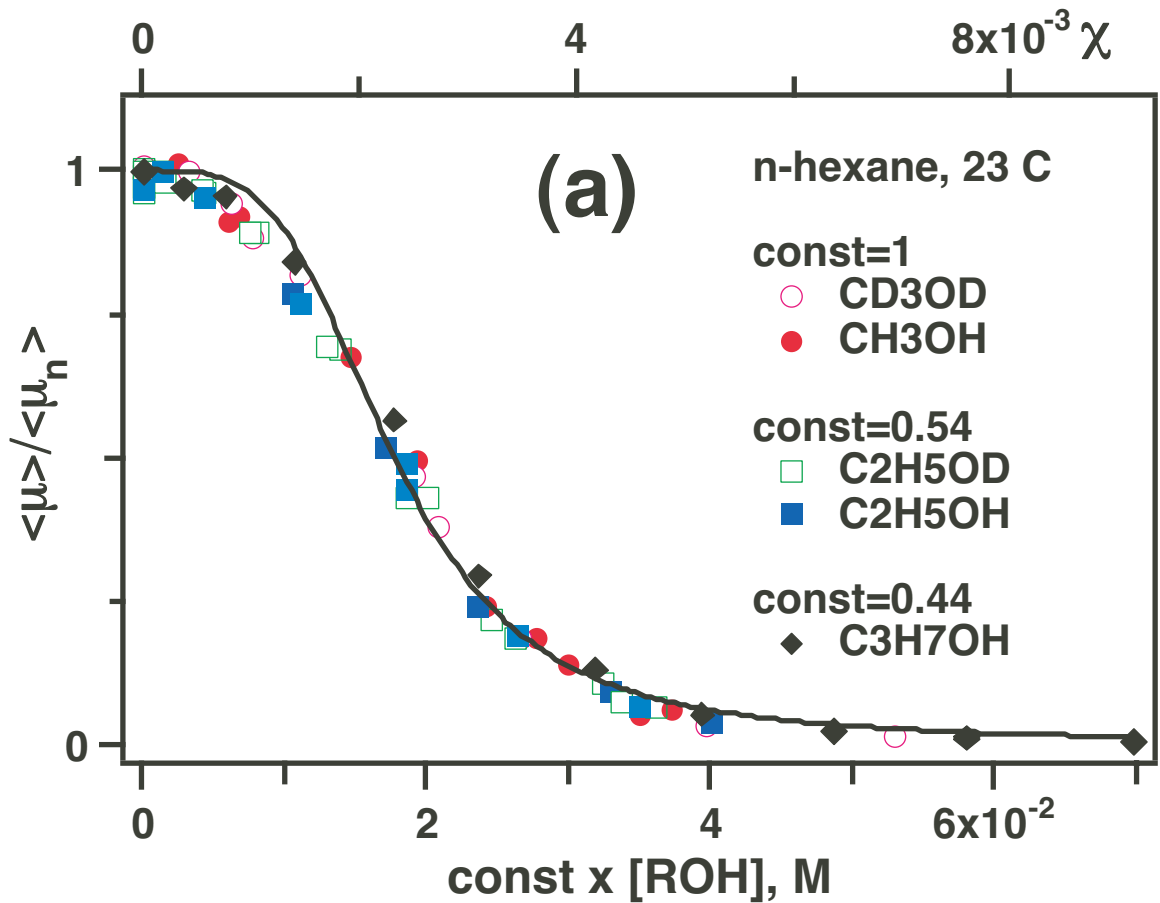
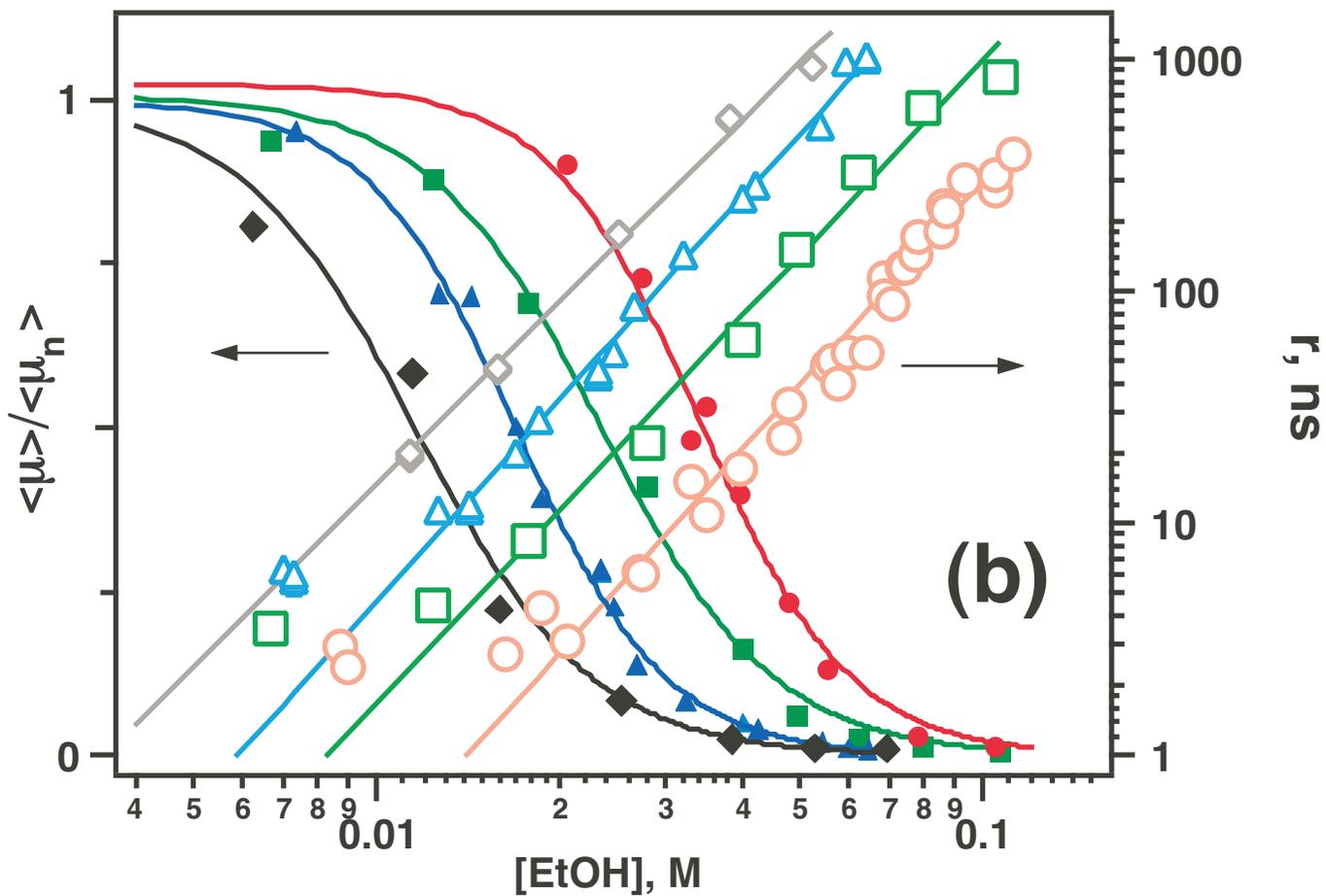

Figure 6: Shkrob & Sauer

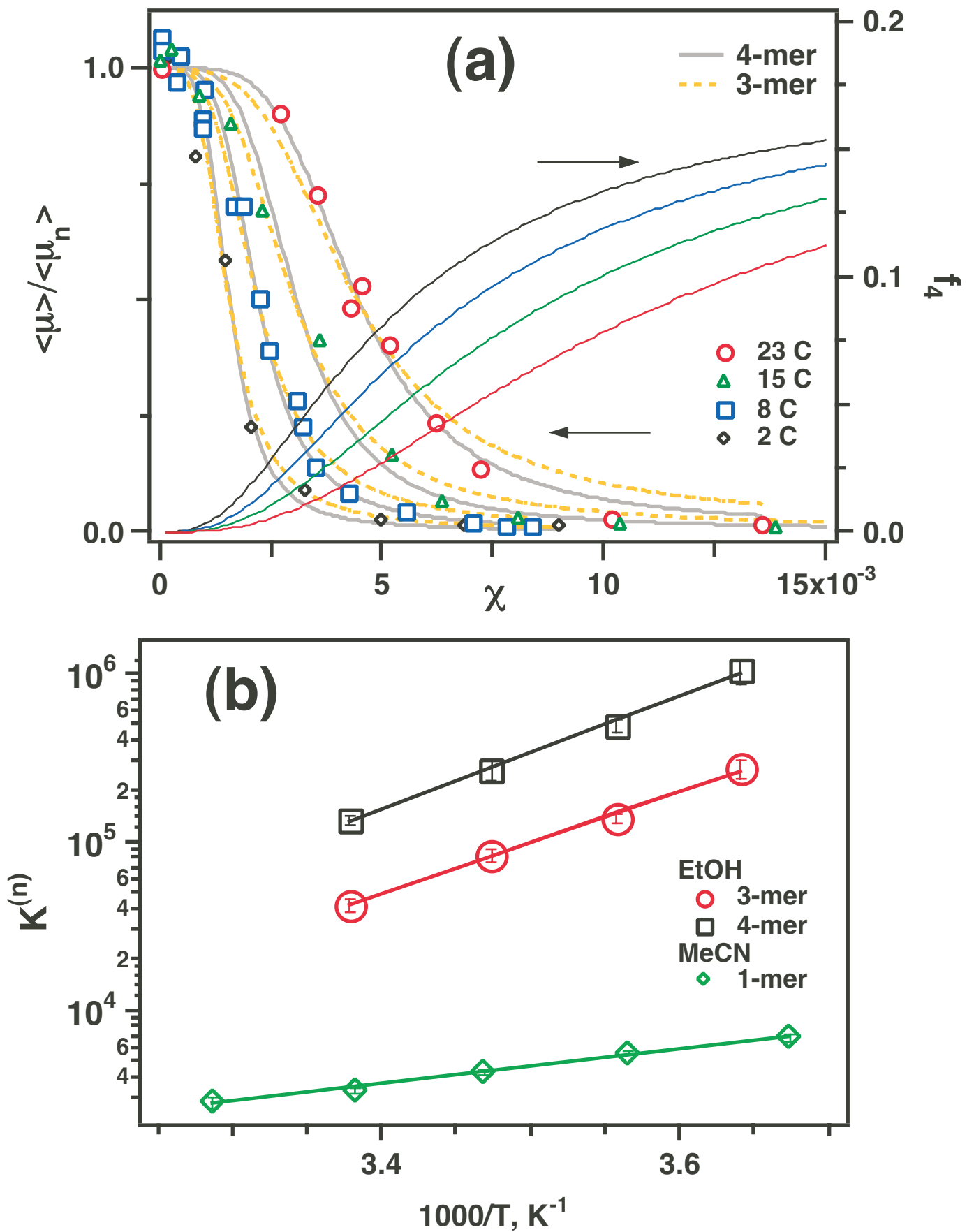

Figure 7; Shkrob & Sauer

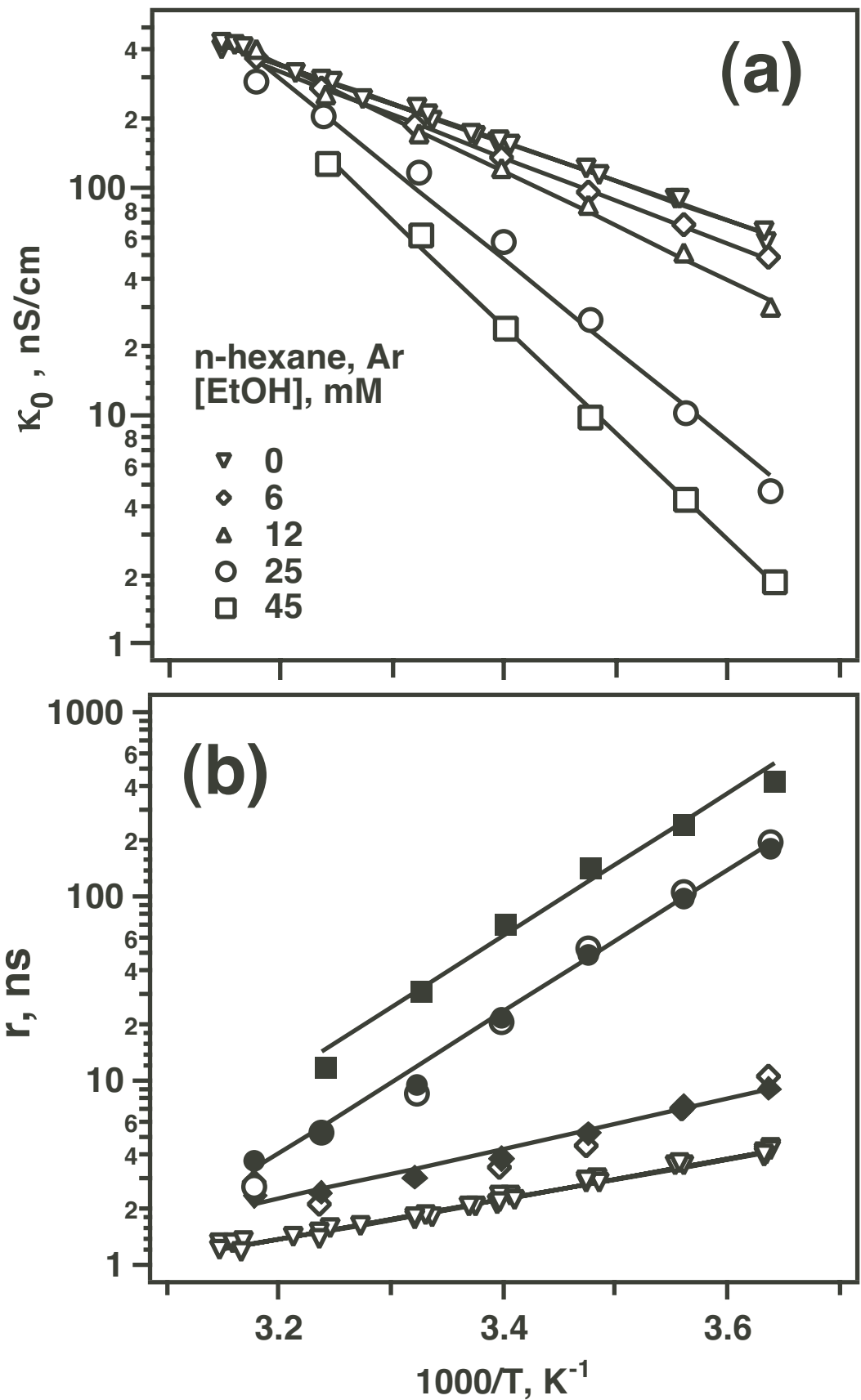

Figure 8; Shkrob & Sauer

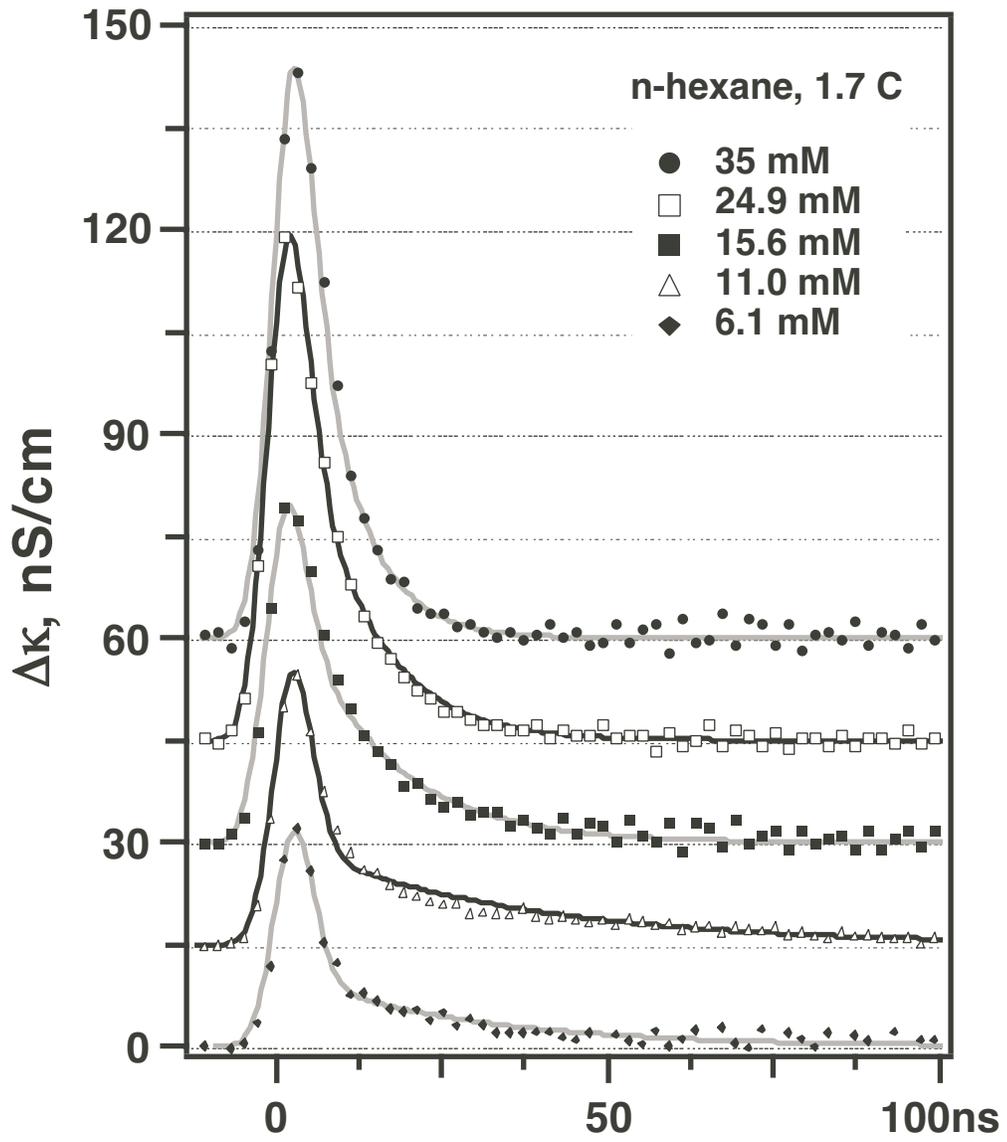

**Figure 9; Shkrob & Sauer**

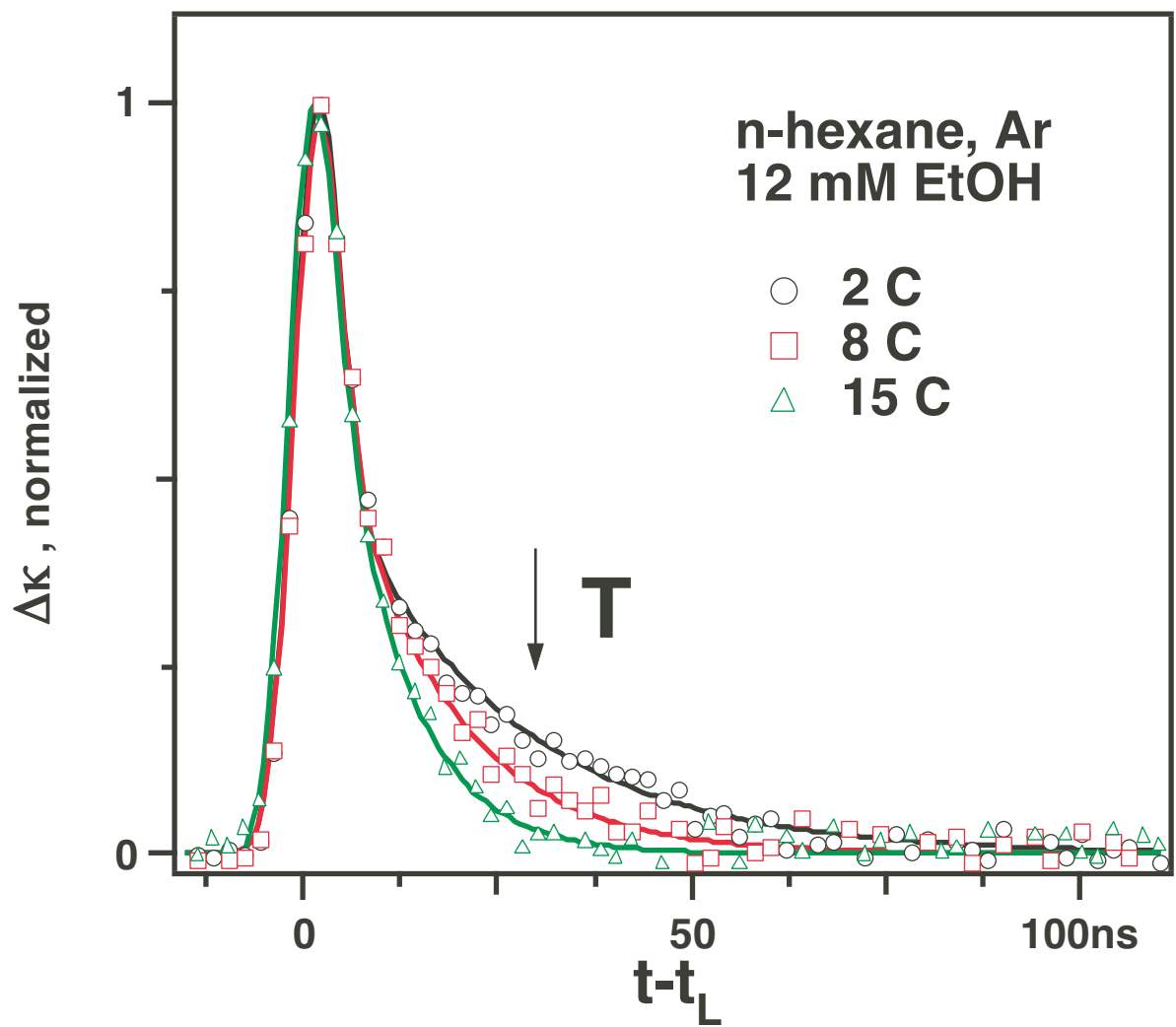

Figure 10; Shkrob & Sauer

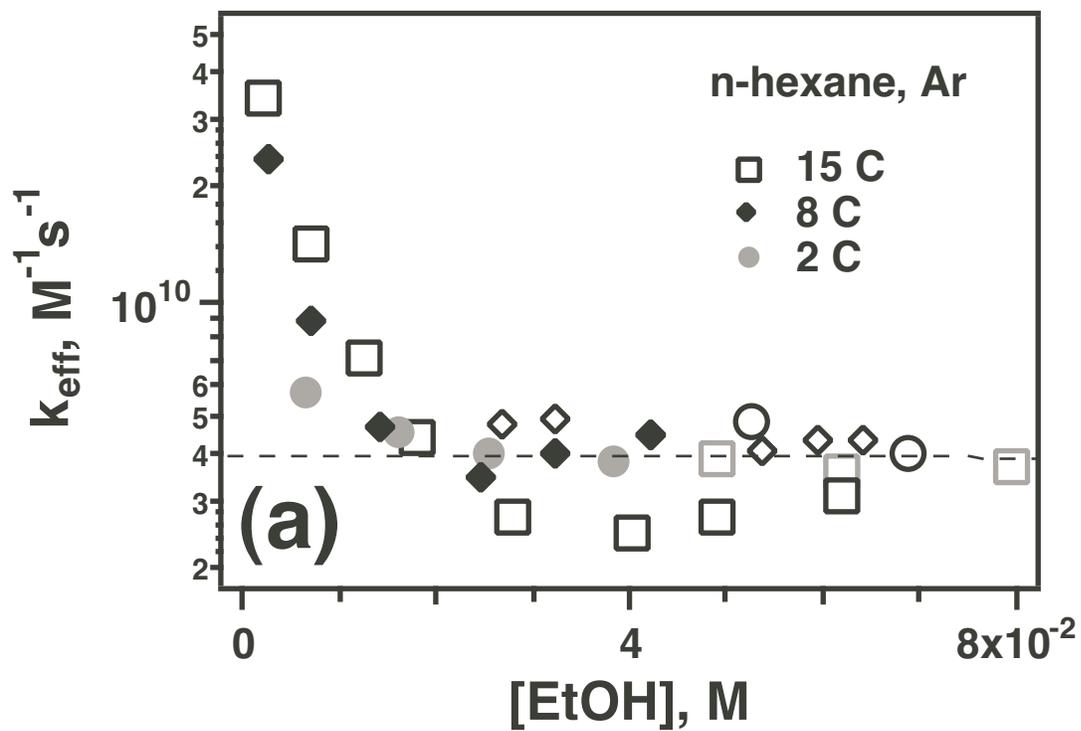
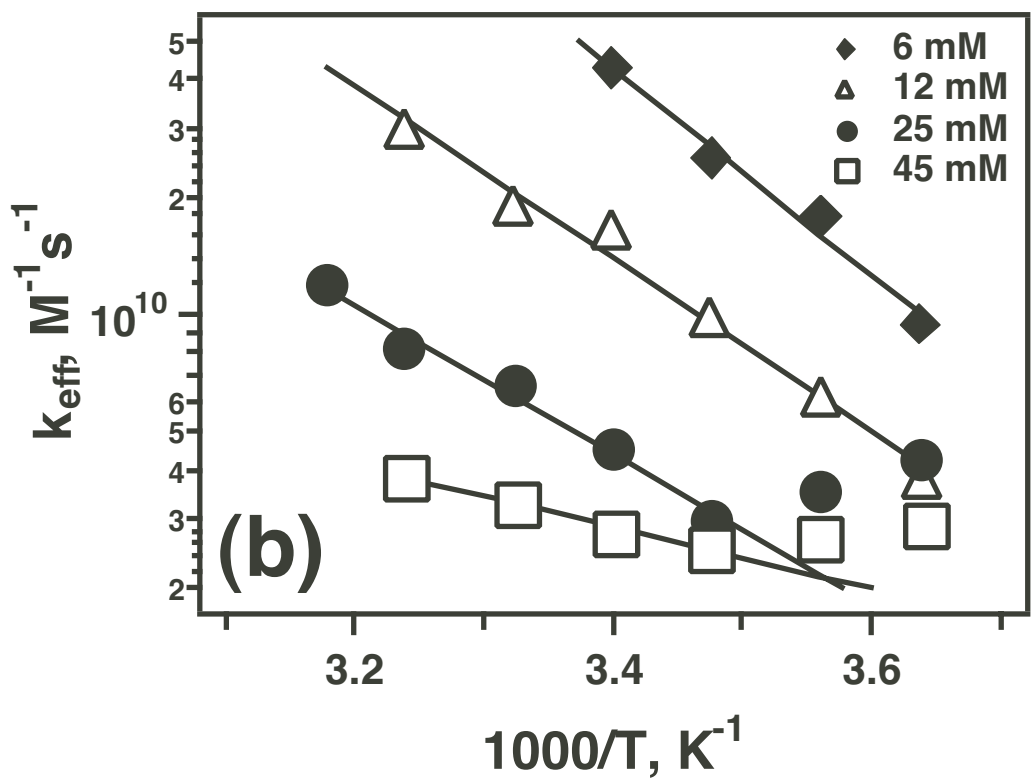

Figure 11; Shkrob & Sauer

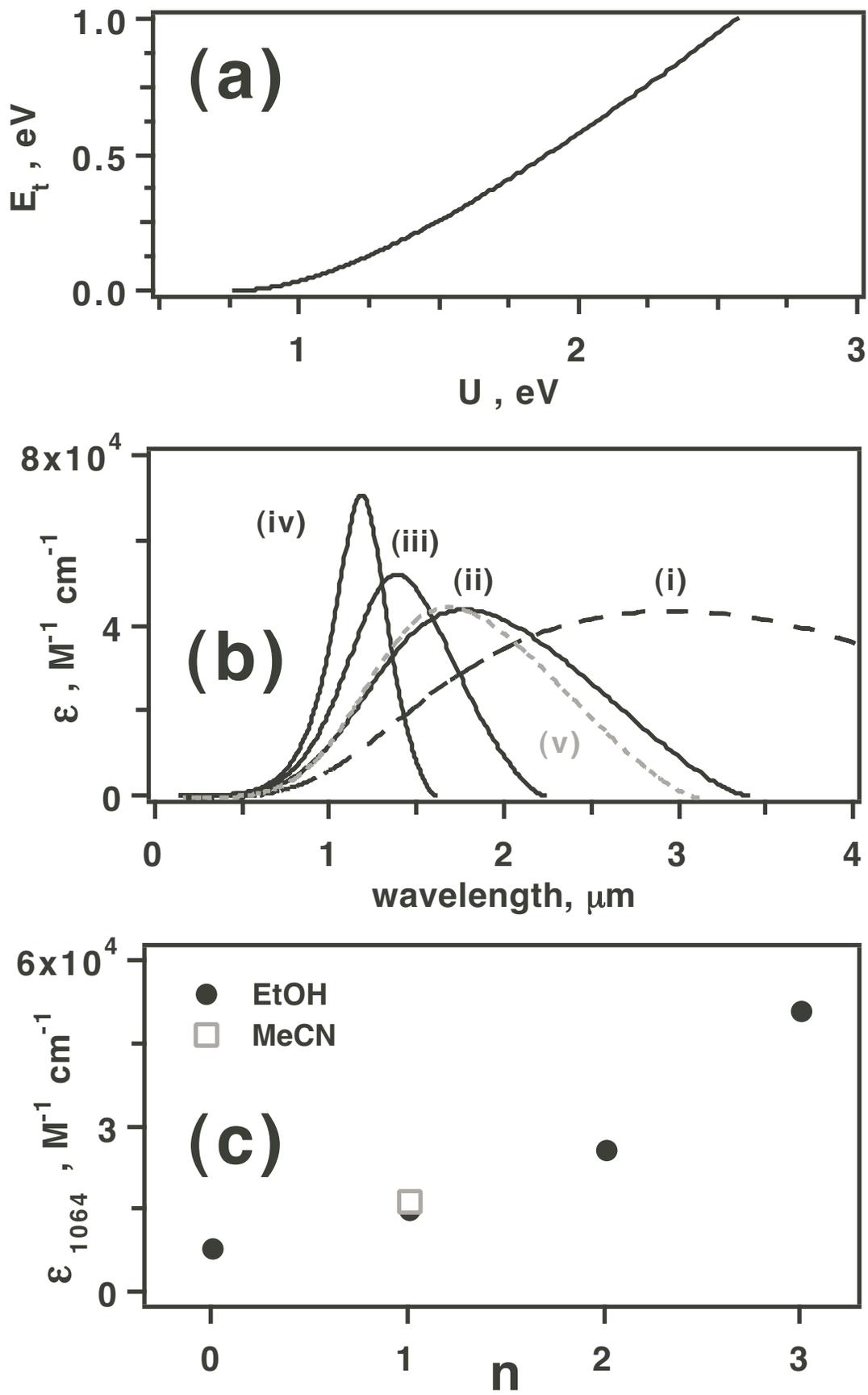

Figure 12; Shkrob & Sauer

# Electron Trapping by Polar Molecules in Alkane Liquids: Cluster Chemistry in Dilute Solution.

*Ilya A. Shkrob and Myran. C. Sauer, Jr.*

Radiation and Photochemistry Group, Chemistry Division, Argonne National Laboratory, 9700 South Cass Avenue, Argonne, Illinois 60439

*Tel* 630-2529516, *FAX* 630-2524993, *e-mail:* shkrob@anl.gov.

# Supporting Information.

**Appendix. Two-trap model.**

In this Appendix, we consider the model in which two trapped species, electron-1 and electron-2 exist in thermodynamic equilibrium with a quasifree electron, (Schemes 1 and 2 below). A generalization of this model for more than two trapped species is straightforward.

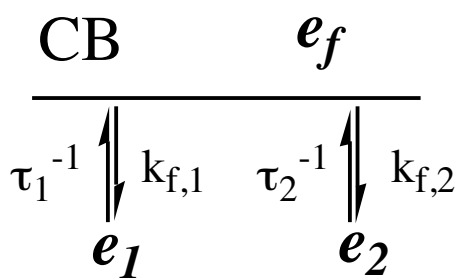
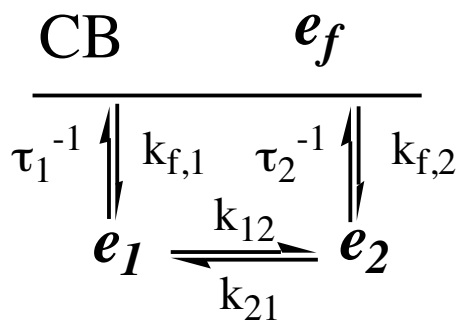

Scheme 1                           Scheme 2

Let $e_m(t)$ ($m = 1,2$) be the concentrations of trapped electrons, $e_m^0$ be their equilibrium concentrations before photoexcitation promoting the electron to the CB, $e_f(t)$ and $e_f^0$ be the corresponding concentrations for quasifree electron, $k_{f,m}$ and $k_m = \tau_m^{-1}$ be the trapping and detrapping rates and $k_{12}$ and $k_{21}$ be the rate constants for direct $1 \to 2$ and $2 \to 1$ transformations (Scheme 2). We will assume that $k_m$, $k_{12,21} \ll \tau_f^{-1} = k_f = k_{f,1} + k_{f,2}$, where $\tau_f$ is the lifetime of the quasifree electron, so that $e_f \ll e_m$ during the photoexcitation. Let

$$J(t) = \left(J/\tau_p \sqrt{\pi}\right) \exp\left(-[t/\tau_p]^2\right) \tag{A1}$$

be the irradiance of a Gaussian pulse with a pulse duration $\tau_p \gg \tau_f$ and a total fluence $J = \int dt\, J(t)$ (unless specified otherwise, the integration over time is from $t = -\infty$ to $t = +\infty$) and $\sigma_m$ be the cross sections for electron photodetachment form electron-1 and electron-2 ($\sigma_m J(t) \ll k_f$). During the photoexcitation pulse, the kinetic equations are given by

$$\frac{de_{1,2}}{dt} = k_{f;1,2} e_f - k'_{1,2}(t) e_{1,2} \mp (k_{12} e_1 - k_{21} e_2), \tag{A2}$$

$$\frac{de_f}{dt} = k'_1(t) e_1 + k'_2(t) e_2 - k_f e_f, \tag{A3}$$

where $k'_m(t) = \sigma_m J(t) + k_m$. Assuming stationary conditions for $e_{qf}^-$, $de_f/dt \approx 0$ and $e_f \ll e_{1,2}$, we obtain

$$e_f = \tau_f \{k'_1(t) e_1 + k'_2(t) e_2\} \tag{A4}$$

and

$$\frac{de_1}{dt} = -\frac{de_2}{dt} = -[P_2 k'_1(t) + k_{12}] e_1 + [P_1 k'_2(t) + k_{21}] e_2, \tag{A5}$$

where $P_m = k_{f,m}/k_f$ are the partition coefficients ($P_1 + P_2 = 1$) and $e_1(t = -\infty) = e_1^0$. Let us first consider the case where the equilibrium is set *before* the photoexcitation of the trapped electrons. Since at equilibrium $de_m^0/dt = 0$, the "equilibrium constant"

$$K = \frac{e_2^0}{e_1^0} = \frac{P_2 k_1 + k_{12}}{P_1 k_2 + k_{21}}. \tag{A6}$$

Introducing the modified partition coefficients $P'_1 = P_1 + k_{21}\tau_2$ and $P'_2 = P_2 + k_{12}\tau_1$, the latter equation may be written as

$$K = P'_2 \tau_2 / P'_1 \tau_1, \tag{A7}$$

2S.

so that

$$e_m^0/e_0 = P_m' \tau_m / (P_1' \tau_1 + P_2' \tau_2) \tag{A8}$$

and

$$e_f^0/e_0 = \tau_f (P_1' + P_2') / (P_1' \tau_1 + P_2' \tau_2), \tag{A9}$$

where $e_0 = e_1 + e_2$ is the total concentration of negatively charged species. Assuming that only quasifree electrons are conducting (with mobility $\mu_f \gg \mu_m$), the equilibrium conductivity $\kappa_0 = Fe_0 \langle \mu \rangle$, where $F$ is the Faraday constant and $\langle \mu \rangle = \mu_f e_f^0/e_0$ is the *apparent* electron mobility given by

$$\langle \mu \rangle = \mu_f \tau_f \frac{k_1 e_1^0 + k_2 e_2^0}{e_0} = \mu_f \tau_f \langle \tau^{-1} \rangle, \tag{A10}$$

where $\langle ... \rangle$ stands for averaging over the equilibrium concentrations of traps. We will first consider Scheme 1 in which the direct $1 \leftrightarrow 2$ transformations are neglected. In such a case,

$$\langle \mu \rangle = \mu_f \tau_f / (P_1 \tau_1 + P_2 \tau_2) \tag{A11}$$

Let us assume that $e_1$ in Scheme 1 is the electron residing in the intrinsic solvent trap ($e_{solv}^-$) and $e_2$ is the electron residing in the solute-associated trap, $\{e^- : S\}_{solv}$. We further assume that $k_{f,2} = k_{f,S}[S]$, where $k_{f,S}$ is the second-order constant. Let $\langle \mu_n \rangle = \mu_f / (k_{f,1} \tau_1)$ be the apparent electron mobility in neat solvent. From eq. (A11), we obtain

$$\langle \mu_n \rangle / \langle \mu \rangle = 1 + K, \tag{A12}$$

where $K = K_{eq}[S]$ and

$$K_{eq} = k_{f,S} \tau_2 / k_{f,1} \tau_1, \tag{A13}$$

is the equilibrium constant of $1 \leftrightarrow 2$ conversion (eq. (A6) and (A7)). Eq. (A12) is identical with eq. (6) given in the text. Exactly this dependence is given by a two-state model in which one considers a formal equilibrium with the solvent trap $S$

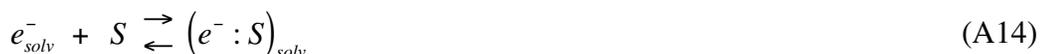

$$e_{solv}^- + S \rightleftarrows (e^- : S)_{solv} \tag{A14}$$

with the equilibrium constant $K_{eq}$, assuming that species $e_{solv}^-$ has mobility $\langle\mu_n\rangle$, while $(e^-:S)_{solv}$ does not contribute to the conductivity signal (note that rate constants of forward and backward reactions (A14) are $P_2 k_1$ and $P_1 k_2$, respectively). We have used such a model both in section 4.1 of this study and in ref. 10 (see section 1S therein) without justification; eq. (A12) justifies this approach. If direct $1 \leftrightarrow 2$ transformations (Scheme 2) are involved, assuming that $k_{12} = k'[S]$ (i.e., considering eq. (A14) as a real rather than a formal equilibrium reaction), we obtain

$$\frac{\langle\mu_n\rangle}{\langle\mu\rangle} = \frac{1 + K_{eq}[S] + (\tau_2 + K_{eq}\tau_1[S])(k_{21} + k'[S])}{1 + k_{21}\tau_2 + k'\tau_1[S]}, \tag{A15}$$

which gives more complex behavior of the apparent mobility $\langle\mu\rangle$ as a function of $[S]$ than that given by formula (A12), although the deviations are substantial only when reactions (A14) occur on the time scale comparable to $\tau_1$ or $\tau_2$.

We turn to the behavior of the photoinduced conductivity signal

$$\Delta\kappa(t) = Fe_0\mu_f \Delta e_f(t) \tag{A16}$$

where $\Delta e_f(t) = e_f(t) - e_f^0$. The quantity of interest is the ratio $r = \Delta A/\kappa_0$, where $\Delta A = \int dt\ \Delta\kappa(t)$ is the area under the signal. This ratio is given by

$$r = \int dt\ \Delta e_f(t)/e_f^0. \tag{A17}$$

For $J \to 0$, kinetic equations (A5) can be solved perturbatively. Let us assume that $e_1 \approx e_1^0 + \delta$ and $e_2 = e_2^0 - \delta$, where $\delta \ll e_0$. Retaining only terms that are linear in $J(t)$ and $\delta$, we obtain

$$\Delta e_f \approx \tau_f\left[\left(\sigma_1 e_1^0 + \sigma_2 e_2^0\right)J(t) + (k_1 - k_2)\delta\right] \tag{A18}$$

and

$$d\delta/dt \approx -k_s\delta + \left(P_1\sigma_2 e_2^0 - P_2\sigma_1 e_1^0\right)J(t), \tag{A19}$$

where

$$k_s = P_1 k_2 + P_2 k_1 \tag{A20}$$

is the inverse settling time $\tau_s$ of the equilibrium reaction (A14). Solving eq. (A18), we obtain

$$\Delta e_f \approx \tau_f \left[ \left( \sigma_1 e_1^0 + \sigma_2 e_2^0 \right) J(t) + \left( k_1 - k_2 \right) \left( P_1 \sigma_2 e_2^0 - P_2 \sigma_1 e_1^0 \right) \int_{-\infty}^{t} d\tau \, J(\tau) \, \exp\left( -k_s [t - \tau] \right) \right]$$

(A21)

Thus, $\Delta\kappa(t)$ can be represented as a weighted sum of the excitation profile $J(t)$ (the "spike") and the same profile convoluted with an exponential whose time constant equals the settling time of equilibrium reaction (A14) (the "slow" component). The analysis of $\Delta\kappa(t)$ kinetics given in section 4.2.2 is based on this general result. Integrating eq. (A21) from $t = -\infty$ to $t = \infty$ we obtain

$$\int dt \, \Delta\kappa(t) \approx JF\mu_f \tau_f \left[ \left( \sigma_1 e_1^0 + \sigma_2 e_2^0 \right) + \frac{\left( k_1 - k_2 \right) \left( P_1 \sigma_2 e_2^0 - P_2 \sigma_1 e_1^0 \right)}{P_1 k_2 + P_2 k_1} \right],$$

(A22)

which can be simplified to

$$\int dt \, \Delta\kappa(t) \approx JF\mu_f \tau_f \frac{\sigma_1 \tau_1 e_1^0 + \sigma_2 \tau_2 e_2^0}{P_1 \tau_1 + P_1 \tau_2}$$

(A23)

Combining eqs. (A11) and (A23), the ratio $r/J$ is expressed as

$$\left( \frac{r}{J} \right)_{J \to 0} \approx \frac{\sigma_1 \tau_1 e_1^0 + \sigma_2 \tau_2 e_2^0}{e_0} = \frac{\sigma_1 \tau_1 + \sigma_2 \tau_2 K}{1 + K}.$$

(A24)

The latter can be written as $r \approx \langle \sigma_t \tau_t \rangle J$ (eq. (7)) which is correct for multiple trapping (see Appendix A of ref. 6). It can be shown that for Scheme 2, $r \approx \langle \sigma \tau' \rangle J$ where

$$\tau'_{1,2} = \tau_{1,2} \frac{1 + (k_{12} + k_{21}) \tau_{2,1}}{1 + k_{12} \tau_1 + k_{21} \tau_2}.$$

(A25)

With the same assumptions made to derive eq. (A12),

$$\tau_s = \frac{\tau_2 + \tau_1 K}{1 + K},$$

(A26)

i.e., the time constant of the "slow" component (second term in eq. (A21)) approaches $\tau_2$ at low $[S]$ ($K \ll 1$) and $\tau_1$ in the opposite limit. Thus, for $\tau_2 \gg \tau_p \gg \tau_1$ the two-trap model predicts that the $\Delta\kappa(t)$ kinetics exhibits an exponential "tail"; as the solute

5S.

concentration decreases, $\tau_s$ increases until it approaches the residence time $\tau_2$ of electron-2. The ratio $\xi_s$ of areas under the "slow" component and the "spike" is given by the ratio of the second and the first terms of eq. (A22), that is

$$\xi_s = \frac{P_1 P_2 (\tau_2 - \tau_1)(\sigma_2 \tau_2 - \sigma_1 \tau_1)}{(\sigma_1 P_1 \tau_1 + \sigma_2 P_2 \tau_2)(P_1 \tau_1 + P_2 \tau_2)}. \tag{A27}$$

For $\tau_2 \gg \tau_1$ and $\sigma_2 \tau_2 \gg \sigma_1 \tau_1$

$$\xi_s \approx \frac{(\tau_2/\tau_1) K}{(1+K)(\sigma_1/\sigma_2 + K)}, \tag{A28}$$

and for $K \ll 1$, the ratio $\xi_s$ increases with increasing $K$, whereas for $K \gg 1$, $\xi_s \propto K^{-1}$ decreases with increasing $K$. The initial increase ($\xi_s \approx (\sigma_2 \tau_2 / \sigma_1 \tau_1) K$) with increasing $K$ occurs only for $\sigma_1 \gg \sigma_2$; in the opposite limit $\xi_s \approx \tau_2/\tau_1 \gg 1$ (i.e., the "prompt" component is negligible) as soon as $K \gg \sigma_1/\sigma_2$. Provided that $\sigma_1 \gg \sigma_2$, for low solute concentrations $[S]$ ($K \ll 1$), the relative weight of the exponential "tail" is low and this weight increases linearly with $[S]$; as the concentration further increases and $K \gg 1$, this weight decreases and the decay becomes faster. Eventually this decay becomes faster than $\tau_p$ and the relative weight also decreases, so the "slow" component cannot be discerned. Though these trends were indeed observed experimentally for $\Delta\kappa(t)$ kinetics obtained in dilute ethanol solutions (section 4.2.2), the overall data cannot be accounted for consistently using the two-trap model. Indeed, for $\sigma_2 \gg \sigma_1$, the second term in eq. (A21) always prevails and the "prompt" component is not observed for any $K$. This unsettling conclusion refers not only to the case of infinitely small fluence $J$ (which is the only regime that can be handled analytically) but also to arbitrary $J$, as may be shown using numerical simulations. While it is possible, in principle, that for alcohol traps $\sigma_2 \ll \sigma_1$ (due to photostimulated proton transfer competing with electron photodetachment), in such a case the ratio $r$ would stay almost linear with $J$ whereas experimentally the "saturation" (see below) sets in for $J > 10^{18}$ photon/cm$^2$ (section 4.2.2). Our simulations suggest that within the confines of the two-state model, it is simply impossible to obtain both the low-concentration $\Delta\kappa(t)$ curves observed experimentally (that exhibit clearly separable "spike" and "slow" components for $[S] \to 0$) and sigmoid power dependencies. Consequently, any scheme intended to explain these data, even at the qualitative level, has to postulate more than two electron species.

For arbitrary laser fluence $J(t)$, eq. (A5) must be solved numerically and the ratio $r$ determined using eqs. (A4) and (A17). Since all equations are linear, the result does not depend on $e_0$. For Scheme 1, the calculation of the ratio $r$ as a function of $J$ (for known pulse duration $\tau_p$ (ca. 4 ns)) requires the knowledge of five parameters: $\sigma_{1,2}$, $\tau_{1,2}$, and the equilibrium constant $K$ (eq. (A6)). Two of these parameters, $\sigma_1 \approx 3.2 \times 10^{-17}$ cm$^2$ and $\tau_1 \approx 8.3$ ps, for electron traps in neat $n$-hexane are known from ref. 24 and our previous study,[6] respectively. The equilibrium constant $K = K_{eq}[S]$ can be estimated from the $\langle \mu \rangle$ data (section 4.2.1). The product $\sigma_2 \tau_2$ can be estimated from the ratio $r/J$ obtained for $K \gg 1$ in the low-fluence regime; in this regime (neglecting direct $1 \leftrightarrow 2$ reactions in Scheme 2), $\langle \sigma \tau \rangle \approx \sigma_2 \tau_2$ (ca. $5.6 \times 10^{-26}$ cm$^2$ s for $(e^- : MeCN)_{solv}$; see section 4.2.1). Thus, the main uncertainty is the cross section $\sigma_2$ since the ratio $\sigma_2/\sigma_1$ is not known (although it can be estimated from the TA experiments within a factor of two). The experimental plots of $r$ vs. $J$ can be simulated using this approach with minimal adjustments to the model parameters. Since no refinement of the model parameters can be made using such simulations, the main goal of these calculations is to demonstrate that the trends observed experimentally in Figs. 11S can be rationalized using the two-trap model of Scheme 1; the exact choice of simulation parameters is not important for such a demonstration.

Fig. 21S(a) shows a family of $r$ vs. $J$ plots for different values of the equilibrium constant $K$. Since $\sigma_2 \tau_2 \gg \sigma_1 \tau_1$, even for relatively low values of $K$ the initial slope of these plots is close to $\sigma_2 \tau_2$. For higher fluence, the ratio "saturates". As seen from the plots, for higher $K$ this saturation is less expressed (see also the normalized plots in Fig. 21S(b) illustrating the decreases in the curvature), because the equilibrium can be continuously shifted within the duration of the pulse. Exactly these patterns are observed experimentally (Fig. 11S and 20S). Plotted as a function of the equilibrium constant $K$ for a given fluence $J$, the ratio $r$ rapidly reaches a plateau at lower fluence (as follows from eq. (A24)) but continues to increase with increasing $K$ at greater fluence (Fig. 22S). Once more, the same behavior was observed in dilute acetonitrile solutions in $n$-hexane (Fig. 4).

When direct $1 \leftrightarrow 2$ reactions are introduced (Scheme 2), the bicomponent kinetics can be readily obtained even for $\sigma_2 \gg \sigma_1$. It may seem that the inclusion of these additional reactions violates the detailed equilibrium principle which relates the equilibrium constant of the $1 \leftrightarrow 2$ reaction with the equilibrium constant for reaction (A14). However, it should be kept in mind that Scheme 1 does not make any provision for interconversion of electron *traps* (such as reaction (5) in the text) occurring separately from reaction (A14). It is precisely the principle of detail equilibrium that allows one to treat this reaction as an additional equilibrium involving *filled* traps.

**Figure captions (1S to 22S).**

**Fig. 1S.**

Speciation plots for neutral alcohol clusters $S_n$ in *n*-hexane using the thermodynamic data from ref. 28 and eq. (10) for (a) 23 °C and (b) 1.8 °C. The net alcohol concentration is given as mole fraction $\chi$ of the alcohol. The monomer fraction (dashed line) is given to the right; equilibrium fractions $f_n = n[S_n]/c$ (solid lines) are plotted to the left (the cluster numbers $n$ are given in the legends).

**Fig. 2S.**

(a) TA spectra from pulse radiolysed Ar-saturated (open symbols) and $CO_2$-saturated (trace (i), filled circles) room temperature *n*-hexane. $CO_2$ is added as electron scavenger. The integration time windows are given in the plot, for trace (i), the integration window is 6 to 10 ns (the absorbance spectrum, which is mainly from the solvent olefin cation does not evolve in the first 500 ns after the electron pulse). This trace was normalized at 500 nm, where most absorbance is from the cation. In Ar-saturated *n*-hexane the absorbance is composite: both $e^-_{solv}$ and the cation contribute to the spectrum at short delay times. At later delay times, the relative contribution from the cation increases as the electron is scavenged by impurity. (b) Decay kinetics of optical absorbance at 0.5 and 1.55 µm in Ar- and $CO_2$-saturated solutions (see the legend in the plot). Only electrons absorb at the longer wavelength, while at the shorter wavelength, only the cations absorb. Upon the addition of electron scavenger, the end-of-the pulse yield of cations increases ca. 2 times. The smooth (black) line drawn through the 1.55 µm curve is single exponential fit (illustrating that electrons mainly decay via scavenging by an impurity).

**Fig. 3S.**

TA kinetics ($\lambda = 1$ µm) from Ar-saturated *n*-hexane solutions at 23 °C containing (a) acetonitrile and (b) ethanol. (Same conditions as in Fig. 2S). The net solute concentrations are indicated in the plots. The dashed trace in (b) is the scaled TA kinetics in neat *n*-hexane (drawn to illustrate slowing down of the decay kinetics in ethanol solutions).

**Fig. 4S.**

Normalized TA spectra from Ar-saturated acetonitrile/n-hexane solutions (23 °C, the same radiolysis conditions as in Fig. 2S). The net molar concentration of the solute is given in (a). The integration window is (a) 11 to 17 ns and (b) 80 to 180 ns. The lower the MeCN concentration the less is the signal from olefin cations in the visible at later delay times. To facilitate the comparison the spectra were normalized at 1.55 μm, where only electron contributes to the absorbance.

**Fig. 5S.**

Normalized TA kinetics (b) and end-of-pulse electron absorbances (a) for $\lambda = 1.55$ μm in room temperature solutions of acetonitrile in *iso*-octane (the concentrations are given in (b)). The bold line in (a) is the linear plot.

**Fig. 6S.**

(a-c) Time evolution of absorbance spectra observed in pulse radiolysis of room temperature Ar-saturated, acetonitrile/*iso*-octane solutions (the concentrations are given in (a)). The spectra are normalized at 1.55 μm, where only trapped electron absorbs the analyzing light. At short delay times, the spectrum is dominated by this trapped electron; at later delay times the signal from an olefin cation (or a dimer anion) can also be observed.

**Fig. 7S.**

A plot illustrating lack of time evolution for absorbance spectra observed in pulse radiolysis of ethanol/*n*-hexane solutions (the time windows are given in the plots). The spectra obtained for (a) 121 mM and (b) 262 mM EtOH are normalized at their respective maxima. The lines are Lorentzian-Gaussian least squares fits.

**Fig. 8S.**

(a) Extrapolated conductivity $\kappa_0$ from the electron (open circles) in acetonitrile/*iso*-octane solutions vs. the net molar concentration of MeCN (23 °C). The line drawn through the symbols is the optimum fit to eq. (6). (b) To the left: The same plot on the extended logarithmic scale; the solid line is a fit to $\langle\mu_n\rangle/\langle\mu\rangle = 1 + K_{eq}c + K_n c^n$ (see the text). To the right: ratio $r$ (open squares) for the same solution (50 ns delay time, $9 \times 10^{18}$ photon/cm$^2$ pulse of 1064 nm photons.)

**Fig. 9S.**

(a) Conductivity kinetics from acetonitrile/*iso*-octane solutions photoionized using 248 nm light (a thin line plotted to the left). Also shown are the $\Delta\kappa$ signals induced by the subsequent excitation of trapped electrons by 1064 nm light (cf. Fig. 3 in the text). The same excitation conditions were used as in Fig. 8S(b). The amplitude of $\Delta\kappa$ follows the decay kinetics of $\kappa - \kappa_i \approx \kappa$, so that the ratio $r$ is independent of the delay time. (b) $\kappa$ (to the left and to the bottom) and $\Delta\kappa$ (to the right and to the top) kinetics in room temperature Ar-saturated *iso*-octane solutions containing MeCN (the net solute concentrations are given in the plot). The smooth lines are exponential fits. The smooth lines drawn through the $\Delta\kappa$ data are Gaussian fits.

**Fig. 10S.**

Results similar to those in Fig. 9S (a) are shown for 17 mM MeCN in Ar-saturated *n*-hexane. The solution temperatures are indicated in (a). Conductivity kinetics (solid line in (a) and dashed lines in (b) are plotted to the right), 1064 nm photon induced kinetics are plotted to the left in (b). The green line in (b) indicates the delay time of the 1064 nm pulse (using the same excitation conditions as in Fig. 9S). The smooth lines drawn through the $\kappa$ data in (a) are exponential fits.

**Fig. 11S.**

Power dependencies of ratio $r$ at $t_L \approx 50$ ns for several concentrations of acetonitrile in room temperature *n*-hexane (the concentrations and photon fluences are indicated in (b)). In the high concentration limit ($K >> 1$, see Fig. 4(a)) the initial slope approaches the product of photodetachment cross section and the lifetime of the $(e^- : MeCN)_{solv}$ state. (b) The same plot after normalization at $8.5 \times 10^{18}$ photons/cm$^2$.

**Fig. 12S.**

Decay kinetics of photoconductivity for a 5 μm solution of anthracene in *n*-hexane containing 12 mM EtOH (the temperatures are indicated in the plot). The signal is mainly from the electrons generated by 2-photon ionization of anthracene by 248 nm laser light. The smooth curves are exponential fits. Note that at the lowest temperature of 2 ºC the kinetics becomes exponential only after 100 ns. This delay is due to slow settling of the electron equilibria (see Figs. 9 and 10) at low temperature, in dilute alcohol solutions.

**Fig. 13S.**

Normalized conductivity signal $\kappa_i$ from ions (anions and cations) in ethanol/*n*-hexane solutions vs. [EtOH] (i.e., the conductivity signals attained at the end of electron scavenging in the solution). The ions were generated by 2-photon ionization of room temperature Ar- and SF$_6$-saturated solutions (filled diamonds and circles, respectively), Ar-saturated solution containing 0.3 mM benzene as a sensitizer (filled squares), and SF$_6$-saturated 5 μm anthracene solutions at 2 ºC (filled triangles), 8 ºC (open squares) and 15 ºC (filled downward triangles). Also shown (to the right) is the mobility of F⁻ (see Fig. 14S(b)). Addition of alcohol reduces anion mobility causing the decrease in the conductivity signal from the ions. The line is a guide for the eye.

**Fig. 14S.**

(a) Time-of-flight conductivity traces (the electrode spacing is 0.8 mm, the voltage is 5 kV) from 0.65 mM triethylamine in $SF_6$-saturated *n*-hexane (23 ºC) as a function of ethanol concentration (molar concentrations are indicated in the plot). $SF_6$ rapidly scavenges the electron within the duration of the 248 nm pulse yielding a fluoride anion, F⁻. A 100 µm slit is placed near the cathode so that the conductivity signal is from migrating F⁻ anions (the conductivity at the flat region is ca. 100 pS/cm). The cations rapidly migrate in the electric field and discharge at the cathode; a fraction of these cations decay via. neutralization in the bulk (this reaction is over in 0.5 ms). The higher the alcohol concentration, the longer is the time of flight for the anions. (b) The plot of drift mobility for F⁻ vs. [EtOH]. See also Fig. 13S.

**Fig. 15S.**

First order constant $\Delta k$ for electron decay in room temperature, Ar-saturated solutions of (a) ethanol and (b) methanol in *n*-hexane vs. the molar concentration (to the bottom) or mole fraction (to the top) of these two alcohols. Filled symbols are for protiated, open symbols are for deuterated alcohols. See section 4.2.2 for more detail.

**Fig. 16S.**

Dependencies of 1064 nm photon induced conductivity signals from very dilute (< 5 mM) ethanol/*n*-hexane solutions at (a) 2 ºC and (b) 23 ºC vs. the delay time $t_L$ of the IR pulse (to the right). The decay kinetics of conductivity signals from the electron (plotted to the left) are juxtaposed for comparison. At both temperatures, the amplitudes of the $\Delta\kappa$ signals decrease in proportion to $\kappa_e(t_L) = \kappa - \kappa_i$, suggesting no trapping by higher alcohol clusters in the course of the slow "scavenging" reaction (Fig. 15S). In this plot and in Figs. 17S(b) and 18S, the fluence of 1064 nm photons is $9\times10^{18}$ photon/cm².

**Fig. 17S.**

Decay kinetics of (a) 248 nm photon induced conductivity signals and (b) 1064 nm induced signals (for $t_L$ =70 ns) from Ar-saturated *n*-hexane solutions containing 6 mM EtOH. The temperatures of the solutions are indicated in the plot. The smooth curves in (a) are exponential fits; the smooth curves drawn through the symbols in (b) are weighted sums of a Gaussian (with the time profile of the excitation pulse) and the same Gaussian convoluted with an exponential function. To facilitate the comparison, $\Delta\kappa$ kinetics are vertically spaced.

**Fig. 18S.**

Same as Fig. 17S(b): the concentration dependence of $\Delta\kappa$ kinetics at 14.9 °C. Ethanol concentrations are indicated in the plot.

**Fig. 19S.**

Power plot for the ratio $r$ (compare with Fig. 11S(a)) for several concentrations of EtOH in room temperature, Ar-saturated $n$-hexane. The net solute concentrations are indicated in the plot. Note the logarithmic vertical scale.

**Fig. 20S.**

(a) The dependence of parameter $K = \langle\mu_n\rangle/\langle\mu\rangle - 1$ on [EtOH] for Ar-saturated $n$-hexane at 23 °C (circles). The solid line is a $K \propto c^4$ plot. (b) The dependence of ratio $r$ (at $t_L \approx 50$ ns) vs. $K$ for ethanol in $n$-hexane for fluences $J = 8.1 \times 10^{18}$ (circles) and $5.4 \times 10^{17}$ photon/cm$^2$ of 1064 nm light (squares); the same series are shown in (a). The lines are guides for the eye. Compare with Fig. 4(b). The open and filled symbols (b) correspond to different integration methods for the $\Delta\kappa$ signal.

**Fig. 21S.**

(a) Plots of ratio $r$ vs. fluence $J$ of 1064 nm photons for simulation parameters given in the Appendix. (b) Same plot, after normalization at $J=8\times10^{18}$ photon/cm$^2$. The constant $K$ is indicated in plot (a). Compare with Figs. 11S and 19S.

**Fig. 22S.**

The same calculation as in Fig. 21S: the ratio $r$ is plotted vs. $K$ for different fluences of 1064 nm photons (indicated in the plot). Compare with Figs. 4 and 20S.

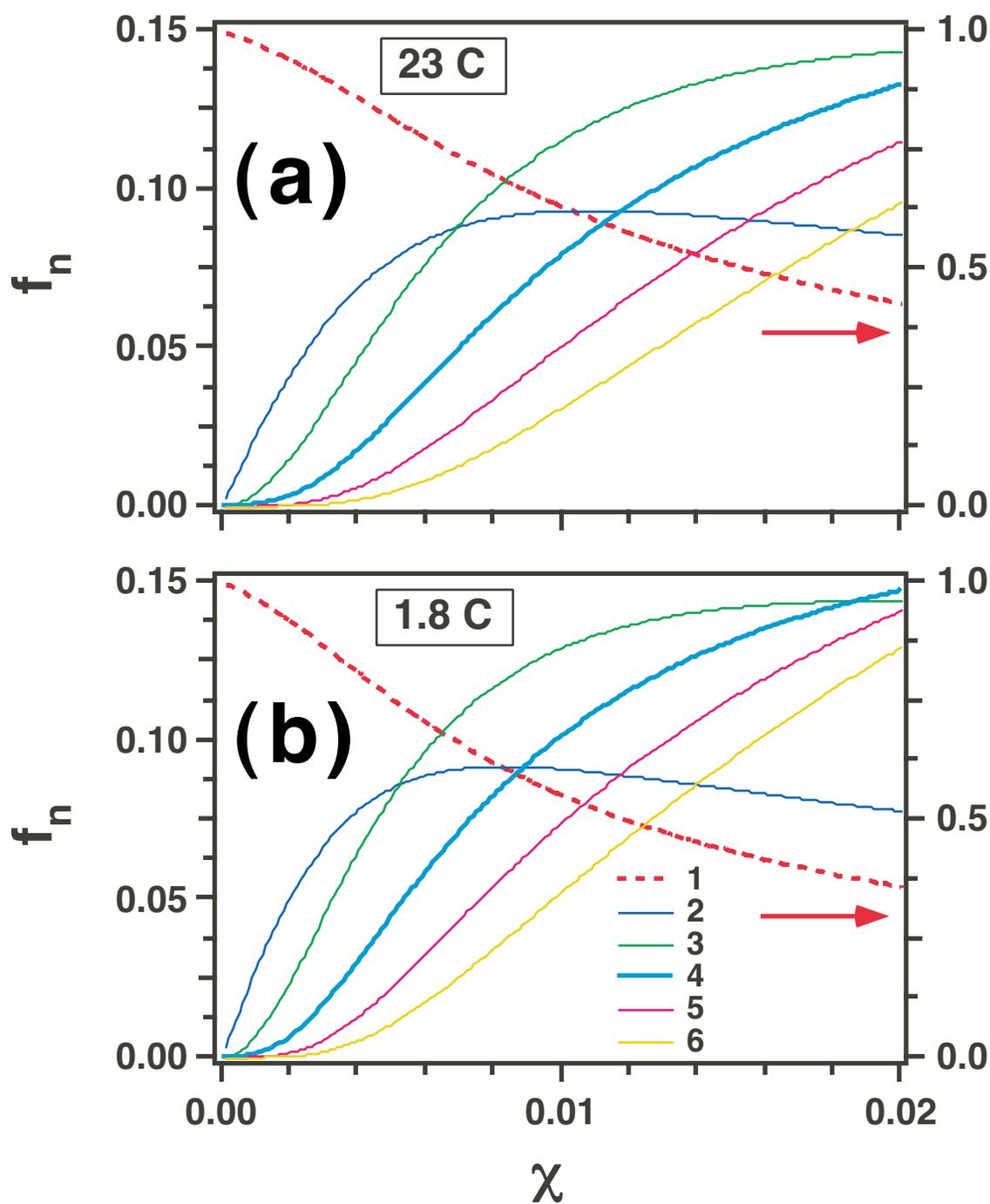

Figure 1S; Shkrob & Sauer

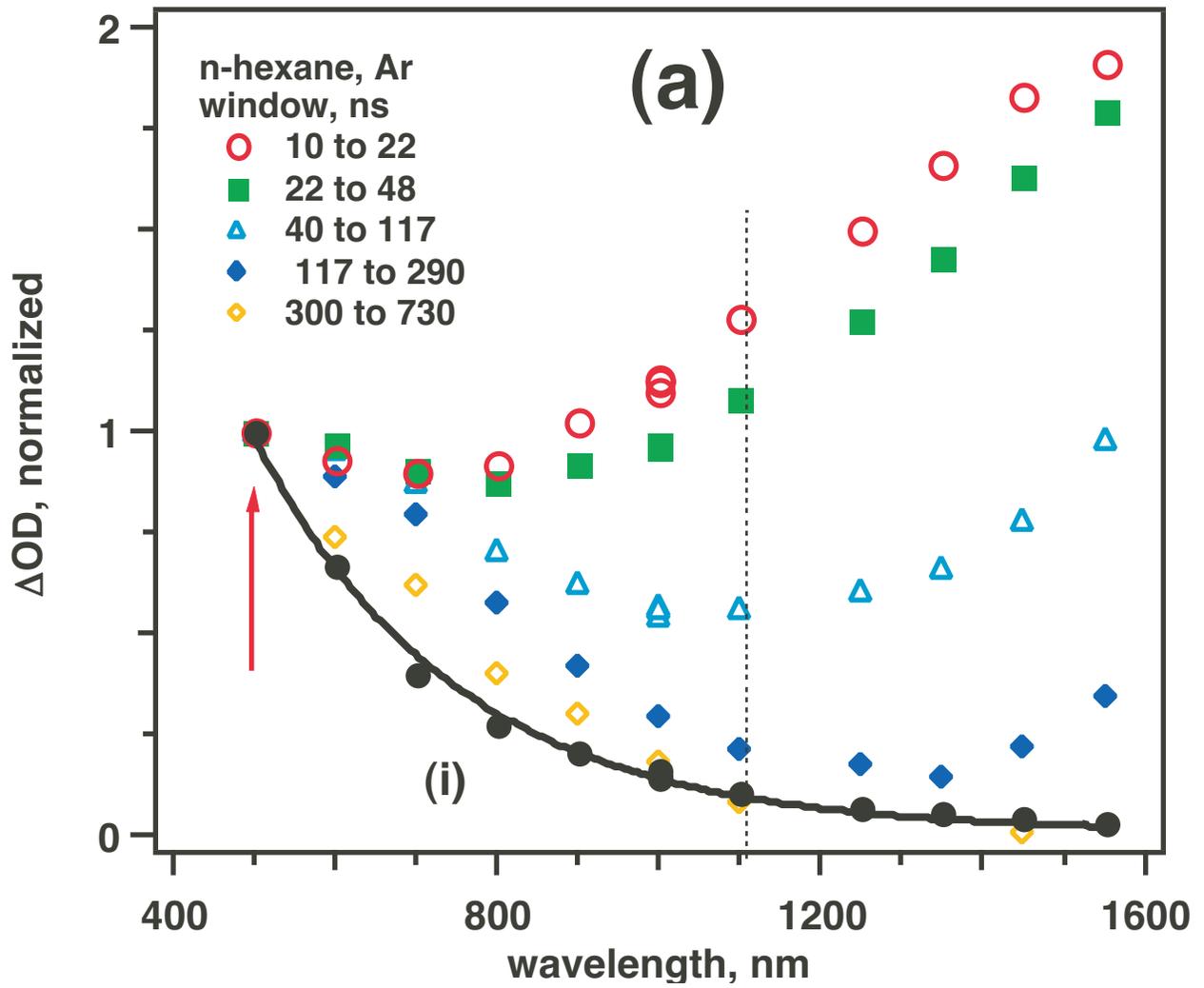
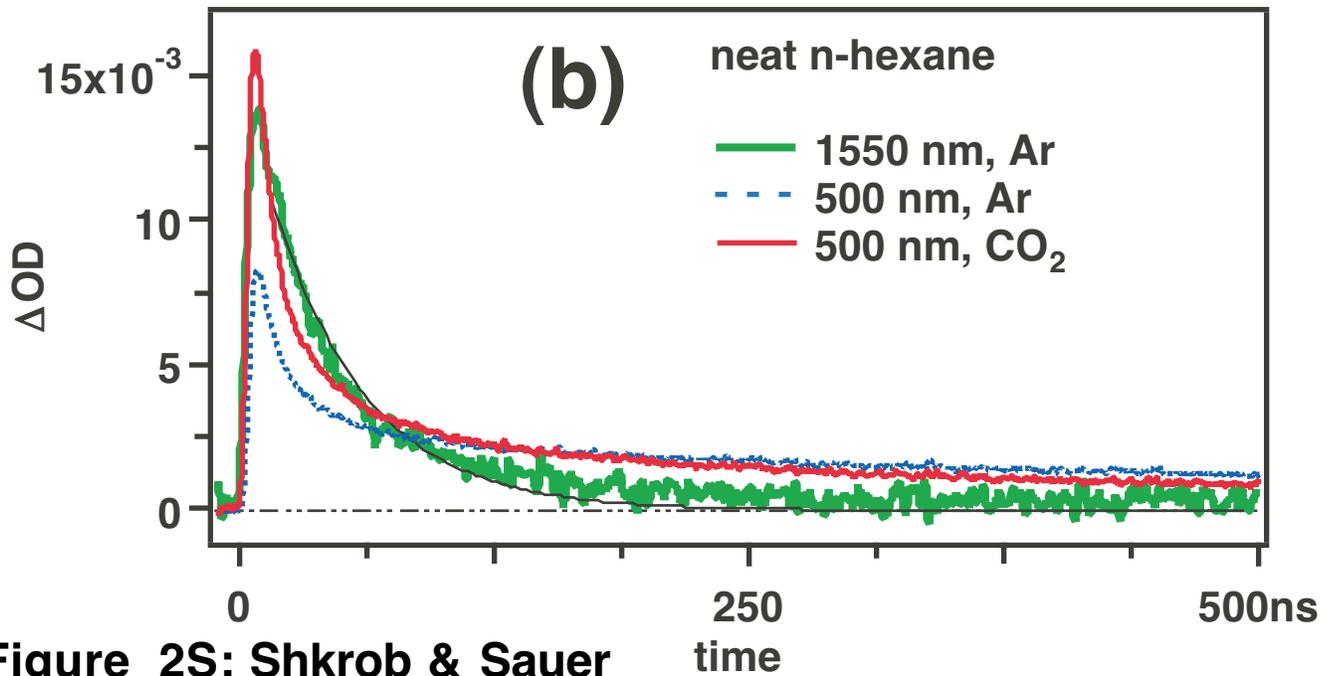

Figure 2S; Shkrob & Sauer

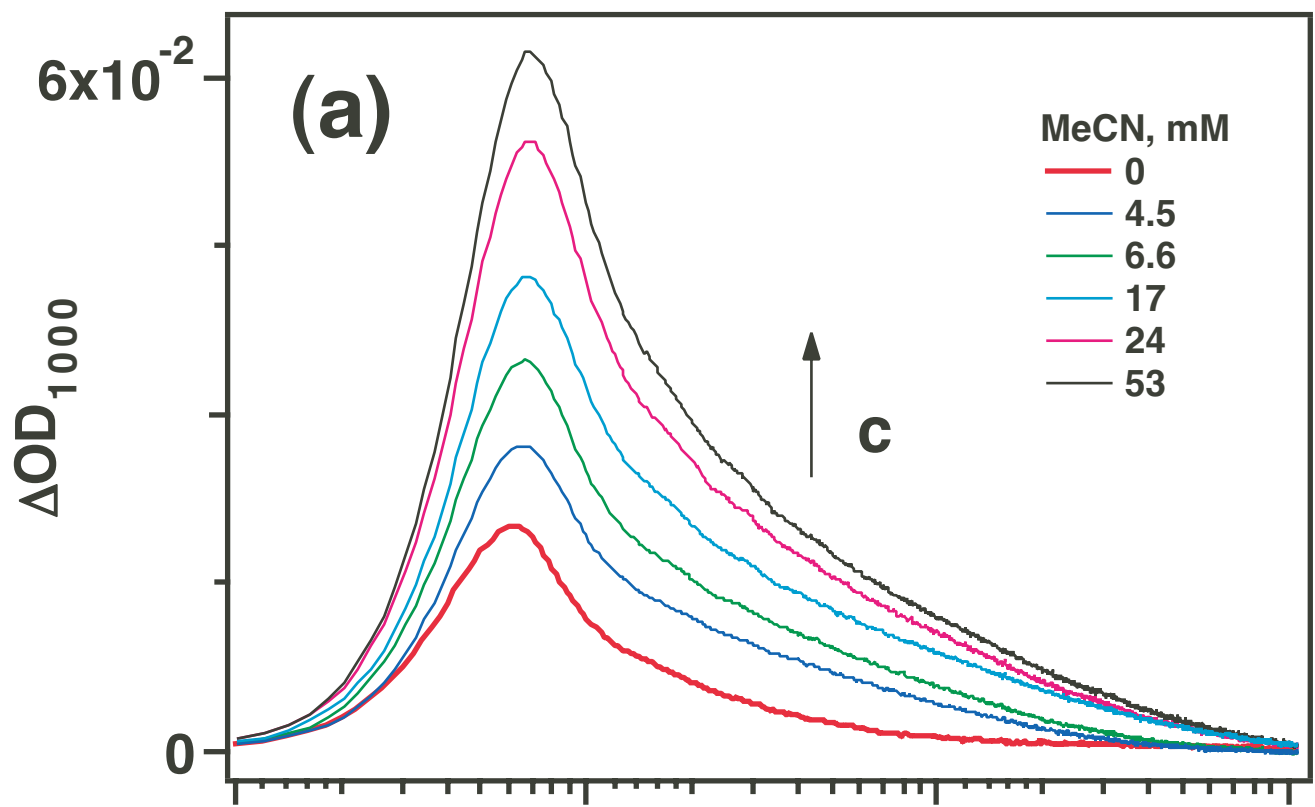
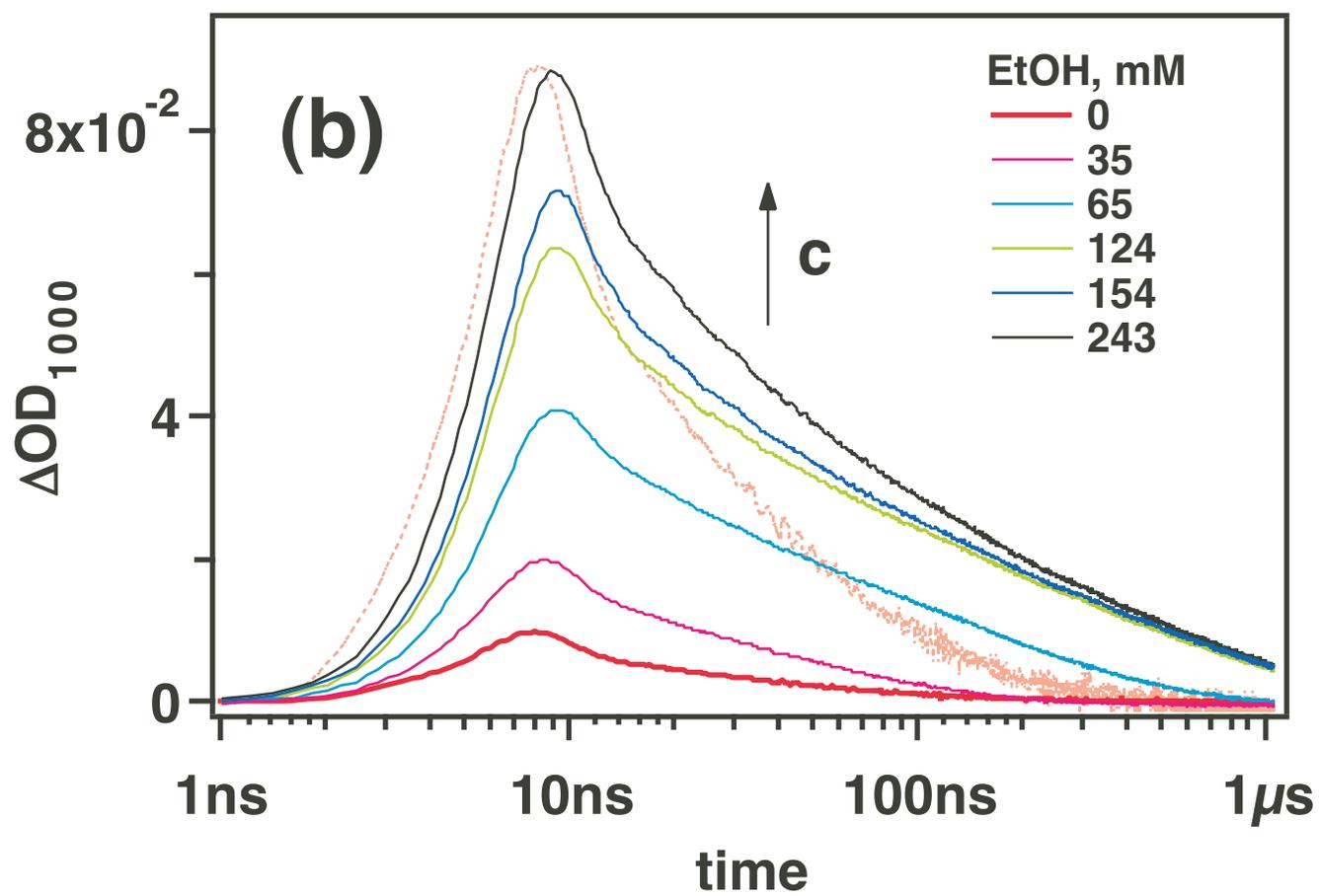

Figure 3S; Shkrob & Sauer

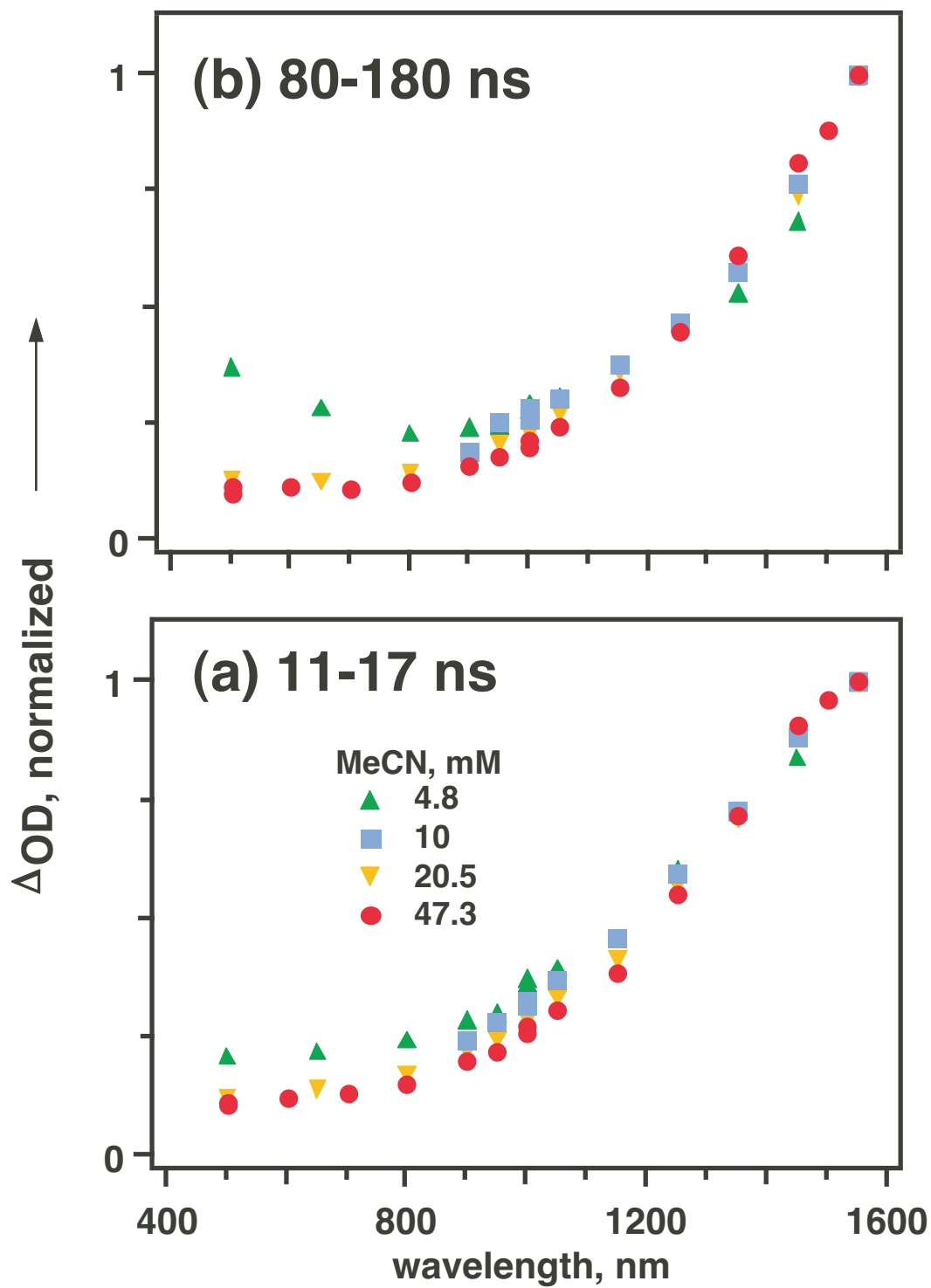

Figure 4S; Shkrob & Sauer

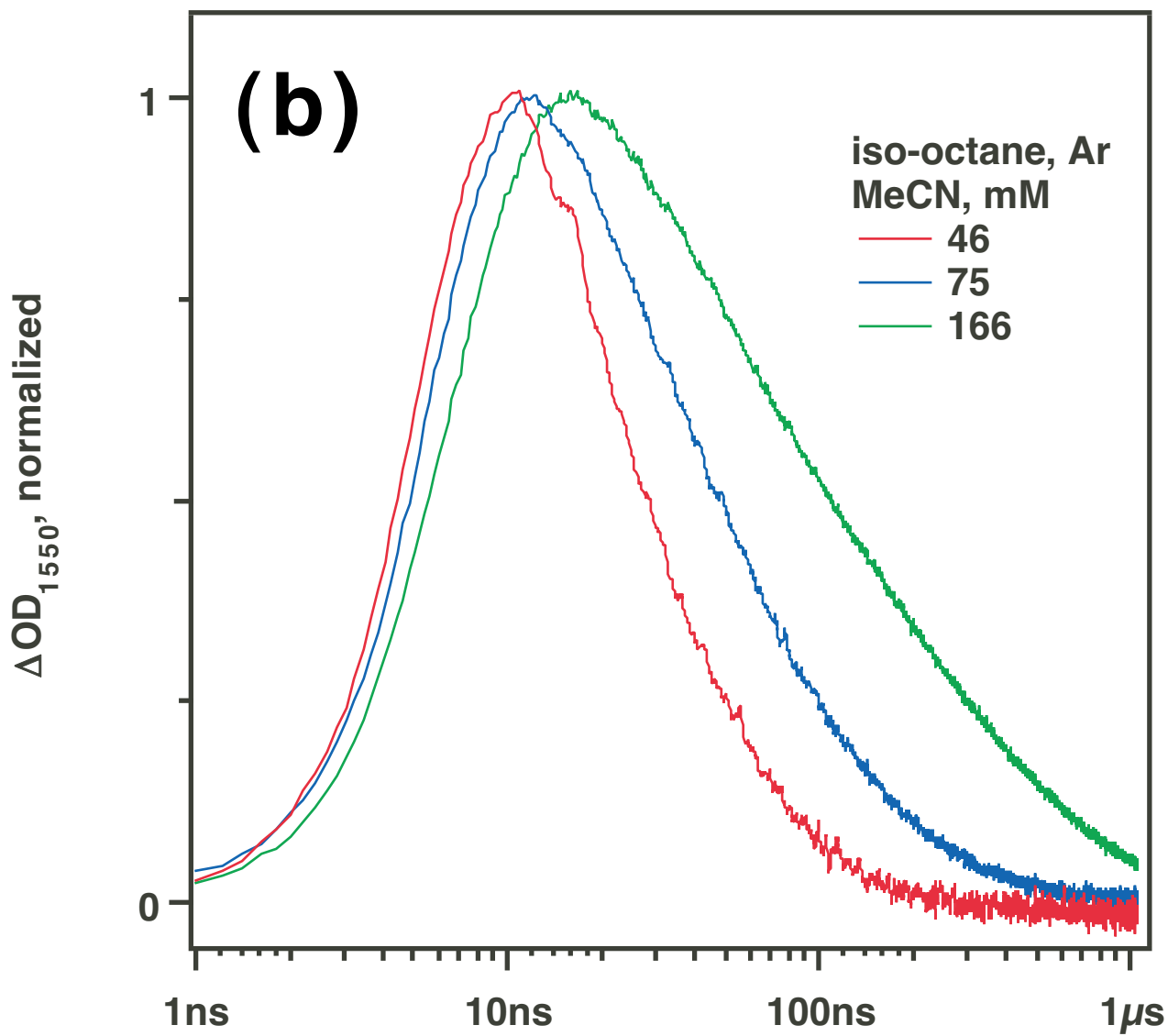

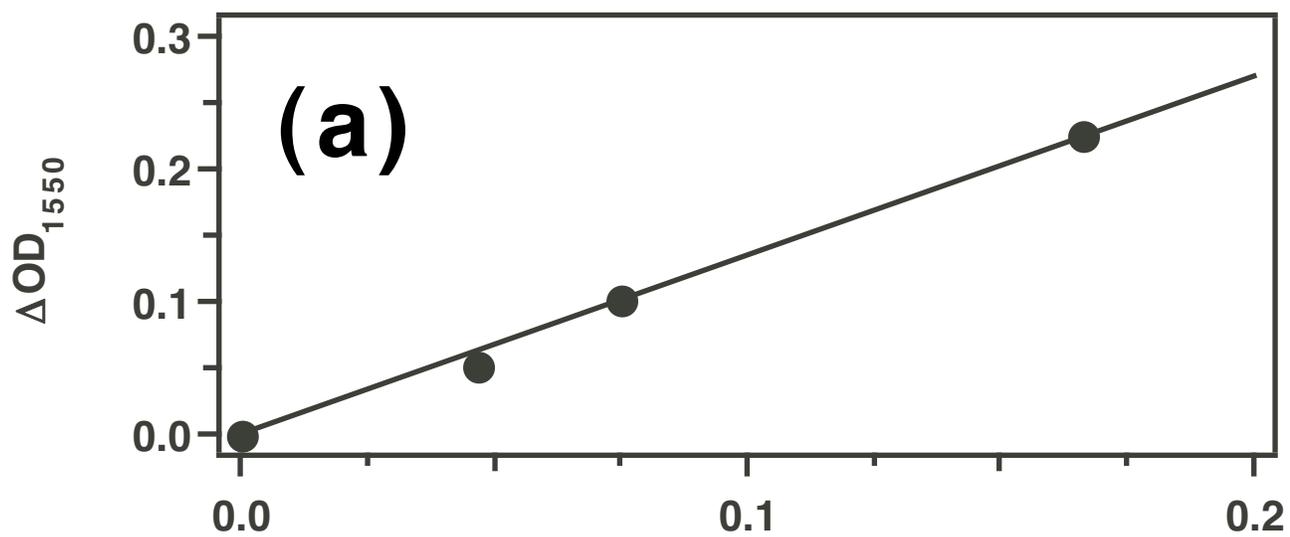

Figure 5S; Shkrob & Sauer

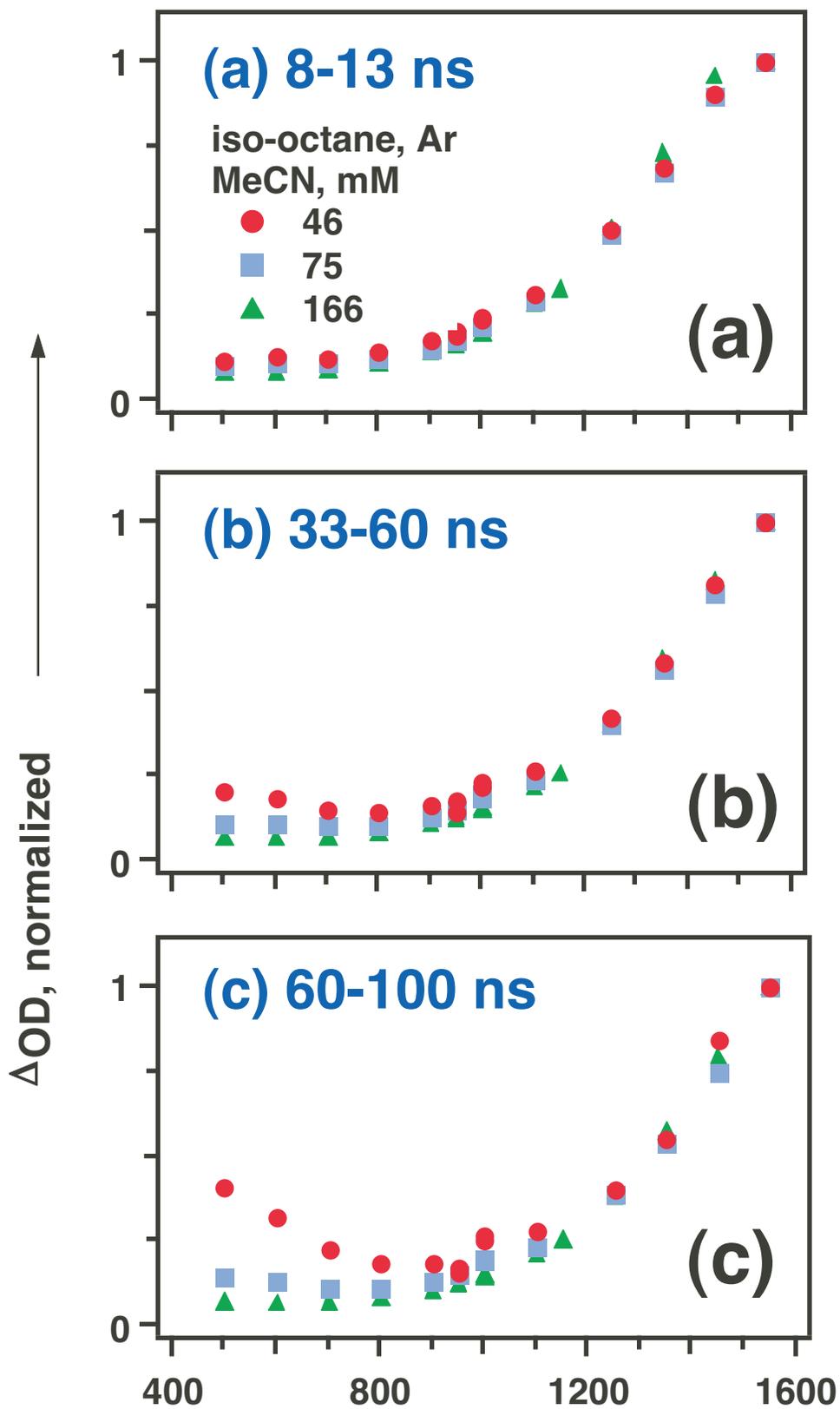

Figure 6S; Shkrob & Sauer

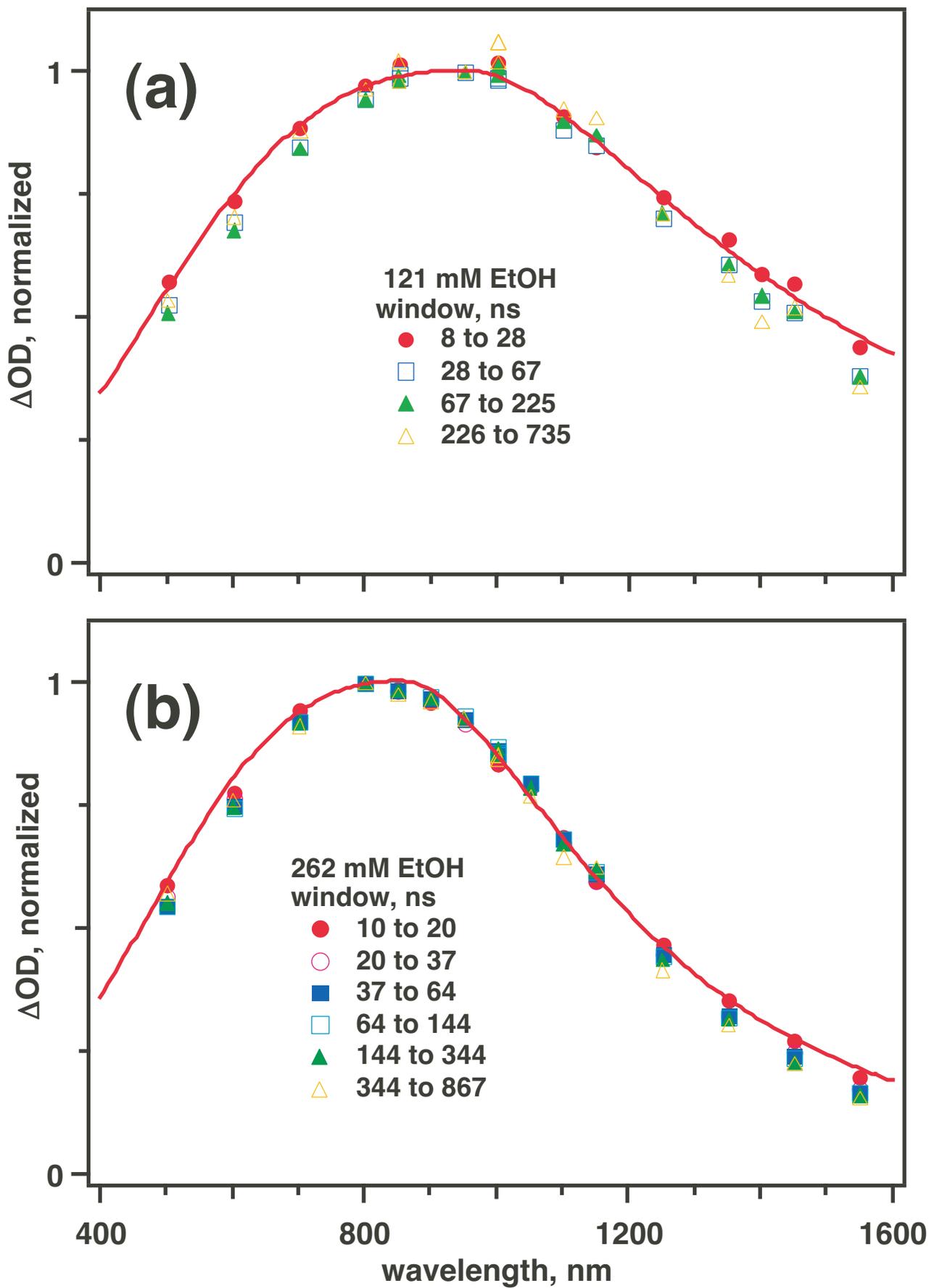

Figure 7S; Shkrob & Sauer

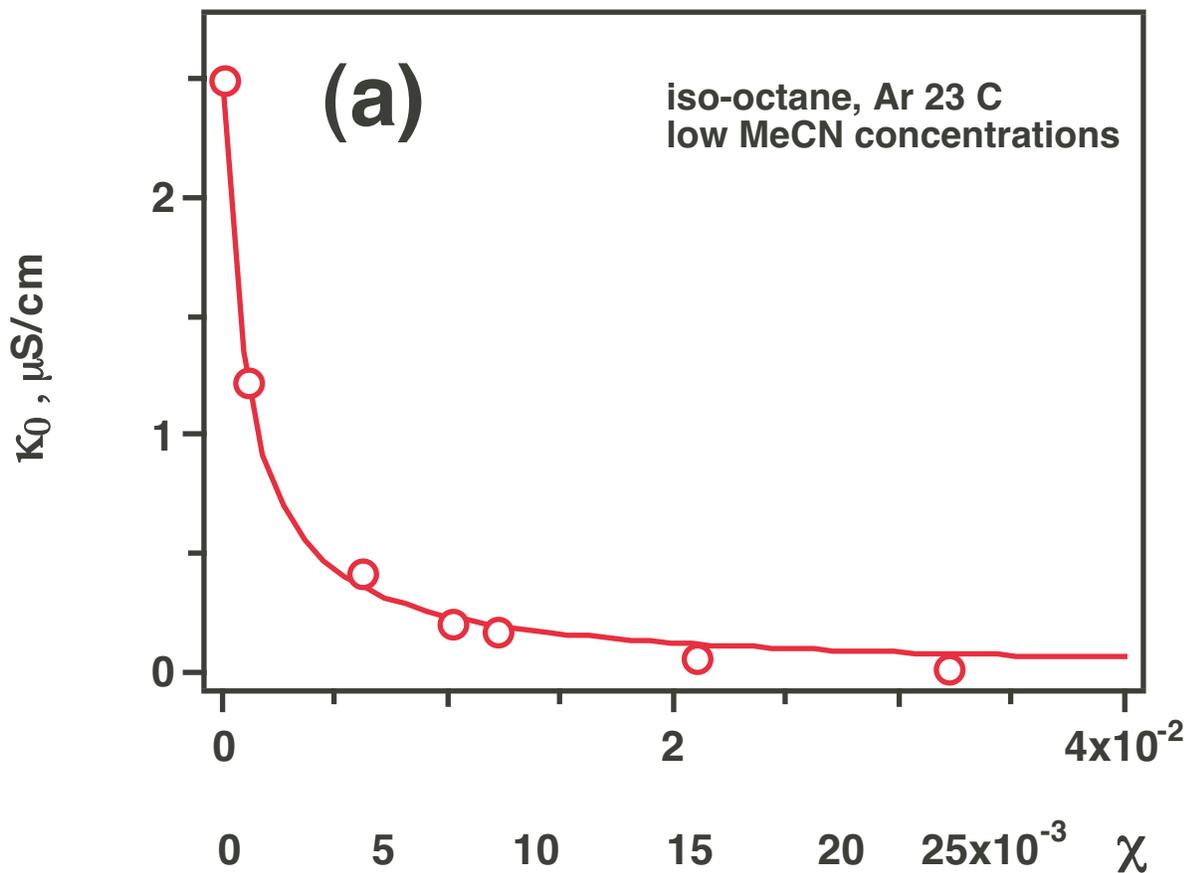
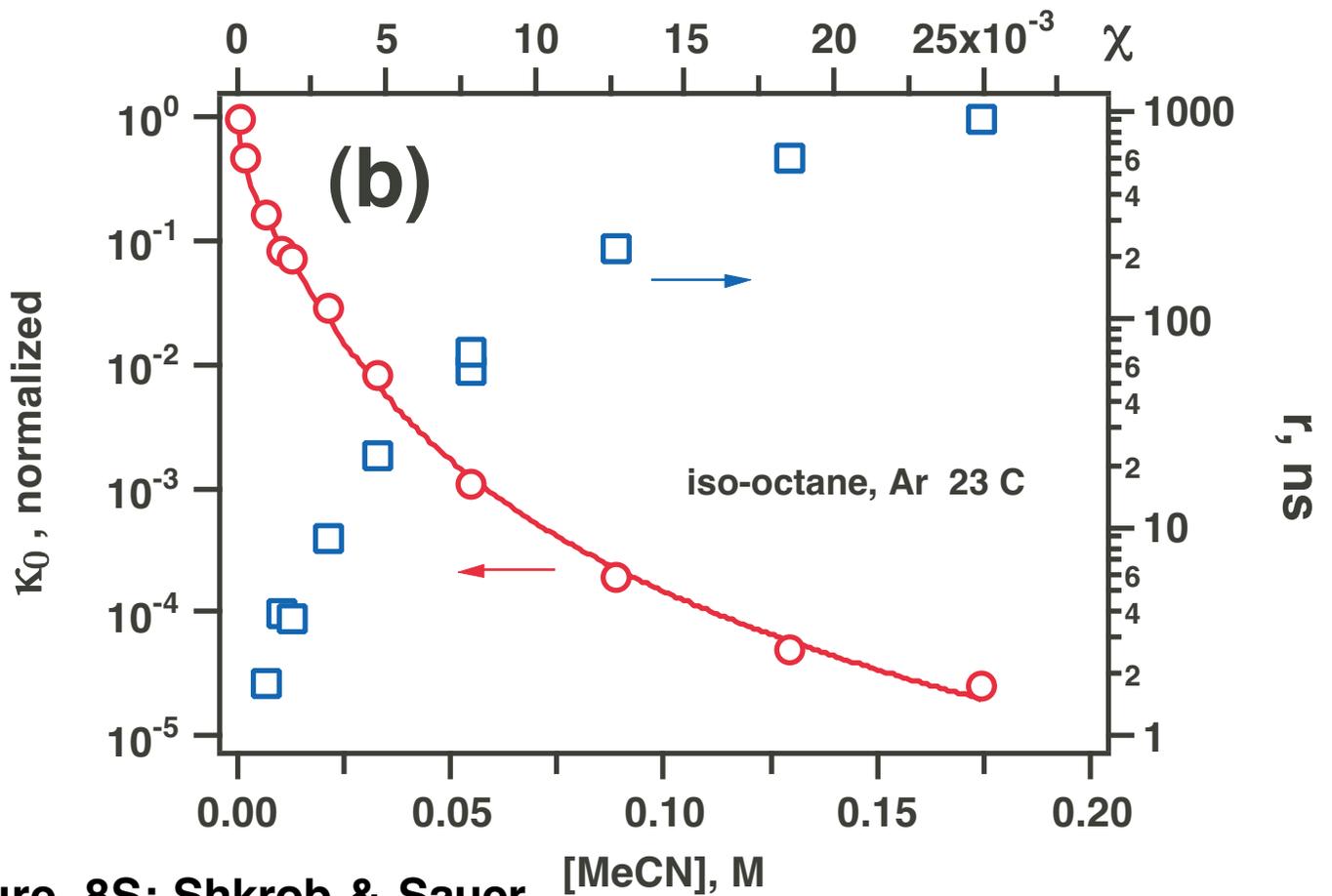

**Figure 8S; Shkrob & Sauer**

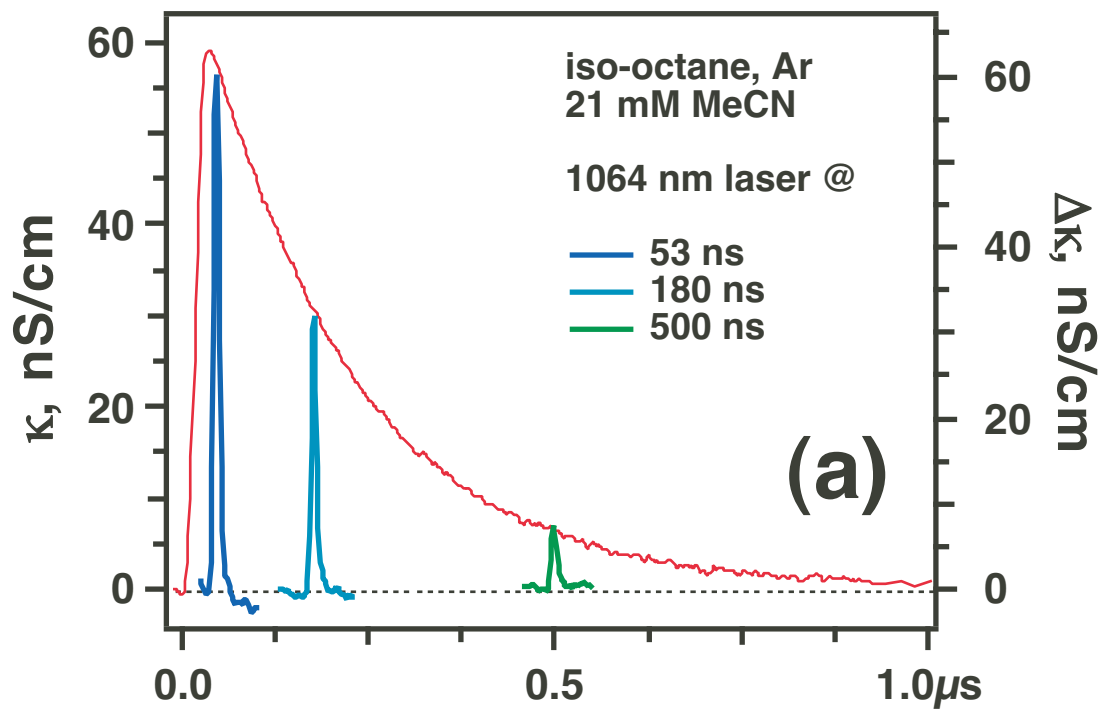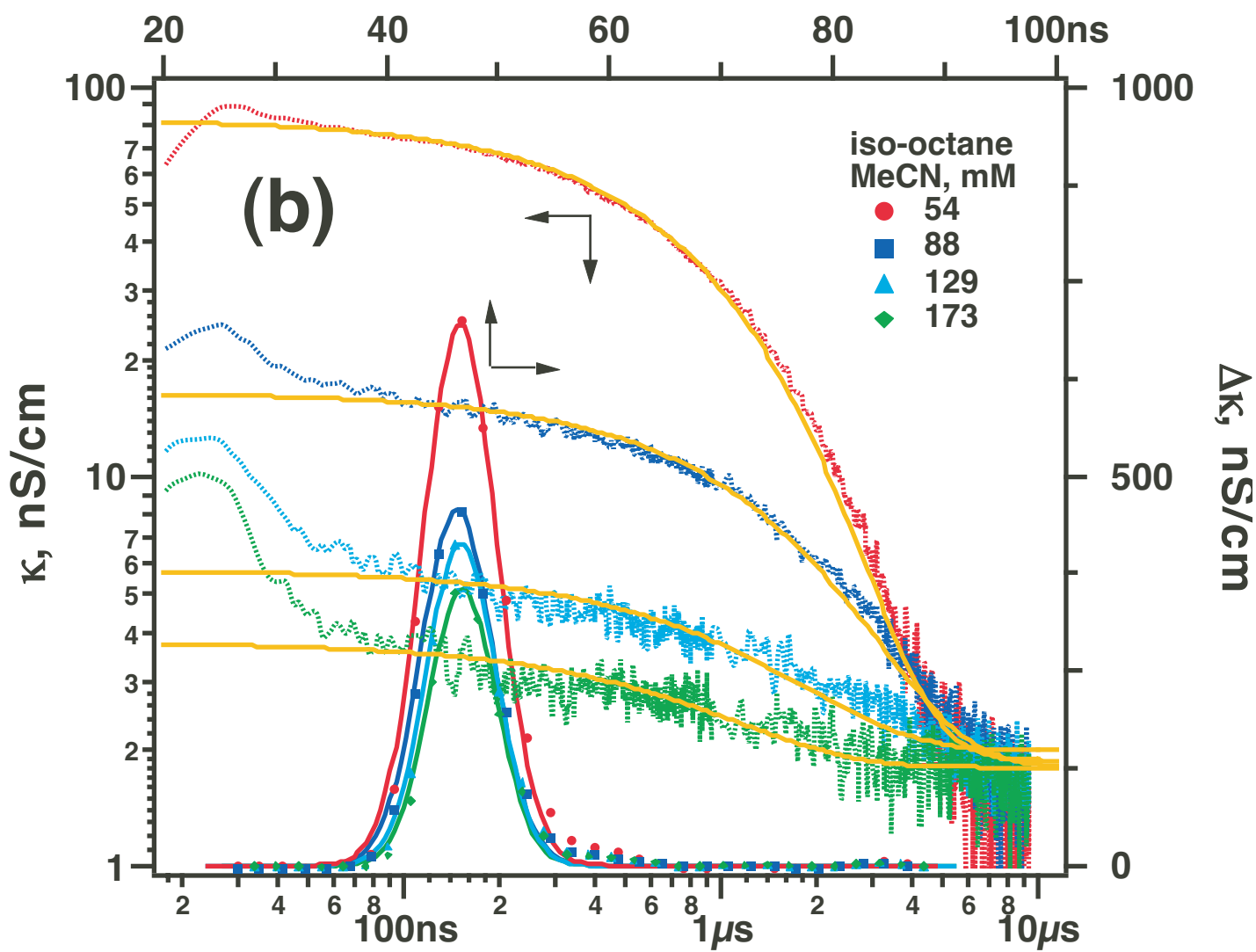

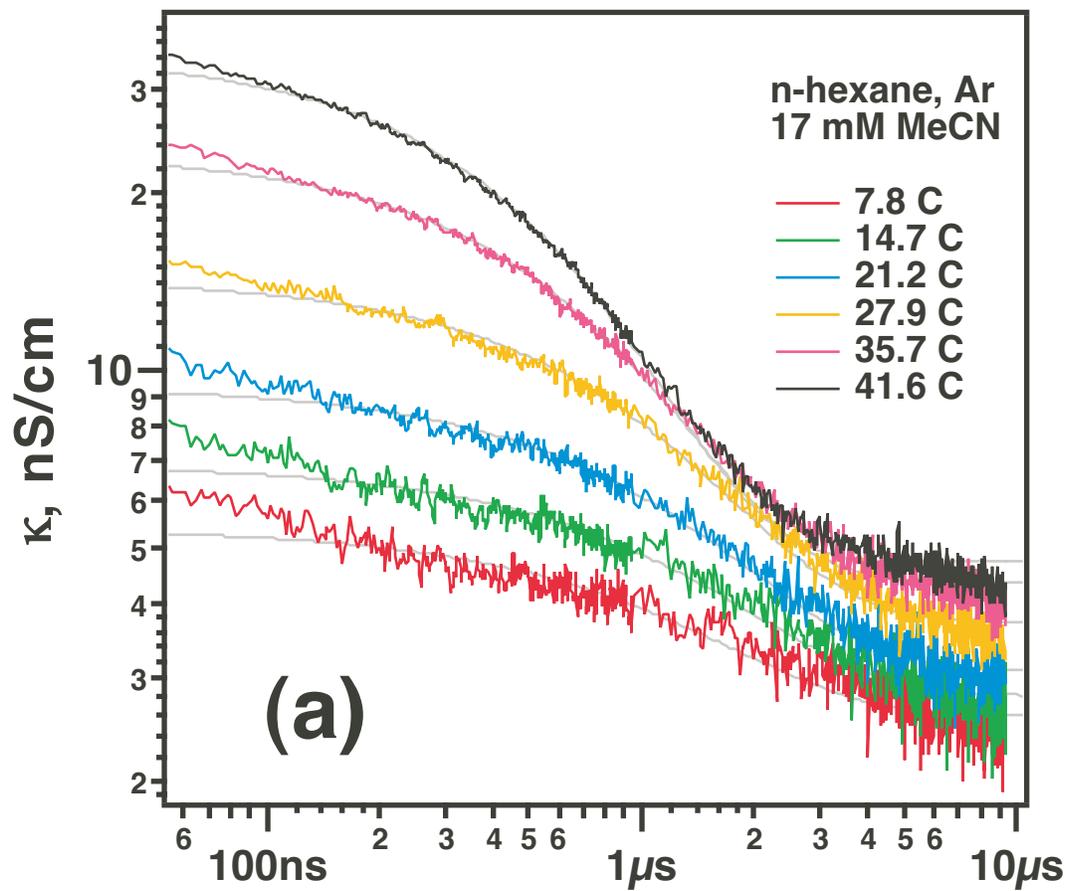

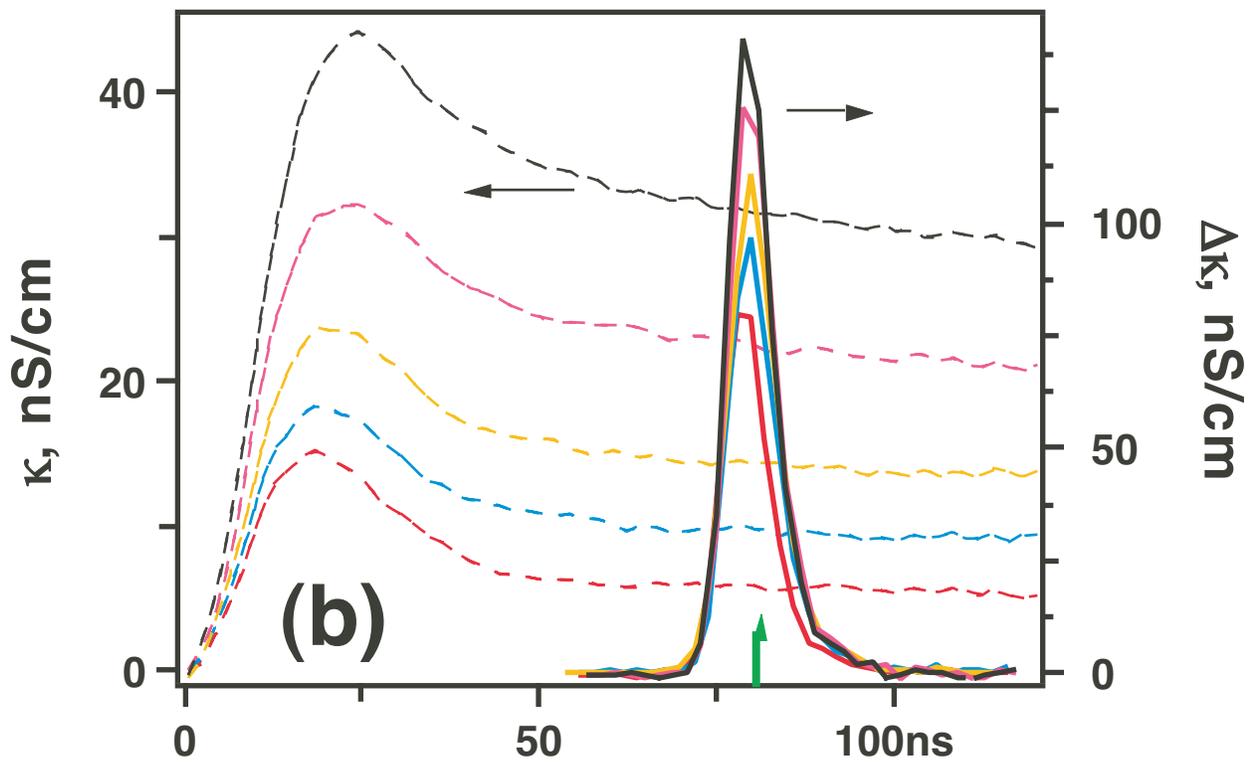

**Figure 10S; Shkrob & Sauer**

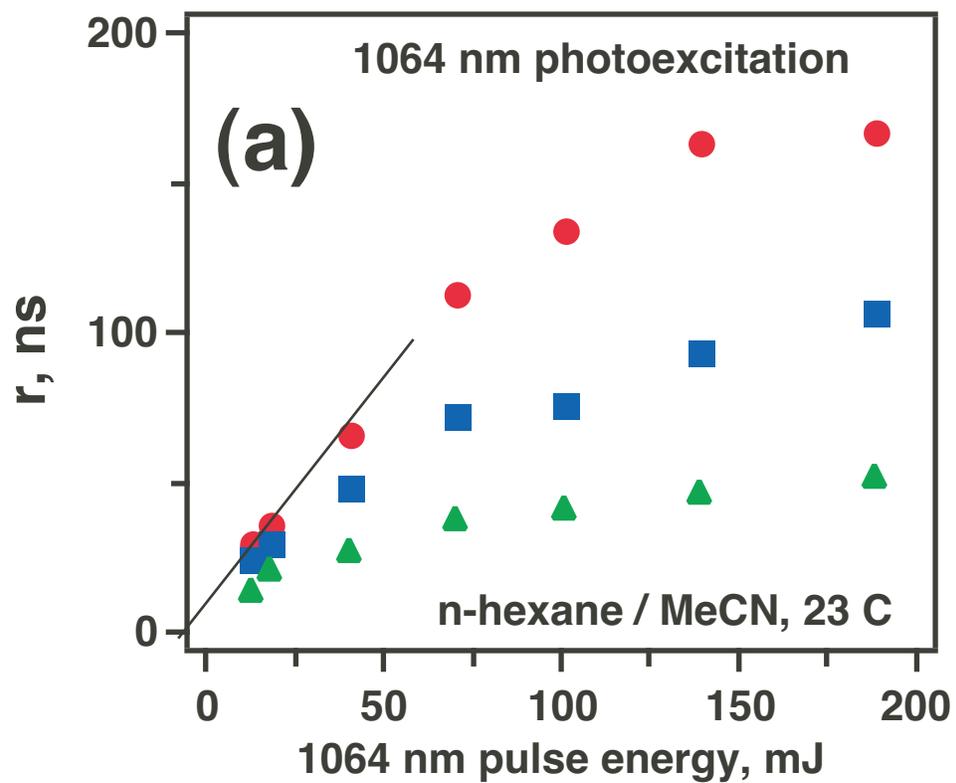

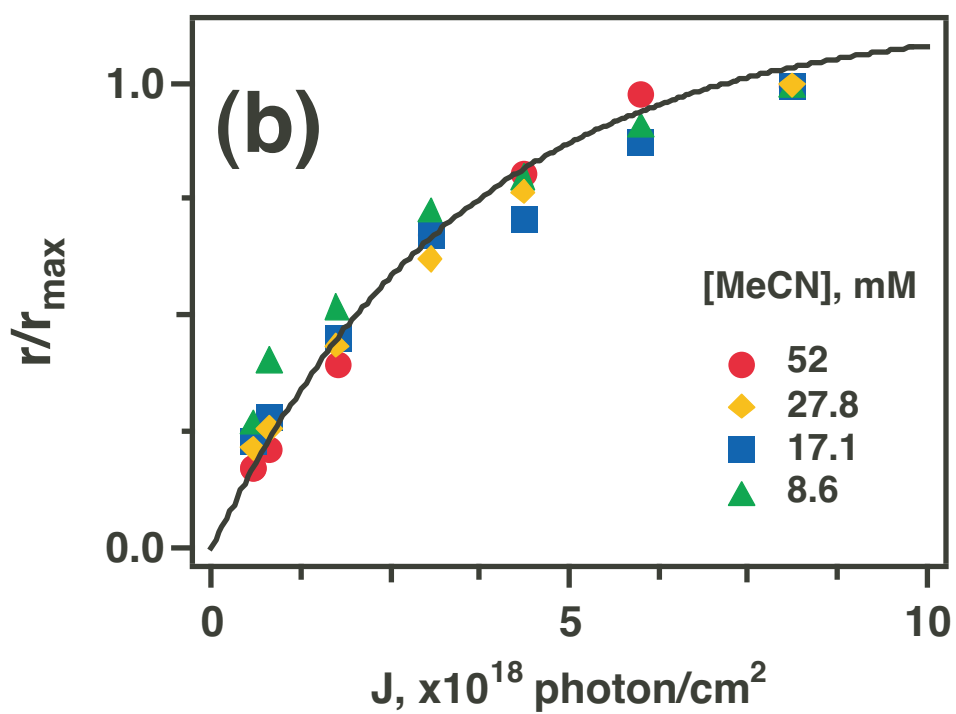

Figure 11S; Shkrob & Sauer

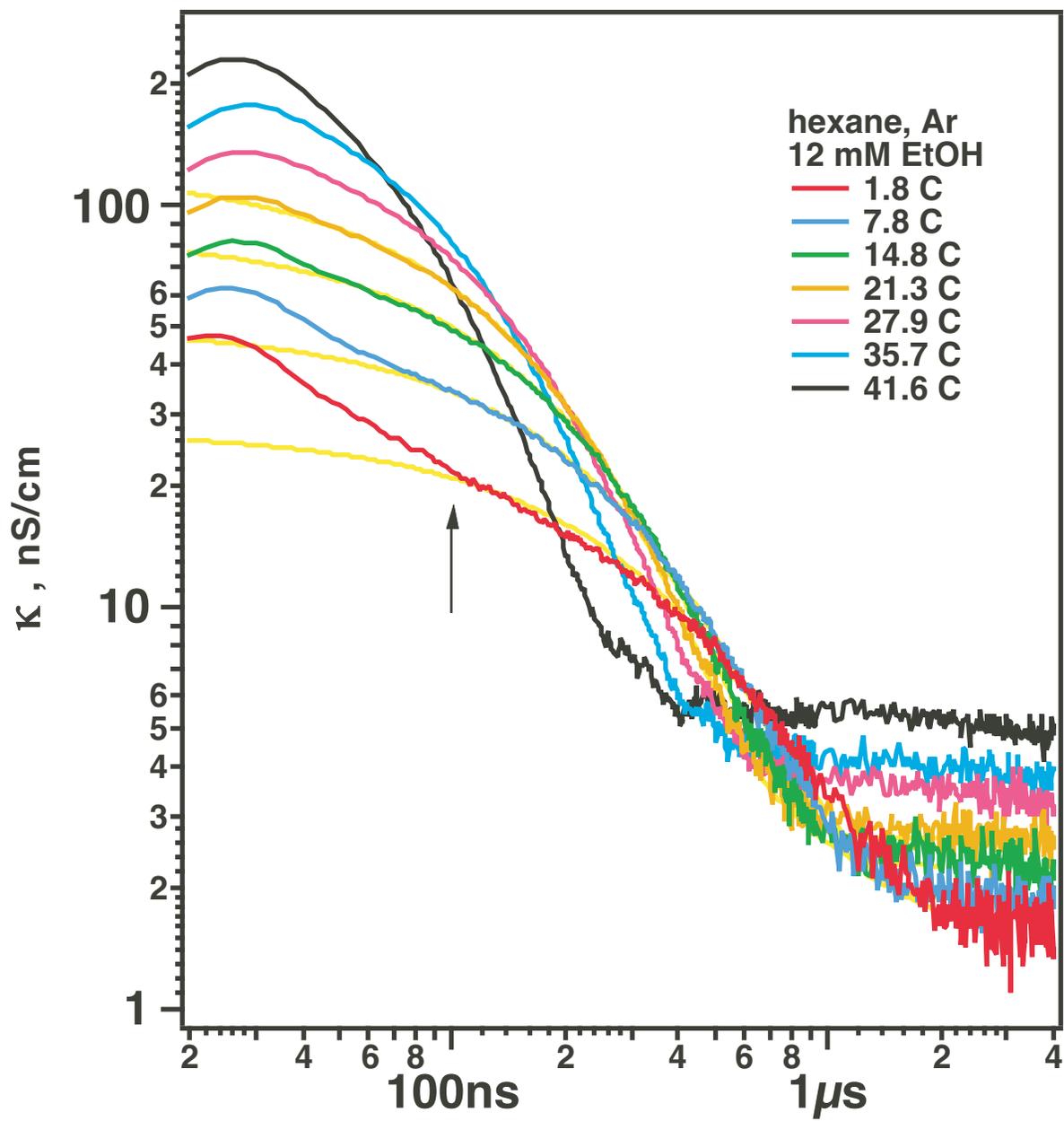

**Figure 12S; Shkrob & Sauer**

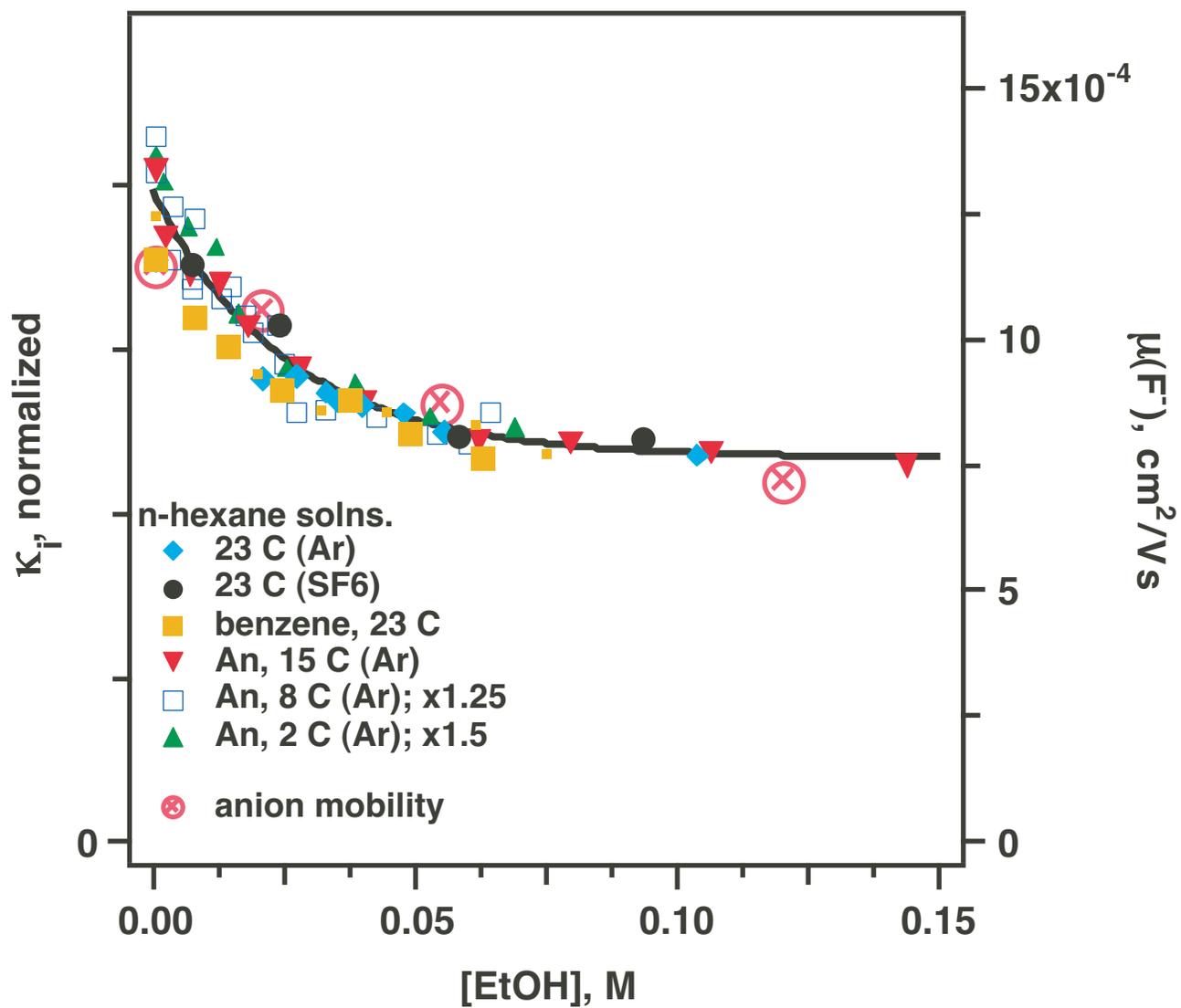

Figure 13S; Shkrob & Sauer

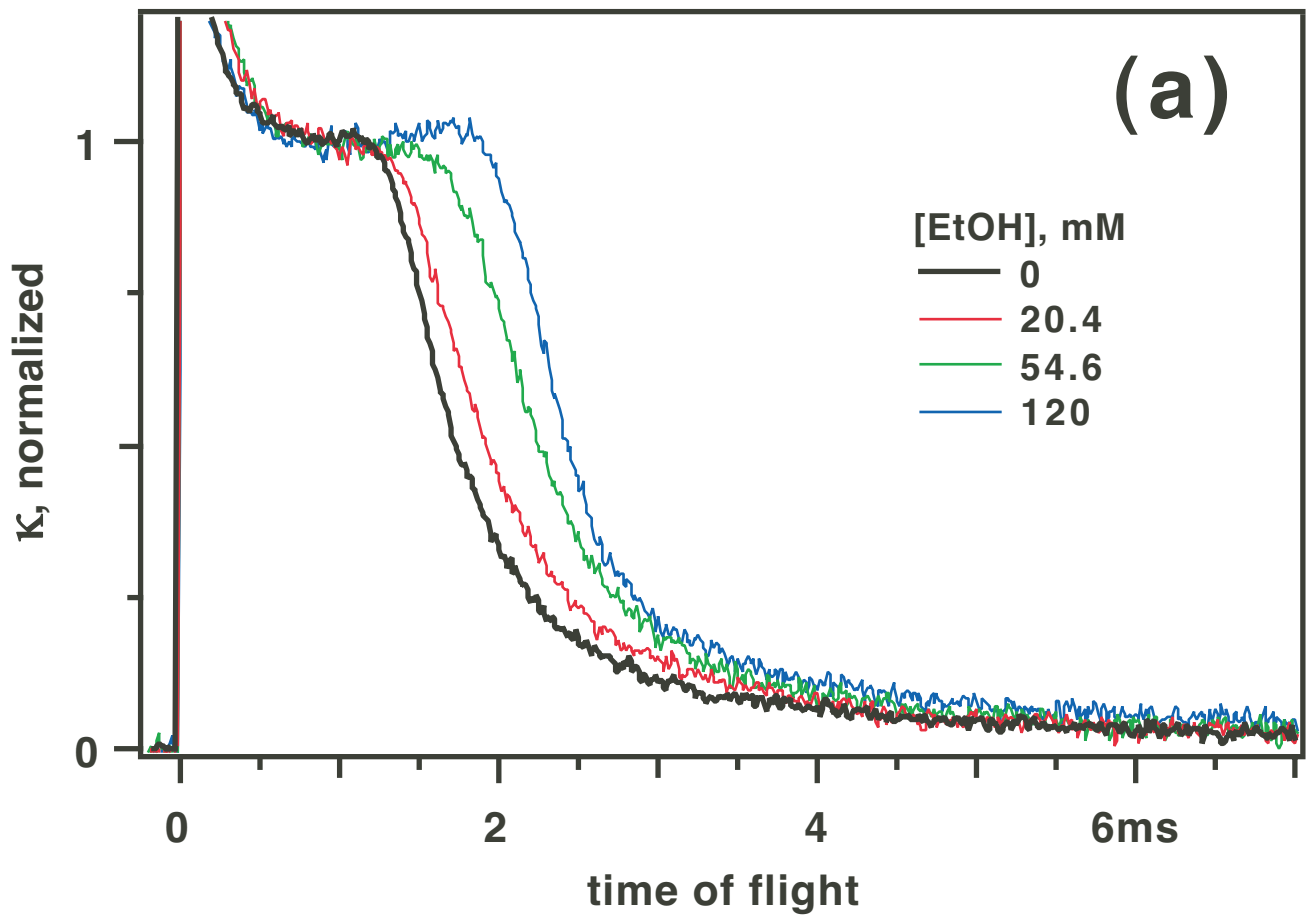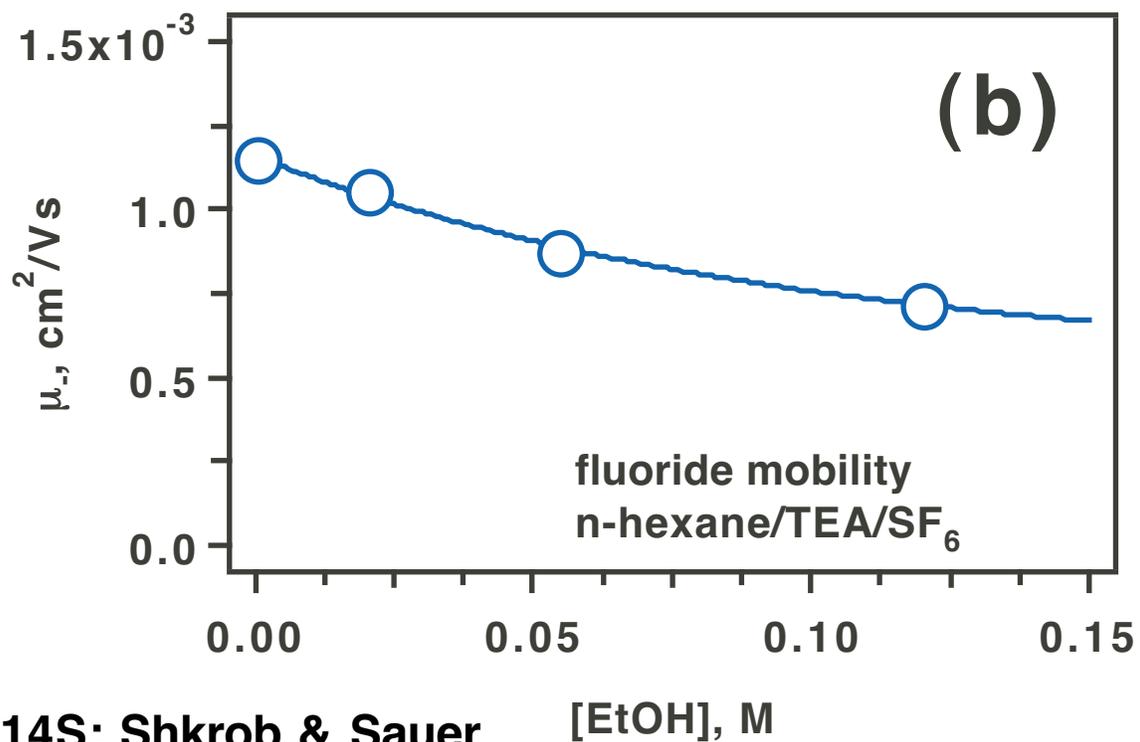

Figure 14S; Shkrob & Sauer

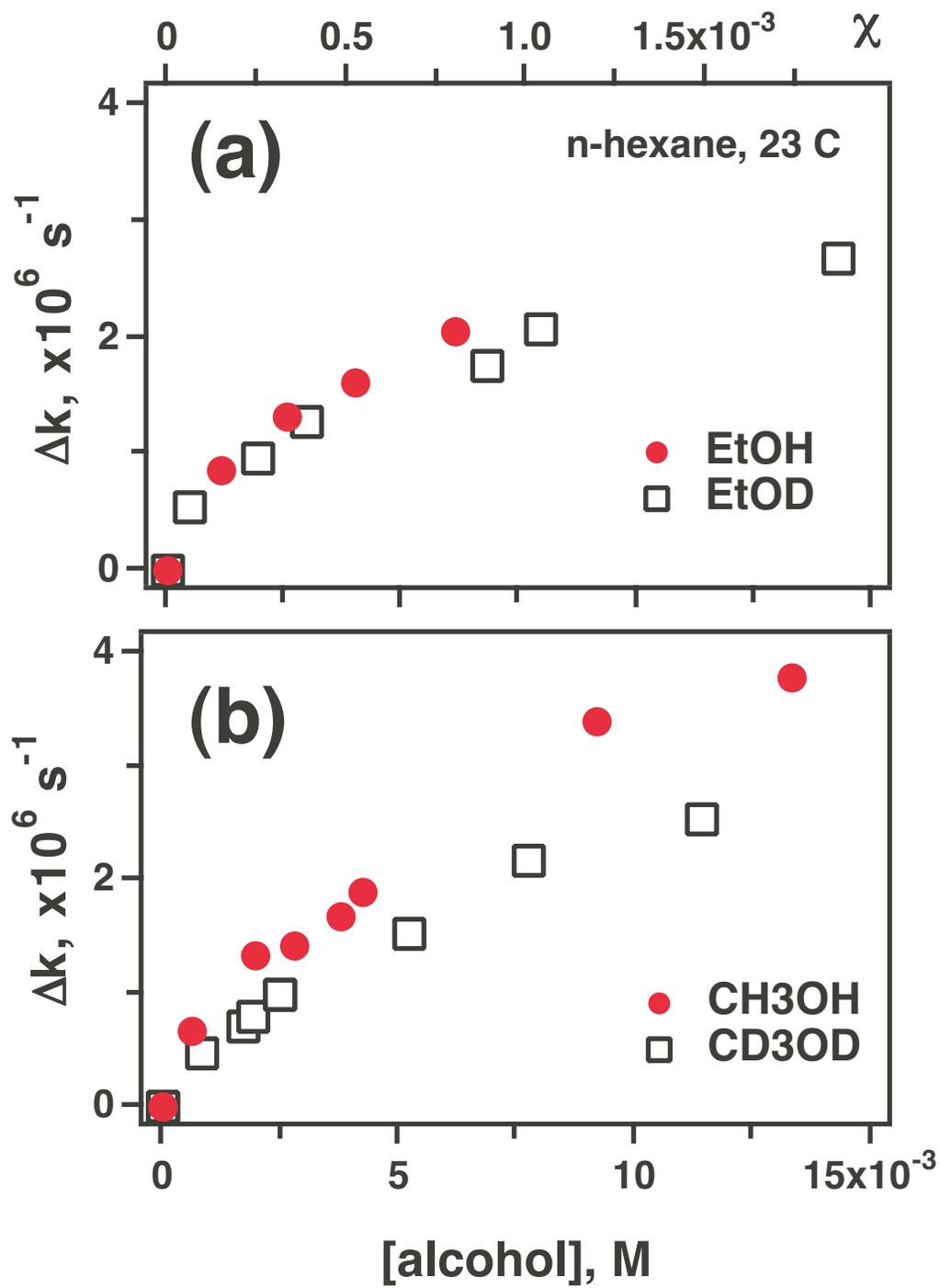

Figure 15S; Shkrob & Sauer

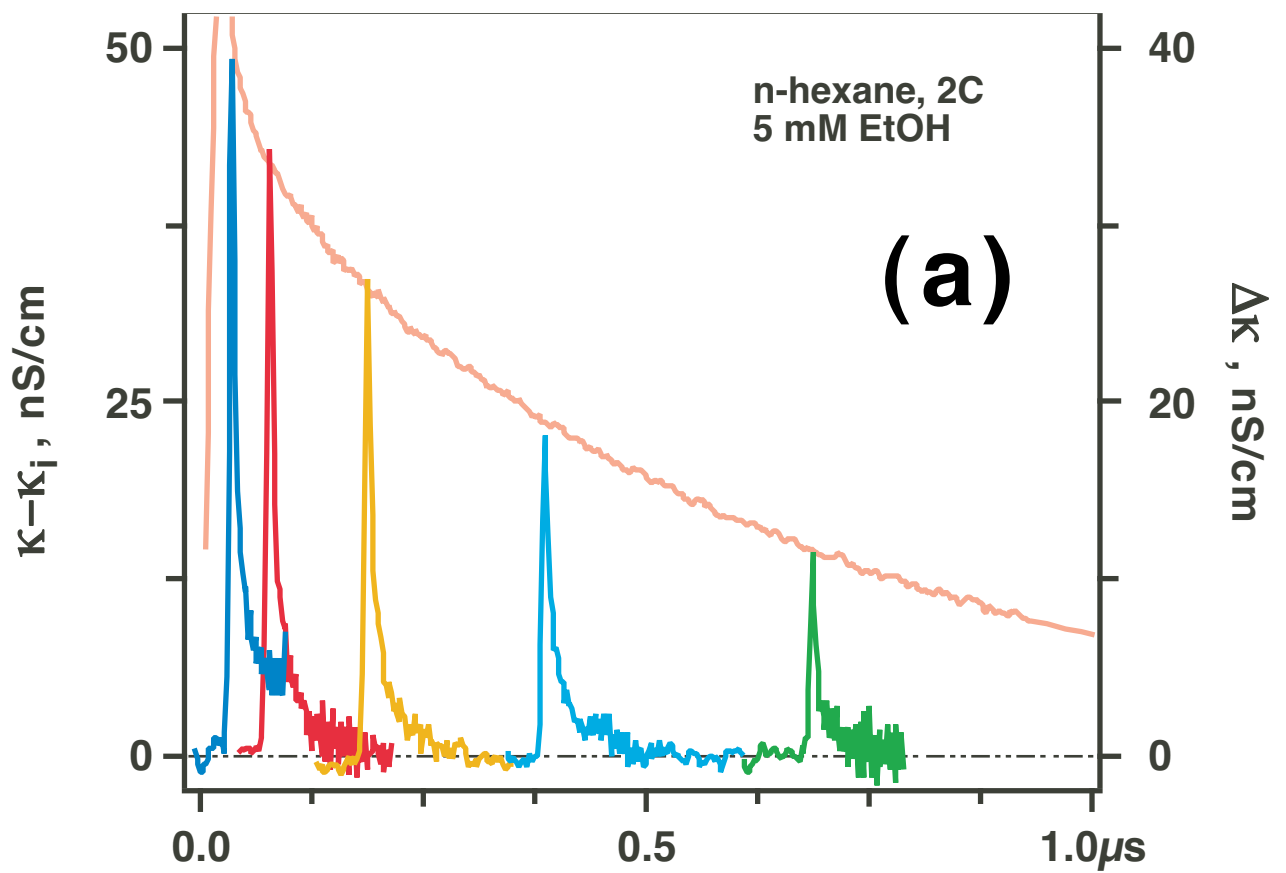
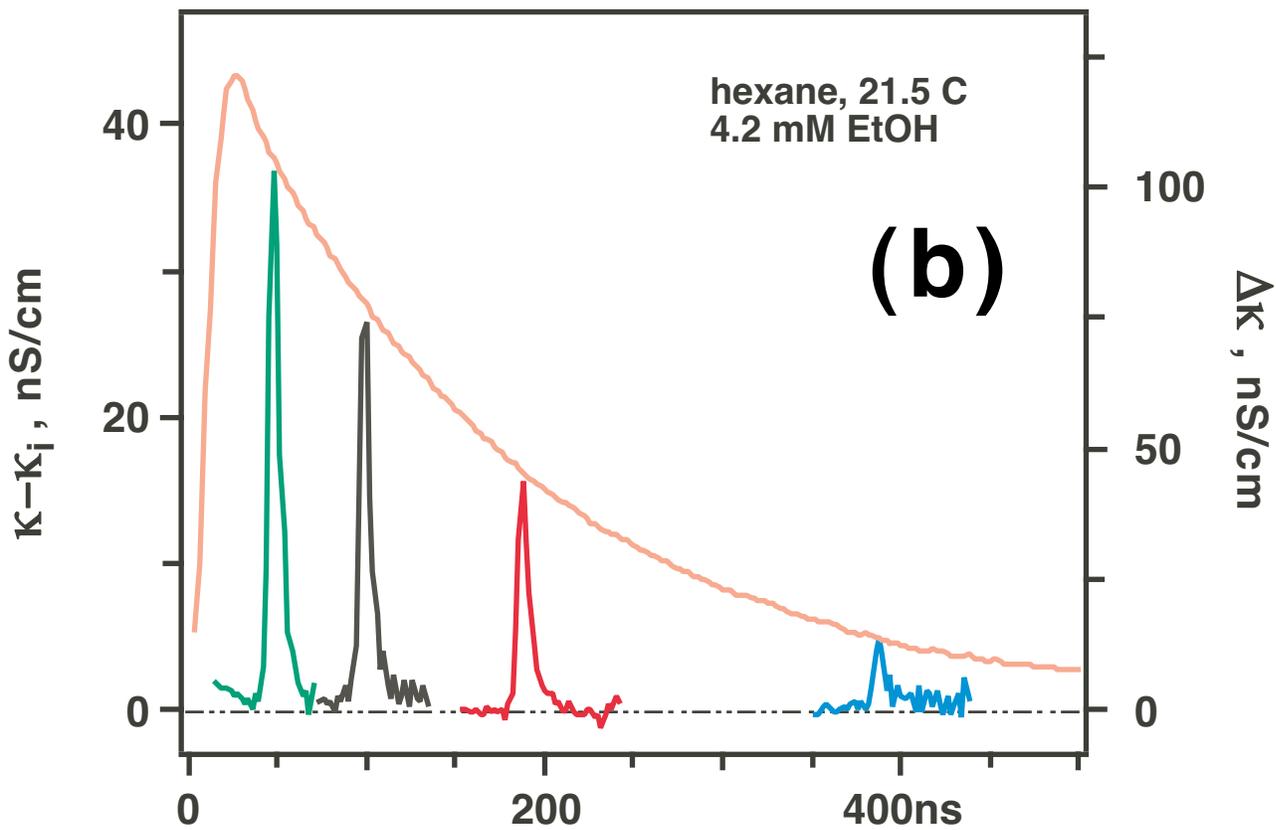

Figure 16S; Shkrob & Sauer

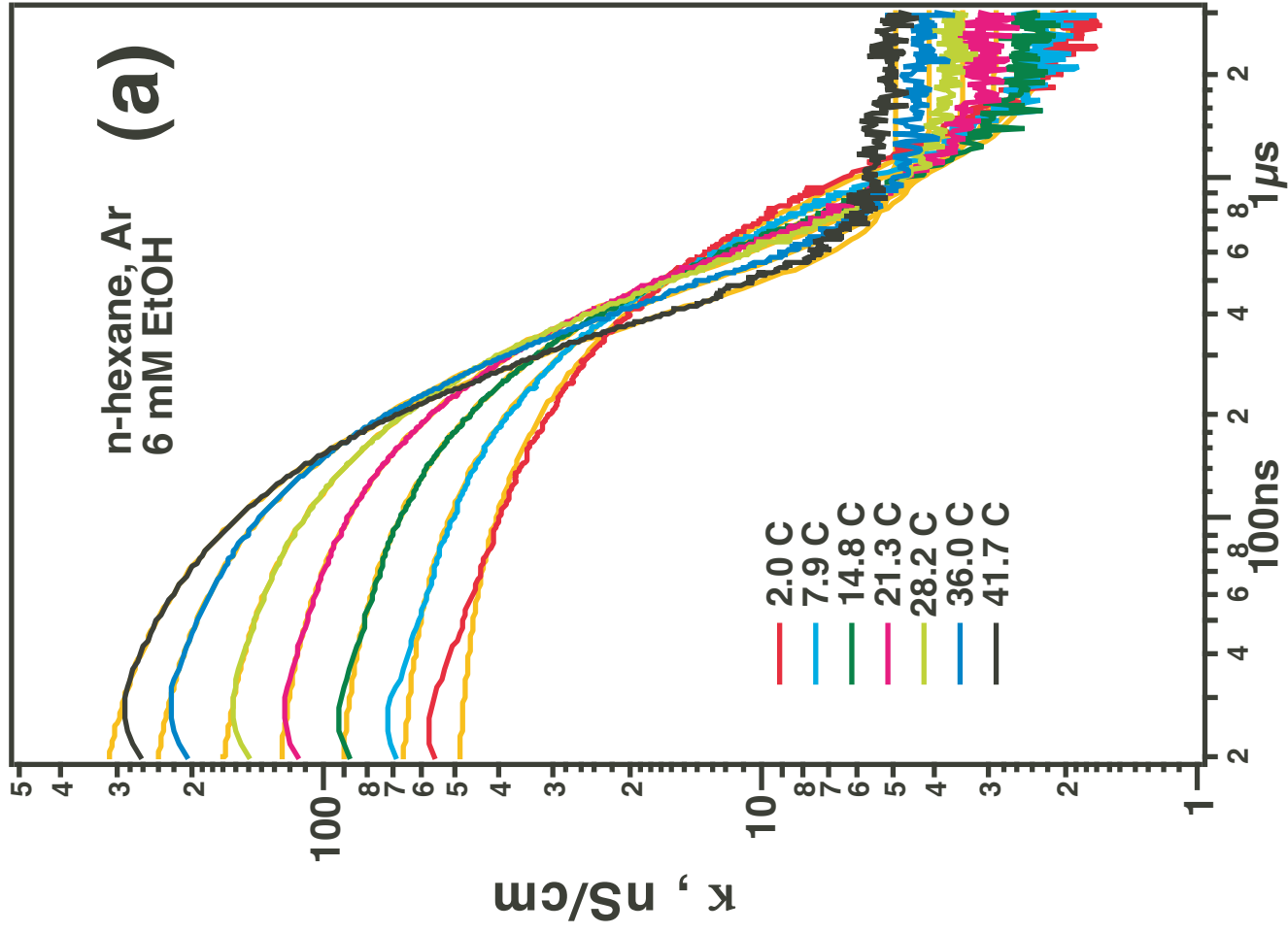
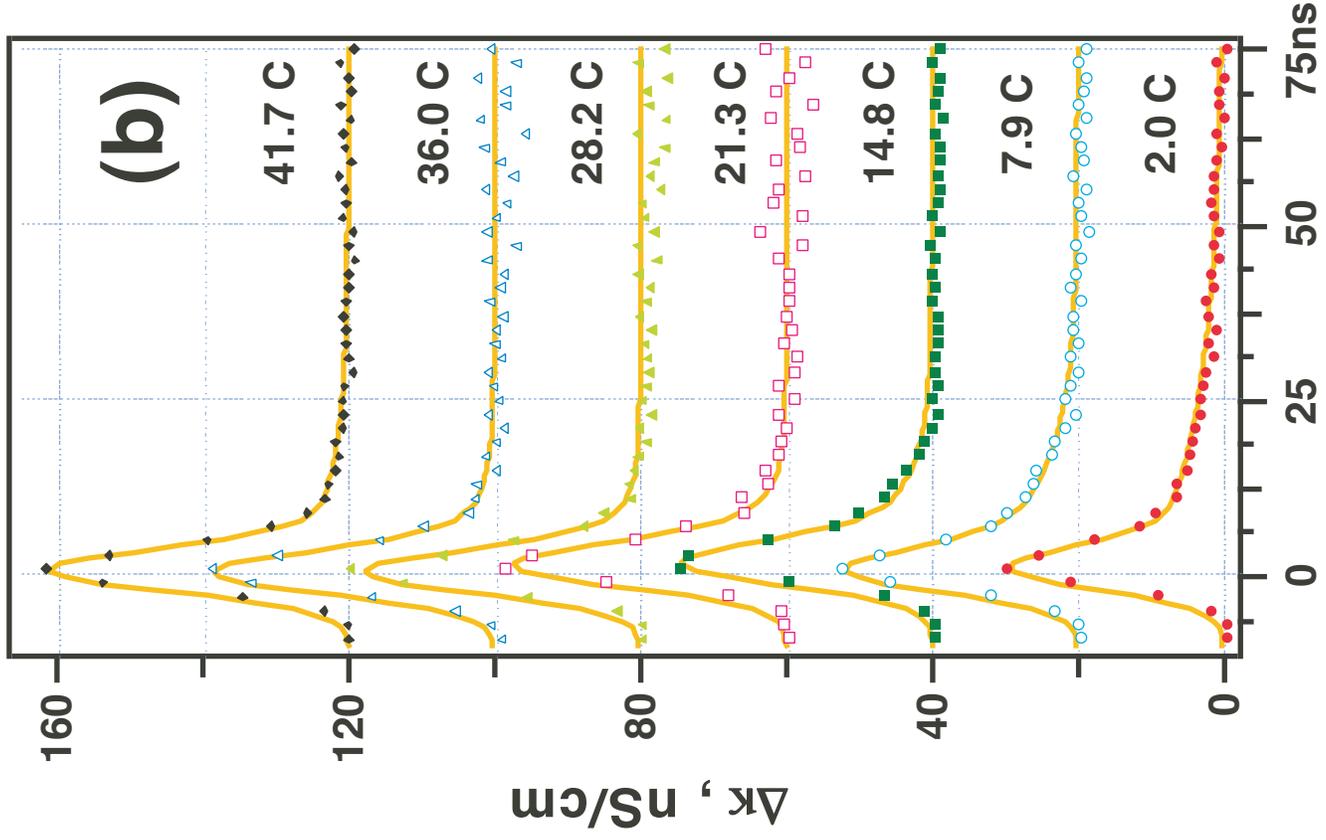

Figure 17S; Shkrob & Sauer

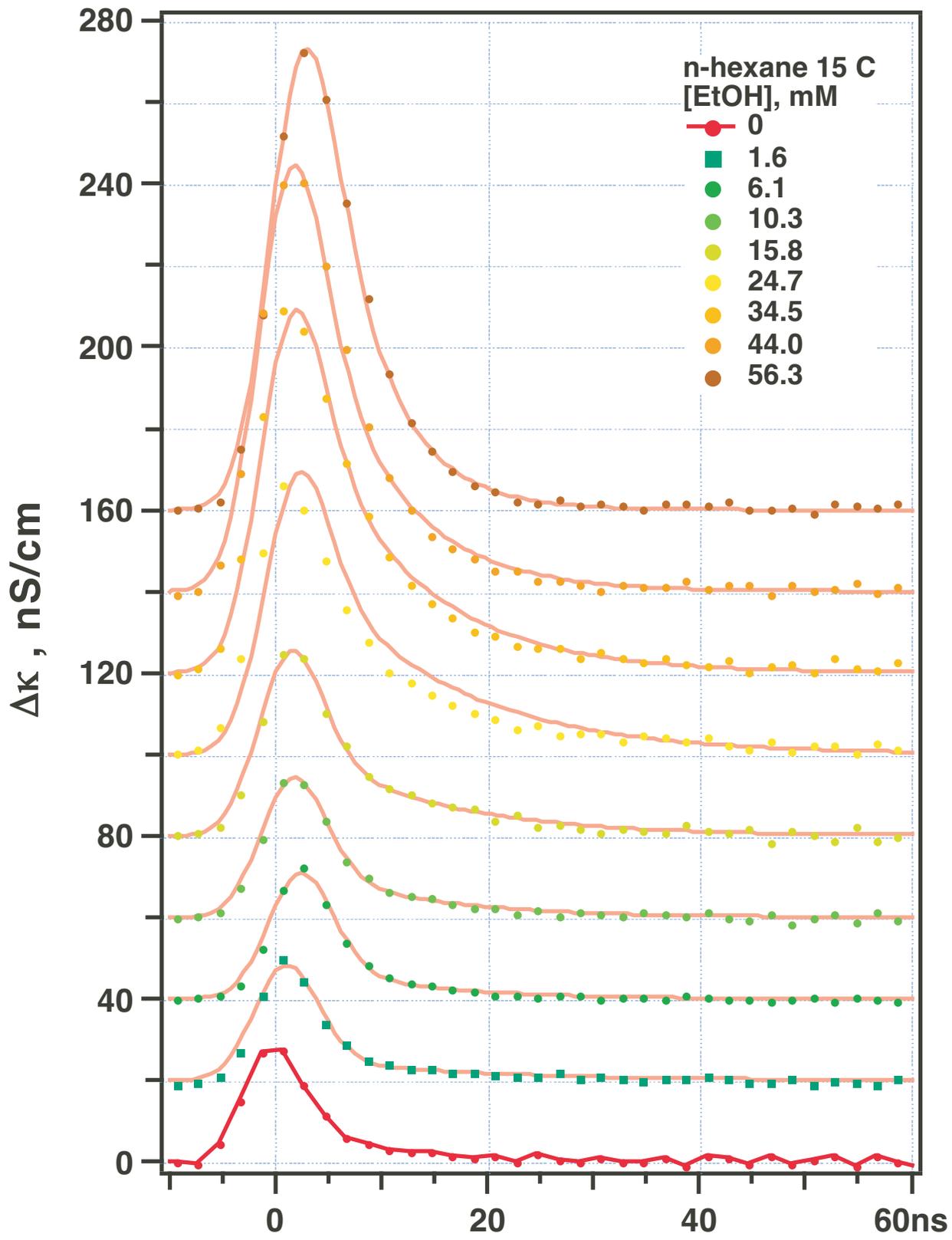

Figure 18S; Shkrob & Sauer

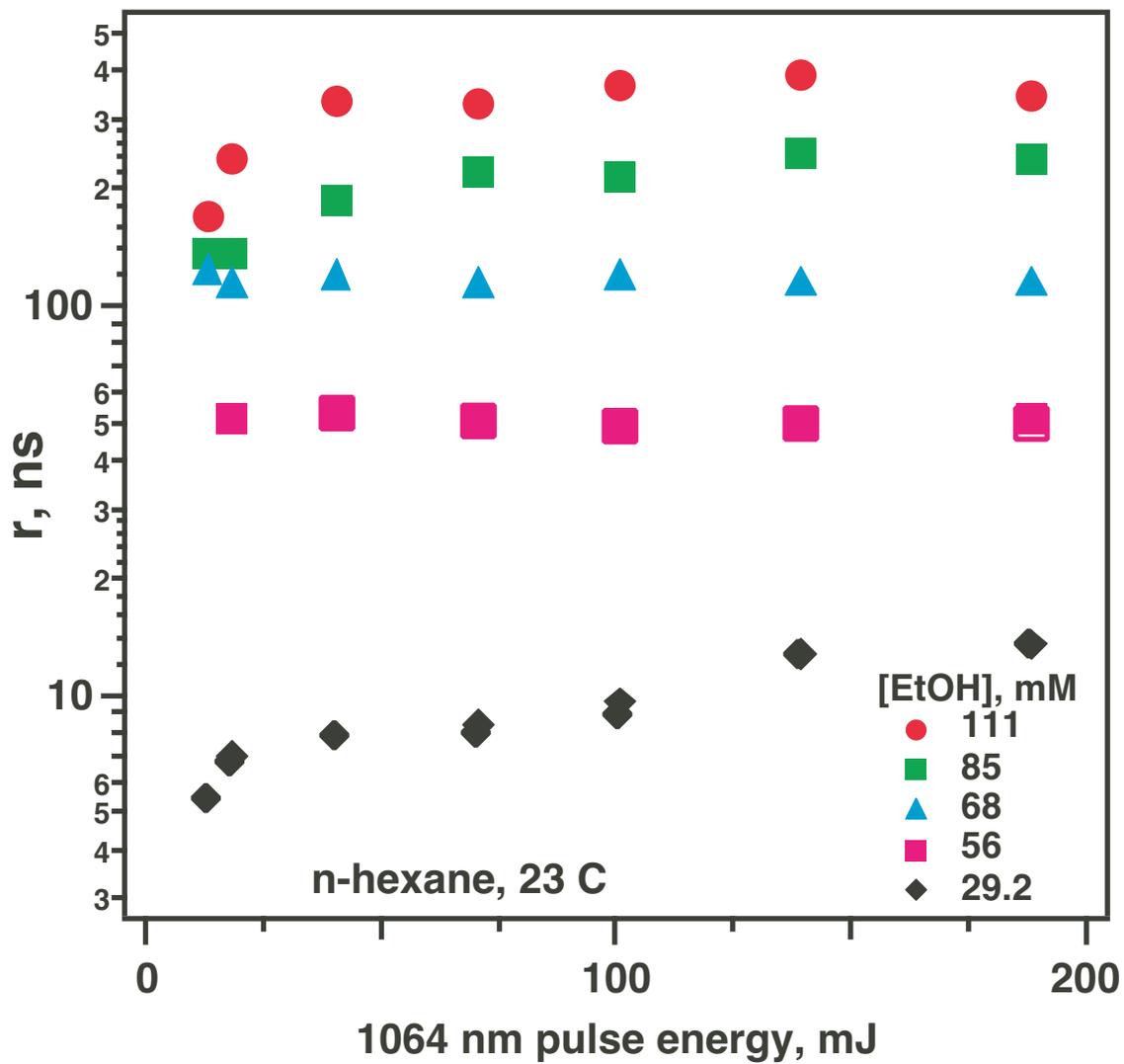

Figure 19S; Shkrob & Sauer

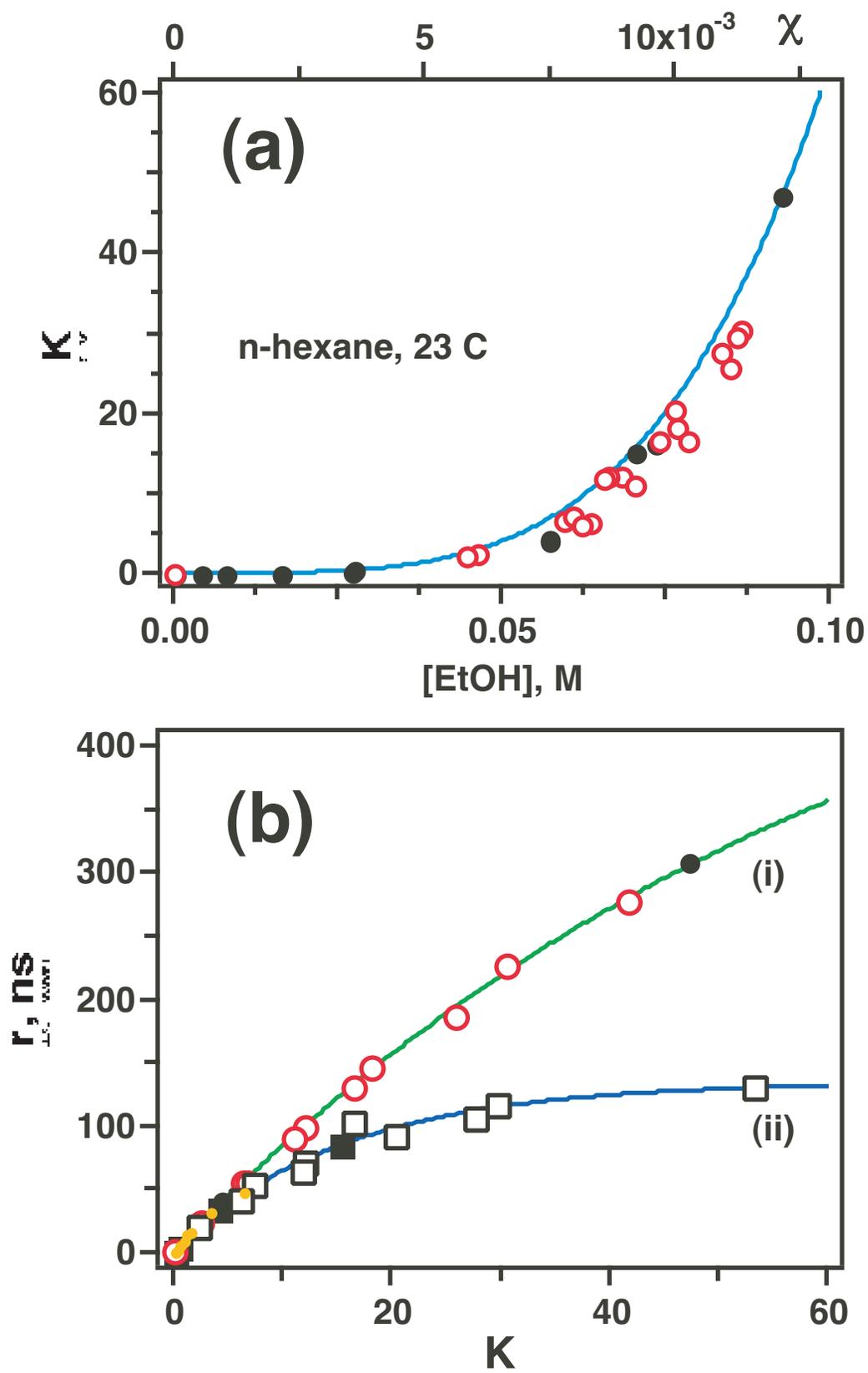

Figure 20S; Shkrob & Sauer

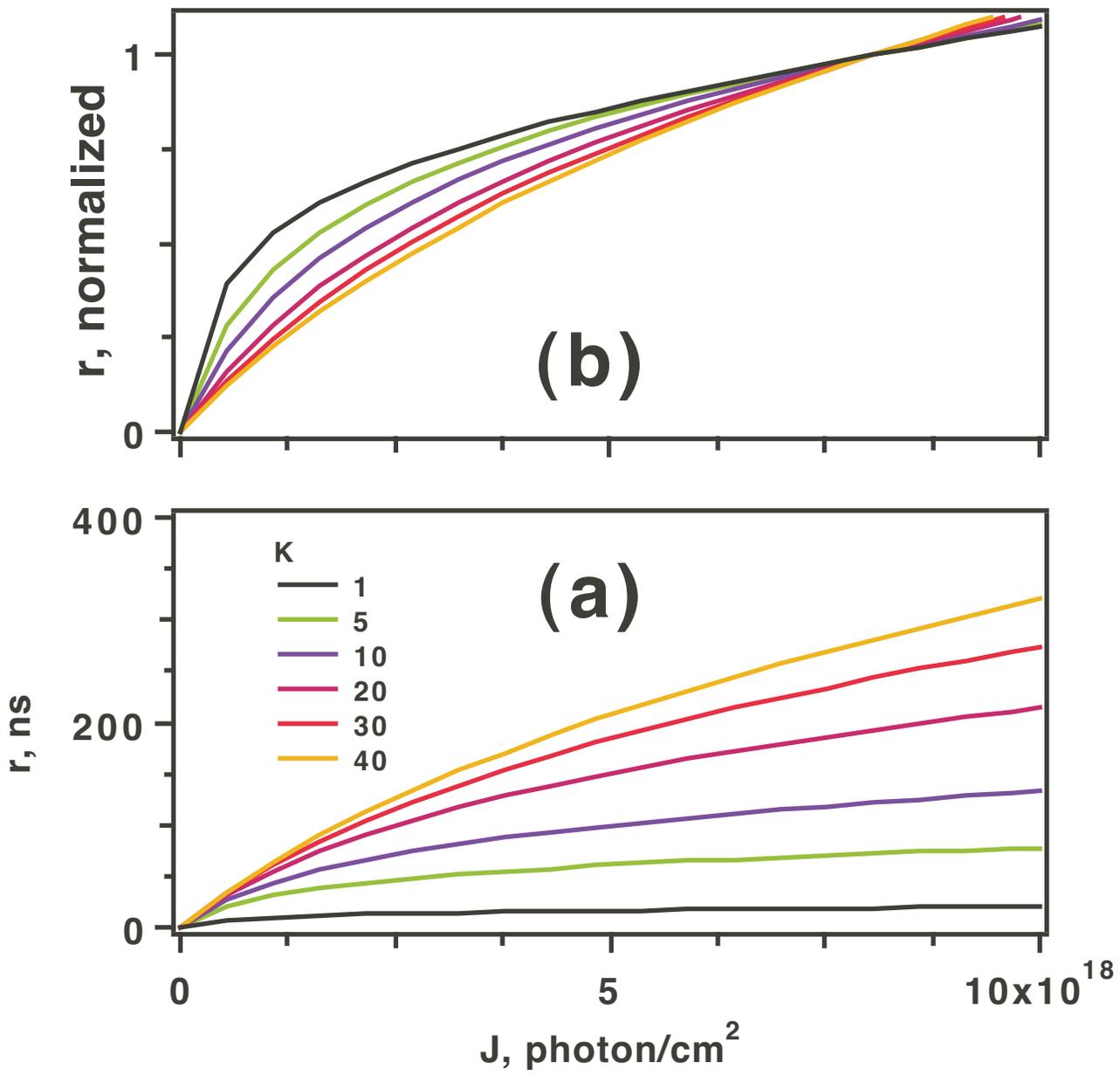

Figure 21S; Shkrob & Sauer

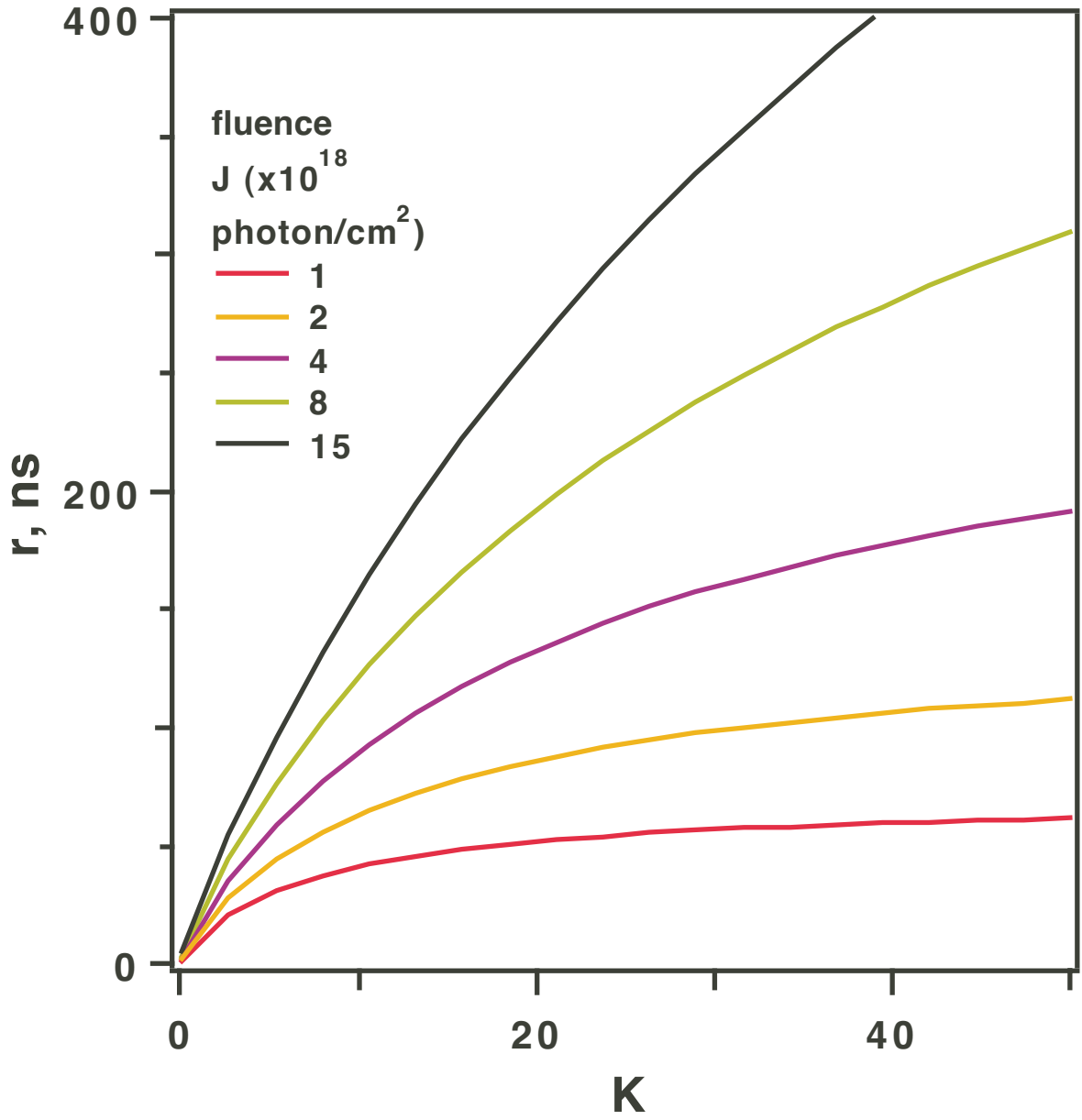

Figure 22S; Shkrob & Sauer